\documentclass[journal]{IEEEtran}
\usepackage{cite}

\ifCLASSINFOpdf
\else
\fi

\usepackage{xcolor}
\usepackage{graphicx}
\usepackage{amsmath, bbm}
\usepackage{amssymb}
\usepackage{amsthm}
\usepackage{booktabs}
\usepackage{multirow}
\usepackage{array,multirow}
\usepackage{cite}
\usepackage{textcomp}
\usepackage{hyperref}
\usepackage{algorithm,algorithmic}
\usepackage{cancel}
\usepackage{tikz}
\usetikzlibrary{arrows, arrows.meta}
\usepackage{setspace}

\newtheorem{theorem}{Theorem}

\DeclareMathAlphabet\mathbfcal{OMS}{cmsy}{b}{n}

\begin{document}
%
\title{Scalable Image Coding for Humans and Machines}

\author{Hyomin~Choi,~\IEEEmembership{Member,~IEEE,}
        and~Ivan~V.~Baji\'{c},~\IEEEmembership{Senior Member,~IEEE}
\thanks{H. Choi and I. V. Baji\'{c} are with the School of Engineering Science, Simon Fraser University, Burnaby, BC, V5A 1S6, Canada. E-mail: chyomin@sfu.ca, ibajic@ensc.sfu.ca}}

\markboth{Submitted for peer review}
{Shell \MakeLowercase{\textit{et al.}}: Bare Demo of IEEEtran.cls for IEEE Journals}

\maketitle

\begin{abstract}
At present, and increasingly so in the future, much of the captured visual content will not be seen by humans. Instead, it will be used for automated machine vision analytics and may require occasional human viewing. Examples of such applications include traffic monitoring, visual surveillance, autonomous navigation, and industrial machine vision. To address such requirements, we develop an end-to-end learned image codec whose latent space is designed to support scalability from simpler to more complicated tasks. The simplest task is assigned to a subset of the latent space (the base layer), while more complicated tasks make use of additional subsets of the latent space, i.e., both the base and enhancement layer(s). 
For the experiments, we establish a 2-layer and a 3-layer model, each of which offers input reconstruction for human vision, plus machine vision task(s), and compare them with relevant benchmarks. The experiments show that our scalable codecs offer 37\%--80\% bitrate savings on machine vision tasks compared to best alternatives, while being comparable to state-of-the-art image codecs in terms of input reconstruction. 
\end{abstract}


\IEEEpeerreviewmaketitle

\section{Introduction}
\IEEEPARstart{A}{dvances} in artificial intelligence (AI) are having a major impact on both image/video coding and computer vision. New research areas are emerging at their intersection, where the goal is to develop compression systems to support both human and machine vision~\cite{duan2020video}. Related standardization activities -- Video Coding for Machines (VCM)~\cite{vcm_call_for_evidence} and JPEG-AI~\cite{JPEG-AI_use_cases} -- have recently been initiated. 

Traditionally, input compression for human vision and feature compression for machine vision have been approached separately or sequentially, through paradigms such as \emph{compress-then-analyze} or \emph{analyze-then-compress}~\cite{duan2020video}. Examples of the former include traditional computer vision, where it is common to train and test vision models on JPEG images, as well as compressed-domain analytics such as~\cite{dfe_compressed_images, edge_detection_dct, nightingale2014deriving, khatoonabadi2012video, choi2017corner, choi2017hevc, alvar2018cantell, alvar2018canfind}, where analysis is performed on conventionally-compressed bitstreams without full decoding or input reconstruction.   
Examples of the latter include Compact Descriptors for Visual Search (CDVS)~\cite{cdvs_std}, where machine vision-relevant features such as  SIFT~\cite{sift} are first extracted from the input, then compressed. In this case, however, reconstructing the input image would require significant additional bitrate.  

Recent deep neural network (DNN)-based image coding methods~\cite{balle2018variational, minnen2018joint, cheng2020image, hu2020coarse, lee2018context, xie2021enhanced} offer competitive rate-distortion (RD) performance against traditional codecs, and their 
perceptual performance
is even more impressive~\cite{clic_dataset,Dandan2021_advances}. DNN-based codecs map the input image into a latent space, which is then 
quantized and arithmetically coded~\cite{witten1987arithmetic}. Meanwhile, DNN computer vision models for  classification~\cite{he2016deep, wang2017residual}, object detection~\cite{ren2015faster, liu2016ssd, Redmon2018_yolov3}, and segmentation~\cite{he2017mask} pass the input image through a sequence of latent spaces while performing analysis. What is interesting about these latent spaces is that they are at least as compressible as the input, which we prove in the Appendix. Hence, combining image compression and DNN-based analysis is theoretically justified. 

\begin{figure}[t]
    \centering
    \begin{minipage}[b]{\linewidth}
    \centering
    \includegraphics[width=\textwidth]{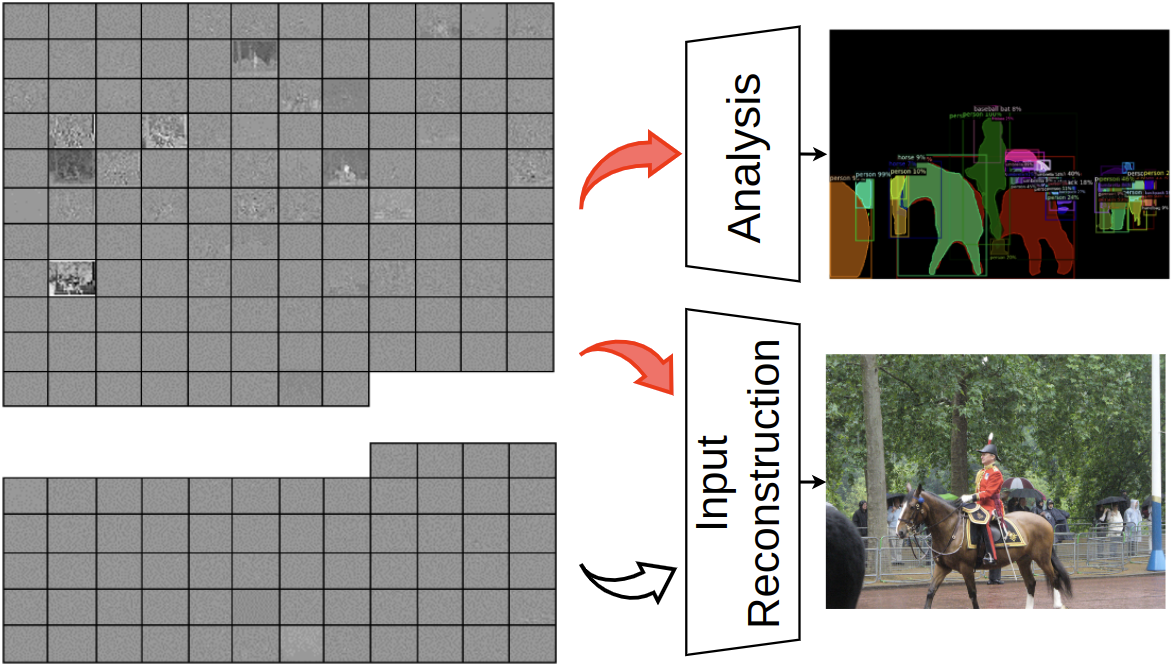}
    \end{minipage}
\caption{An example of latent-space scalability: channels of a feature tensor (left) and the tasks they support (right).} 
\label{fig:intro_proposal}
\end{figure}

Another important trend has been the development of DNN models that support multiple tasks, including input reconstruction, using compressible representations~\cite{torfason2018towards, Choi2018NearLosslessDF, saeed_multi_task_learning, alvar2020bit, alvar2020pareto}. 
In these methods, 
however, the entire latent space must be reconstructed to support any of the tasks. 
The most recent proposals~\cite{hu2020towards, liu2021semantics, choi_lss_icip2} focus on  multi-task scalability. For instance,~\cite{hu2020towards} presented a scalable framework to support facial landmark reconstruction as the base task and input reconstruction as the enhancement task. 
However, both tasks rely on generative~\cite{goodfellow2014generative} decoding, which makes it hard to guarantee reconstruction fidelity. 
Liu \textit{et al.}~\cite{liu2021semantics} present scalable image compression supporting coarse-to-fine classification as well as input reconstruction as the enhancement task. However, in this approach, multiple latent spaces are compressed, leading to inefficiency. The approach proposed in this paper compresses a single latent space, and still achieves scalability among multiple tasks, as illustrated in Fig.~\ref{fig:intro_proposal}. 

This paper is an extension of our recent preliminary work~\cite{choi_lss_icip2}, which presented a 2-layer model based on the YOLOv3~\cite{Redmon2018_yolov3} backbone to support object detection and input reconstruction. In~\cite{choi_lss_icip2}, the enhancement portion of the latent space did not need to be reconstructed, but it still needed to be entropy decoded. Those initial ideas are extended in the present paper in the following ways:
\begin{itemize}
    \item We extend the 2-task model from~\cite{choi_lss_icip2} so that base and enhancement layers are now in separate bitstreams, and the enhancement layer does not need to be decoded to support the base task.
    \item We develop a 3-layer model based on a Feature Pyramid Network (FPN)~\cite{FPN} backbone, which supports object detection, object segmentation, and input reconstruction.
    \item We provide theoretical analysis of latent-space compressibility (Appendix).
    \item We provide additional analysis and mutual information measurements showing latent-space task-specific information steering facilitated by the proposed loss function.
    \item We provide a more extensive experimental validation, including an ablation study and comparison with a wider range of benchmarks.
\end{itemize}

In Section~\ref{sec:literature_review}, we present preliminaries related to DNN-based multi-task image compression. Section~\ref{sec:proposed_method} presents the proposed method for multi-task image coding with latent-space scalability. 
Experimental results are presented in Section~\ref{sec:experimental_results}, followed by conclusions in Section~\ref{sec:conclusion}.

\section{Preliminaries and related work}
\label{sec:literature_review}

\begin{figure}[t]
    \centering
    \begin{minipage}[b]{0.45\linewidth}
    \centering
    \includegraphics[width=\textwidth]{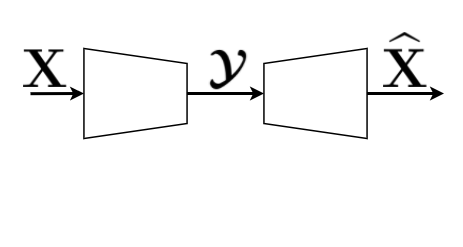}
    \centerline{(a)}\medskip
    \end{minipage}
    \begin{minipage}[b]{0.45\linewidth}
    \centering
    \includegraphics[width=\textwidth]{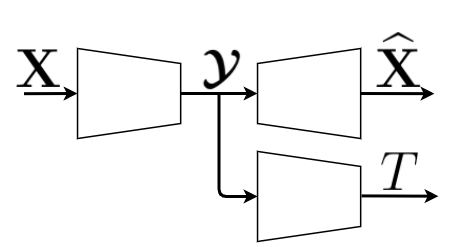}
    \centerline{(b)}\medskip
    \end{minipage}
    \hfill
    
    \centering
    \begin{minipage}[b]{0.45\linewidth}
    \centering
    \includegraphics[width=\textwidth]{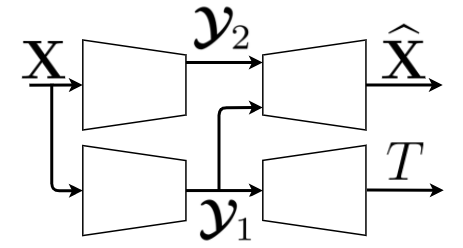}
    \centerline{(c)}\medskip
    \end{minipage}
    \begin{minipage}[b]{0.45\linewidth}
    \centering
    \includegraphics[width=\textwidth]{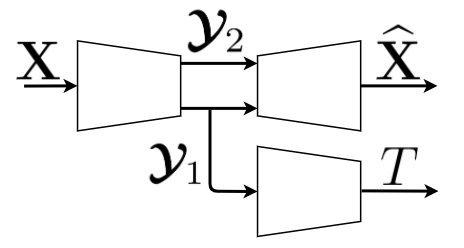}
    \centerline{(d)}\medskip
    \end{minipage}
\caption{Various frameworks of DNN-based compression system. (a) Input compression-only. (b) Multi-task with a single bitstream. (c) Multi-task with separately optimized scalable bitstream. 
(d) Multi-task with jointly optimized scalable bitstream with latent-space scalability.}
\label{fig:various_archi}
\end{figure}

DNN-based image compression approaches~\cite{balle2018variational, minnen2018joint, cheng2020image, hu2020coarse, lee2018context} map the input image $\mathbf{X}$ into a latent representation $\mathbfcal{Y} \in \mathbb{R}^{N \times M \times C}$, as shown in Fig.~\ref{fig:various_archi}(a), where $C$ is the number of latent feature channels and  $N \times M$ is the resolution of each channel. The latent representation $\mathbfcal{Y}$ is compressed, and later decoded to produce the reconstructed input 
$\widehat{\mathbf{X}} \approx \mathbf{X}$. 
Several multi-task DNNs~\cite{torfason2018towards, Choi2018NearLosslessDF, saeed_multi_task_learning} are based on a similar approach, but in addition to input reconstruction $\widehat{\mathbf{X}}$, they enable computer vision inference $T$ (e.g., classification, object detection, etc.) from the latent space without input reconstruction, as shown in Fig.~\ref{fig:various_archi}(b). However, with these approaches, the entire latent representation $\mathbfcal{Y}$ needs to be reconstructed in order to produce either $\widehat{\mathbf{X}}$ or $T$, which is inefficient, as will be discussed in Section~\ref{subsec:motivation}. In particular, it will be seen that the information needed for $T$ is a subset of that needed for $\widehat{\mathbf{X}}$, which is the main motivation behind our proposed latent-space scalability.  

Huo \textit{et al.} in~\cite{hu2020towards}  proposed a scalable multi-task compression framework for face images, whose structure is shown in Fig.~\ref{fig:various_archi}(c). Here, $\mathbfcal{Y}_1$ is the edge map, and $\mathbfcal{Y}_2$ is color information. $\mathbfcal{Y}_1$ is used for face landmark detection $T$, while both $\mathbfcal{Y}_1$ and $\mathbfcal{Y}_2$ are used for reconstructing the input face image $\widehat{\mathbf{X}}$. While this approach gains efficiency from scalability (the base layer $\mathbfcal{Y}_1$ is used for both task), its main drawbacks are that it is specific to face images, and uses generative decoding, which makes $\widehat{\mathbf{X}}$ plausible, but difficult to guarantee fidelity of input reconstruction ($\widehat{\mathbf{X}} \approx \mathbf{X}$). 
Most recently, Liu \textit{et al.}~\cite{liu2021semantics} proposed ``semantic-to-signal'' scalable image compression, which can support multiple classification tasks in a coarse-to-fine manner, but requires coding and reconstruction of multiple latent spaces. 

The approach proposed in this paper, whose preliminary version is described in our recent work~\cite{choi_lss_icip2}, is illustrated in Fig.~\ref{fig:various_archi}(d) (and in more detail in Fig.~\ref{fig:intro_proposal}). A single latent space is split into two parts, $\{\mathbfcal{Y}_1, \mathbfcal{Y}_2\}$, where $\mathbfcal{Y}_1$ represents the base layer and $\mathbfcal{Y}_2$ is the enhancement layer. Only $\mathbfcal{Y}_1$ needs to be decoded to perform inference task $T$, whereas both $\{\mathbfcal{Y}_1, \mathbfcal{Y}_2\}$ are used to reconstruct $\widehat{\mathbf{X}}$. The approach is generic and able to support multiple inference tasks $T$, which we demonstrate by constructing and testing a 3-layer system later in the paper. 


\section{Proposed methods}
\label{sec:proposed_method}

\begin{figure}[b]
    \centering
    \includegraphics[width=\columnwidth]{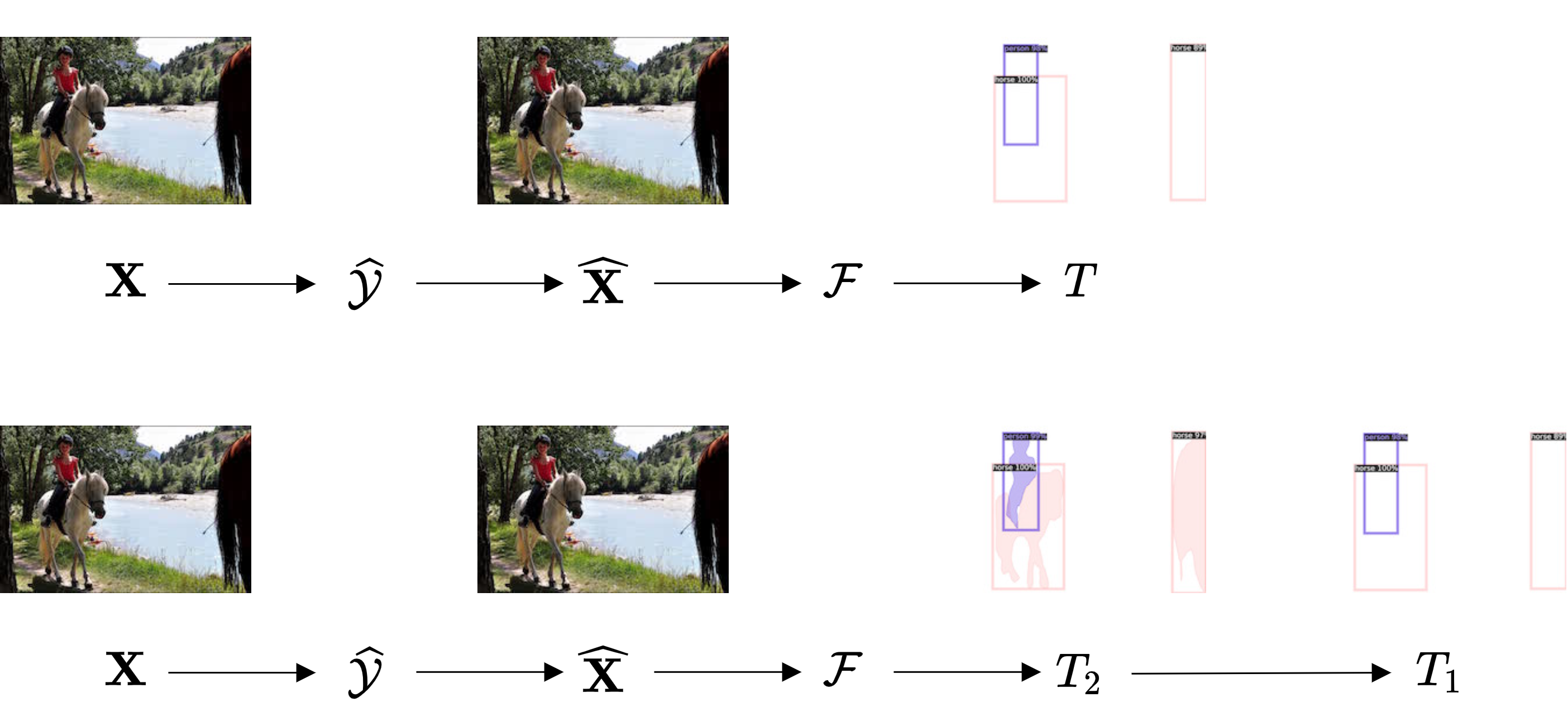}
\caption{Illustration of the data processing inequality for the 2-task model (top) and the 3-task model (bottom).}
\label{fig:DPI_multi-task}
\end{figure}

\begin{figure*}[t]
    \centering
    \begin{minipage}[b]{1.0\linewidth}
    \centering
    \includegraphics[width=\textwidth]{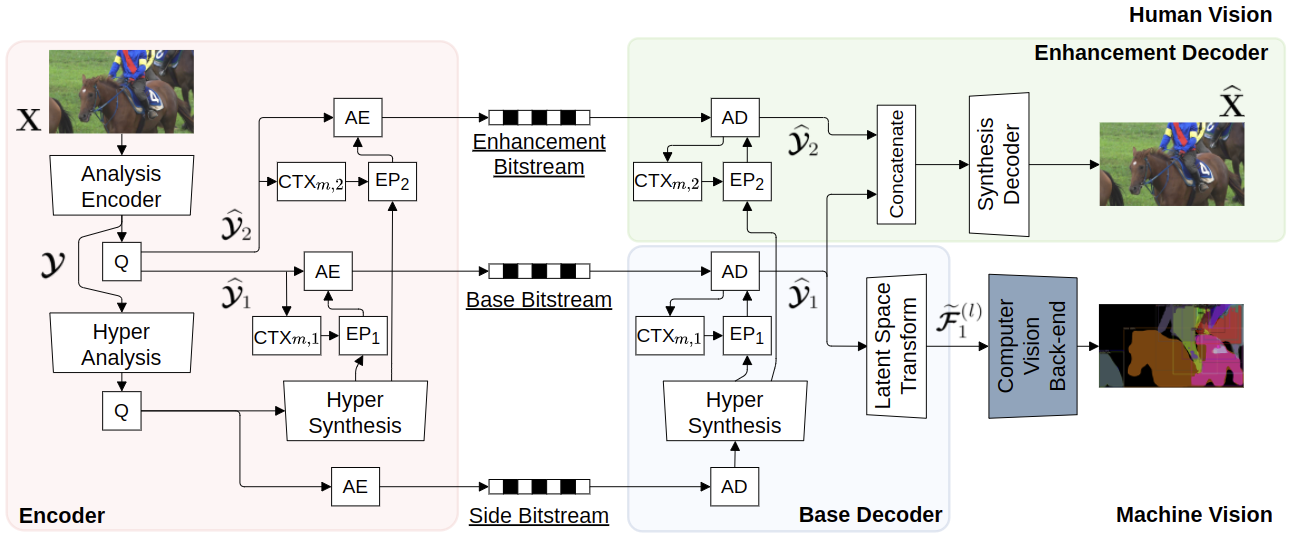}
    \end{minipage}
\caption{An example 2-layer embodiment of the proposed scalable image coding approach. AE/AD represent arithmetic encoder and decoder, respectively. CTX and EP stand for context model and entropy parameters, originally all borrowed from Minnen \textit{et al.}~\cite{minnen2018joint}.}
\label{fig:proposed_frameowrk}
\end{figure*}

\subsection{The backbone}
\label{subsec:motivation}
In developing a scalable DNN-based multi-task compression system, the first step is to choose the backbone that can support the tasks of interest. Since one of the tasks is input reconstruction, we select a recent backbone from~\cite{cheng2020image}, which has shown competitive performance against conventional image codecs. 
Its operation can be described as~\cite{balle2018variational, minnen2018joint, cheng2020image, hu2020coarse, lee2018context}:
\begin{equation}
\begin{split}
 & \mathbfcal{Y} = g_a(\mathbf{X};\phi) \\
 & \widehat{\mathbfcal{Y}} = Q(\mathbfcal{Y}) \\
 & \widehat{\mathbf{X}} = g_s(\widehat{\mathbfcal{Y}};\theta)
\end{split}
\label{eq:transformation_method}
\end{equation}
\noindent where $g_a$ and $g_s$ are the analysis and synthesis transforms, respectively, $\phi$ and $\theta$ are their parameters, $Q$ represents quantization, and the lossless operations of entropy coding and decoding have been omitted from~(\ref{eq:transformation_method}). 

Let $T$ be a machine vision inference task that needs to be derived from the input image $\mathbf{X}$. Since $\widehat{\mathbf{X}} \approx \mathbf{X}$ at high enough rate, this inference can also be derived from $\widehat{\mathbf{X}}$.\footnote{In fact, this is common in practice: computer vision models are usually trained and used on JPEG images, not raw images.} Let $V$ be a DNN that computes $T$, which operates as:
\begin{equation}
\begin{split}
 & \mathbfcal{F} = V^{(\textup{front-end})}(\widehat{\mathbf{X}};\psi) \\
 & T = V^{(\textup{back-end})}({\mathbfcal{F}};\rho)
\end{split}
\label{eq:vision_method}
\end{equation}
\noindent where $\psi$ and $\rho$ are the parameters of the DNN's front-end and back-end, respectively. $V^{(\textup{front-end})}$ consists of the DNN's input layer and a number of subsequent layers, and generates the intermediate feature tensor $\mathbfcal{F}$. $V^{(\textup{back-end})}$ consists of the remaining DNN layers and produces the inference output $T$. 

The processing pipeline described by~(\ref{eq:transformation_method}) and~(\ref{eq:vision_method}) forms a Markov chain $\mathbf{X} \to \widehat{\mathbfcal{Y}} \to \widehat{\mathbf{X}} \to \mathbfcal{F} \to T$. This chain is illustrated in Fig.~\ref{fig:DPI_multi-task}~(top), where the inference task $T$ is object detection. Applying the data processing inequality (DPI)~\cite{Cover_Thomas_2006} to the Markov chain, we get
\begin{equation}
    I(\widehat{\mathbfcal{Y}};\widehat{\mathbf{X}}) \geq I(\widehat{\mathbfcal{Y}};\mathbfcal{F}) \geq I(\widehat{\mathbfcal{Y}};T),
\label{eq:dpi}
\end{equation}
where $I(\cdot \, ; \, \cdot)$ denotes mutual information~\cite{Cover_Thomas_2006}. We can draw two conclusions from this analysis. First, the latent representation $\widehat{\mathbfcal{Y}}$ contains enough information to compute $T$, because it contains enough information to reconstruct $\widehat{\mathbf{X}}$, from which $T$ can be computed. Second, $\widehat{\mathbfcal{Y}}$ carries less information about $T$ than it does about $\widehat{\mathbf{X}}$. This motivates our approach to latent-space scalability - we construct   $\widehat{\mathbfcal{Y}}$ in such a way that only part of it is used to compute $T$, while all of it is used to reconstruct $\widehat{\mathbf{X}}$. Note that the latent space $\widehat{\mathbfcal{Y}}$ is not the same as $\mathbfcal{F}$ (the space from which $V^{(\textup{back-end})}$ computes $T$), so we will also need to construct the latent space transform from $\widehat{\mathbfcal{Y}}$ to $\mathbfcal{F}$ in order to bypass $\widehat{\mathbf{X}}$ when only $T$ is needed. Details of the latent space transforms will be discussed in Section~\ref{subsec:latent_space_transform}.

We will present two embodiments of the above idea: a 2-task (2-layer) system, and a 3-task (3-layer) system. The 2-layer system supports object detection from the base layer and input reconstruction from the full latent space, as explained above. The 3-layer system supports object detection ($T_1$), object segmentation ($T_2$), and input reconstruction ($\widehat{\mathbf{X}}$). The processing chain for such a system can be represented as $\mathbf{X} \to \widehat{\mathbfcal{Y}} \to \widehat{\mathbf{X}} \to \mathbfcal{F} \to T_2 \to T_1$, and is illustrated in Fig.~\ref{fig:DPI_multi-task}~(bottom). Note that it is possible to derive object detection results ($T_1$) from segmentation results ($T_2$) by simply placing bounding boxes around segmented objects and copying the object class. This implies $I(\widehat{\mathbfcal{Y}};T_2) \geq  I(\widehat{\mathbfcal{Y}};T_1)$, and moreover, information required for object detection ($T_1$) is a subset of that required for segmentation  ($T_2$). For this reason, in our 3-layer system, object detection is placed in base layer, segmentation is supported by the base and the first enhancement layer, while the whole latent space supports input reconstruction.

The processing chains in Fig.~\ref{fig:DPI_multi-task} are shown only to illustrate the DPI, and explain the resulting design of the latent space(s) in our systems. In practice, there is no need to reconstruct the input image $\widehat{\mathbf{X}}$ if one wants only to perform object detection or segmentation. This can be achieved directly from the corresponding portion of the latent space through the appropriate latent space transform, as discussed in Section~\ref{subsec:latent_space_transform}.

\subsection{Task-scalable image coding}
\label{subsec:framework}

\begin{figure}[t]
    \centering
    \begin{minipage}[b]{0.8\linewidth}
    \centering
    \includegraphics[width=\textwidth]{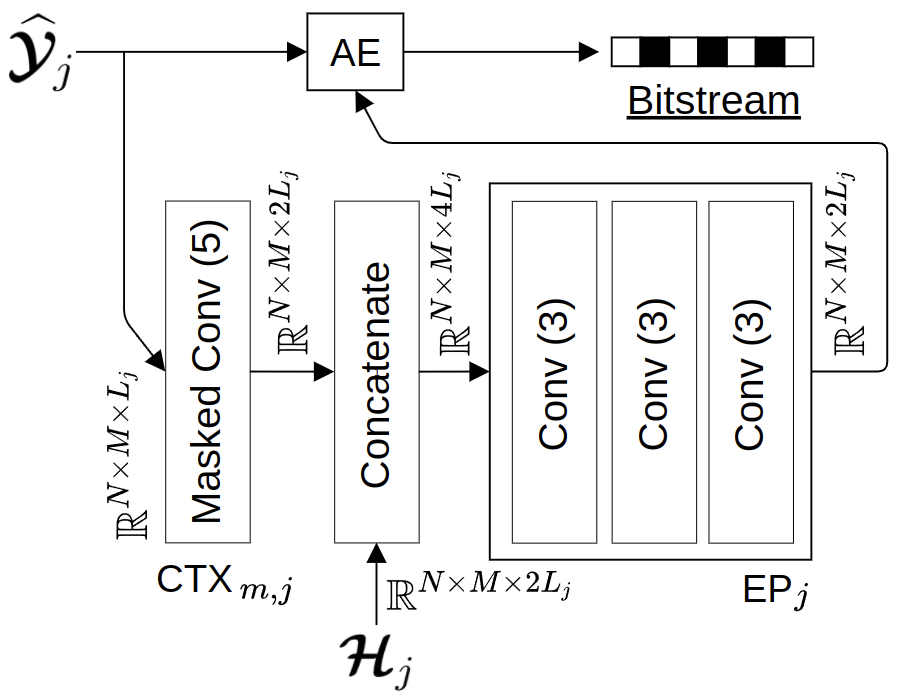}
    \end{minipage}
\caption{Entropy coding-related blocks with tensor dimension specification.
}
\label{fig:proposed_submodule}
\end{figure}

Fig.~\ref{fig:proposed_frameowrk} shows an embodiment of our 2-task (2-layer) system. In particular, we adopt the baseline network~\cite{cheng2020image} using mean and scale hyperprior to capture spatial dependencies in $\mathbfcal{Y}$, estimated by entropy parameter (EP) module and context model (CTX) for arithmetic encoder/decoder (AE/AD). Also, we adopt the configurations for Analysis Encoder, Synthesis Decoder, and Hyper Analysis/Synthesis without attention layers from~\cite{cheng2020image}. 
The Analysis Encoder transforms the input image $\mathbf{X}$ into $\mathbfcal{Y} \in \mathbb{R}^{N \times M \times C}$, 
with $C=192$ as in~\cite{minnen2018joint, cheng2020image}. The latent representation $\mathbfcal{Y}$ is split into sub-latents $\{\mathbfcal{Y}_1, \mathbfcal{Y}_2\}$, where $\mathbfcal{Y}_1 = \{\mathbf{Y}_1, \mathbf{Y}_2,...\mathbf{Y}_i\}$ represents the base layer and $\mathbfcal{Y}_2 = \{\mathbf{Y}_{i+1}, \mathbf{Y}_{i+2},...,\mathbf{Y}_C\}$ represents the enhancement layer. 
The quantized sub-latent  $\widehat{\mathbfcal{Y}}_j$ is coded using its own 
context model (CTX$_{m,j}$) and entropy parameter module (EP$_j$). The side bitstream, which contains coded hyper-parameters, is used by both layers. 

Details of entropy coding related-blocks along with dimensions of relevant tensors are presented in Fig.~\ref{fig:proposed_submodule}. Let $\widehat{\mathbfcal{Y}}_j\in \mathbb{R}^{N \times M \times L_j}$ be the $j$-th quantized sub-latent, where 
$L_j$ is the number of channels in the $j$-th sub-latent. CTX$_{m,j}$ is the context model for $\widehat{\mathbfcal{Y}}_j$ associated with a masked convolutional layer~\cite{minnen2018joint, cheng2020image} with $5 \times 5$ kernels (Masked Conv (5) in the figure), to produce an output tensor with $2L_j$ channels. This output tensor is concatenated with $\mathbfcal{H}_j \in \mathbb{R}^{N \times M \times 2L_j}$, which is the part of Hyper Synthesis output $\mathbfcal{H} \in \mathbb{R}^{N \times M \times 2C}$ corresponding to $\widehat{\mathbfcal{Y}}_j$. The concatenated tensor is input to EP$_j$, which consists of three convolutional layers with $3\times3$ kernels (Conv (3) in the figure), to estimate Gaussian means and variances 
for subsequent entropy coding. Quantized hyperpriors for all sub-latents (input to Hyper Synthesis) are encoded as side bitstream. 
We experimentally observed that the side bitstream accounts for 2--3\% of total bits, on average. 

At the decoder, hyperpriors are reconstructed from the side bitstream. The base-layer sub-latent $\widehat{\mathbfcal{Y}}_1$ is reconstructed from the the base bitstream using the hyperpriors. $\widehat{\mathbfcal{Y}}_1$ is sufficient for object detection. In case input reconstruction is also needed, $\widehat{\mathbfcal{Y}}_2$ is reconstructed from the enhancement bitstream using the hyperpriors, and the input image is reconstructed from $\widehat{\mathbfcal{Y}}_1 \cup \widehat{\mathbfcal{Y}}_2$.

Although this example involves a 2-task (2-layer) system for simplicity of explanation, it is possible to construct  scalable systems with more layers by following the same principles. 
In fact, in Section~\ref{sec:experimental_results}, we also demonstrate a 3-task (3-layer) system that supports object detection (base layer), object segmentation (first enhancement layer) and input reconstruction (second enhancement layer). In this system, the latent space $\widehat{\mathbfcal{Y}}$ is partitioned into three parts: $\{\widehat{\mathbfcal{Y}}_1,\widehat{\mathbfcal{Y}}_2,\widehat{\mathbfcal{Y}}_3\}$, where $\widehat{\mathbfcal{Y}}_1$ supports object detection, $\widehat{\mathbfcal{Y}}_1 \cup \widehat{\mathbfcal{Y}}_2$ supports object segmentation, and the whole latent space $\widehat{\mathbfcal{Y}}_1 \cup \widehat{\mathbfcal{Y}}_2 \cup \widehat{\mathbfcal{Y}}_3$ supports input reconstruction.

\subsection{Task-dependent latent space transform}
\label{subsec:latent_space_transform}

\begin{figure}[t]
    \centering
    \begin{minipage}[b]{0.8\linewidth}
    \centering
    \includegraphics[width=\textwidth]{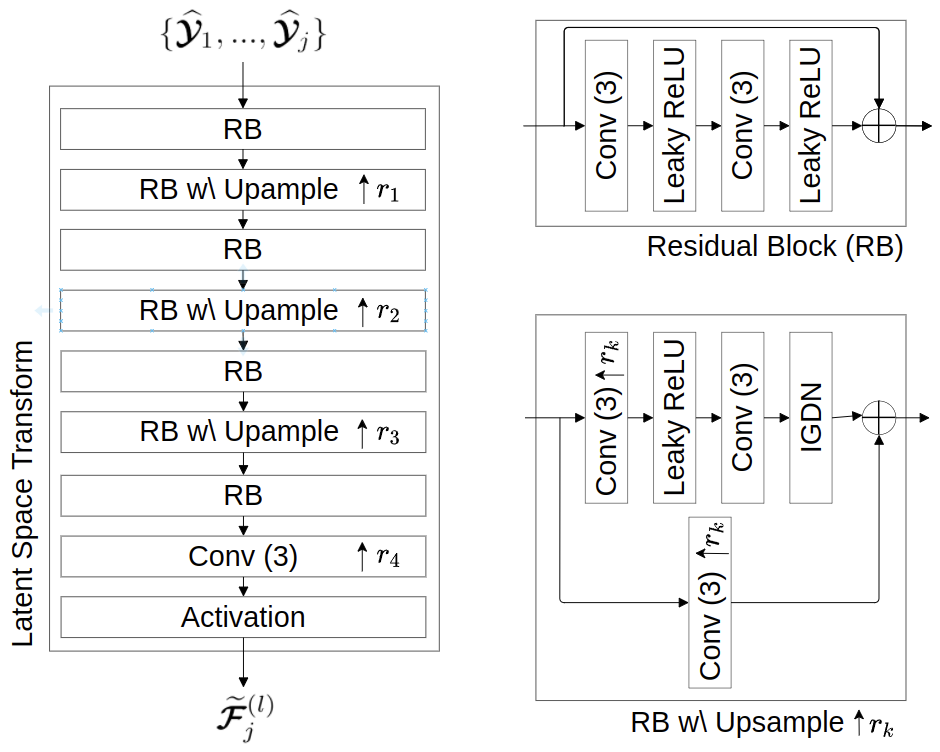}
    \end{minipage}
\caption{Architecture of the latent space transform (LST) block}
\label{fig:latent_transform_module}
\end{figure}

The base layer $\widehat{\mathbfcal{Y}}_1$ 
will support the base computer vision task, say object detection, for example. However, $\widehat{\mathbfcal{Y}}_1$, in general, does not match any of intermediate latent spaces of any of the object detectors. Hence, we introduce the latent space transform (LST) to map $\widehat{\mathbfcal{Y}}_1$ to an intermediate latent space of an object detector we want to use at the decoder. 
Specifically, let $\mathbfcal{F}_1^{(l)}$ be the feature tensor at the output of the $l$-th layer of a chosen object detector $V_1$, that is,   $\mathbfcal{F}_1^{(l)} = V_1^{(\textup{front-end})} (\widehat{\mathbf{X}}, \psi)$, as  in~(\ref{eq:vision_method}). 
Then the goal of the base LST is to map $\widehat{\mathbfcal{Y}}_1$ to $\mathbfcal{F}_1^{(l)}$. More generally, for the $j$-th task, we use $\{\widehat{\mathbfcal{Y}}_1, ..., \widehat{\mathbfcal{Y}}_j\}$ from the encoded latent space, and the goal of $j$-th LST is to map $\{\widehat{\mathbfcal{Y}}_1, ..., \widehat{\mathbfcal{Y}}_j\}$ to $\mathbfcal{F}_j^{(l)} = V_j^{(\textup{front-end})} (\widehat{\mathbf{X}}, \psi)$, where $V_j$ is the DNN implementing the $j$-th task. 

Fig.~\ref{fig:latent_transform_module} shows the structure of our LST, which consists of residual blocks (RBs) similar to those in the Synthesis Decoder~\cite{cheng2020image}. 
A RB consists of two convolutional layers with $3\times3$ kernels (Conv (3) in the figure) followed by Leaky ReLU. A RB with upsampling factor $r_k$ (shown as $\uparrow r_k$) increases spatial dimension by $r_k$ 
and also applies the inverse GDN~\cite{balle2015density}. 
Upsampling factors $r_k$ are chosen based on the dimensions of $\{\widehat{\mathbfcal{Y}}_1, ..., \widehat{\mathbfcal{Y}}_j\}$ and  $\mathbfcal{F}_j^{(l)}$. The last activation function is set to be the same as for the output of $V_j^{(\textup{front-end})}$, to produce a similar dynamic range for the output tensor. The output of the LST is $\widetilde{\mathbfcal{F}}^{(l)}_j$, an approximation to the desired $\mathbfcal{F}_j^{(l)}$, and the LST is trained to minimize their difference. 

\subsection{Information steering through learning}
\label{subsec:loss_function}

To steer the task-relevant information into the corresponding parts of the encoder latent space $\widehat{\mathbfcal{Y}}$, we construct a loss function 
in the form of a rate-distortion Lagrangian
\begin{equation}
    \mathcal{L} = R + \lambda \cdot D,
    \label{eq:RD_loss}
\end{equation}
where $R$ is the rate estimate, $D$ is the combined distortion of various tasks, 
and $\lambda$ is the Lagrange multiplier. 
As in~\cite{minnen2018joint}, 
\begin{equation}
R=\underbrace{\mathbb{E}_{x\sim p_x}\left [ -\textup{log}_{2}p_{\hat{y}}(\hat{y}) \right ]}_{\textup{latent}}+\underbrace{\mathbb{E}_{x\sim p_x}\left [ -\textup{log}_{2}p_{\hat{z}}(\hat{z}) \right ]}_{\textup{hyper-priors}},
\label{eq:rate_term}
\end{equation}
where $x$ denotes input data, $\hat{y}$ is the quantized latent data and $\hat{z}$ is the quantized hyper-prior. 
Distortion $D$ is computed as 
\begin{equation}
D = \textup{MSE}(\mathbf{X},\widehat{\mathbf{X}}) + \sum_{j=1}^{S-1}\gamma_j \cdot \textup{MSE}\left(\mathbfcal{F}^{(l)}_{j},\widetilde{\mathbfcal{F}}^{(l)}_{j}\right),
\label{eq:distortion_term}
\end{equation}
\noindent where $\gamma_j$ are scale factors for various task distortions and 
$S$ is the total number of tasks (i.e., layers). 
Besides training the model to accurately reconstruct task-relevant latent spaces, distortion terms $\textup{MSE}\left(\mathbfcal{F}^{(l)}_{j},\widetilde{\mathbfcal{F}}^{(l)}_{j}\right)$ also serve to steer the task-relevant information into the corresponding portions of the encoder latent space $\widehat{\mathbfcal{Y}}$. This is because $\widetilde{\mathbfcal{F}}^{(l)}_{j}$ only depends on $\{\widehat{\mathbfcal{Y}}_1, ..., \widehat{\mathbfcal{Y}}_j\}$ (through LST) and not the rest of the encoder latent space $\widehat{\mathbfcal{Y}} \setminus \{\widehat{\mathbfcal{Y}}_1, ..., \widehat{\mathbfcal{Y}}_j\}$. Hence, information relevant to task $j$ is backpropagated to $\{\widehat{\mathbfcal{Y}}_1, ..., \widehat{\mathbfcal{Y}}_j\}$. Further insight is provided through information flow measurements below.


\subsection{Information flow}
\label{subsec:info_flow}

In this section we provide further insight into the operation of the proposed scalable framework from an information-theoretic perspective. Empirical information measurements are performed on a 2-layer system whose base task is object detection using a YOLOv3~\cite{Redmon2018_yolov3} back-end, and the enhancement task is input reconstruction. The LST maps the base-layer features $\mathbfcal{Y}_1$ to $\widetilde{\mathbfcal{F}}_1^{(13)}$ of YOLOv3. The network was trained for 300 epochs with $\gamma_1=0.006$ in~(\ref{eq:distortion_term}). Other training details are presented in Section~\ref{ssec:experiments_training}.

\begin{figure}[t]
    \centering
    \begin{minipage}[b]{1\linewidth}
    \centering
    \includegraphics[width=\textwidth]{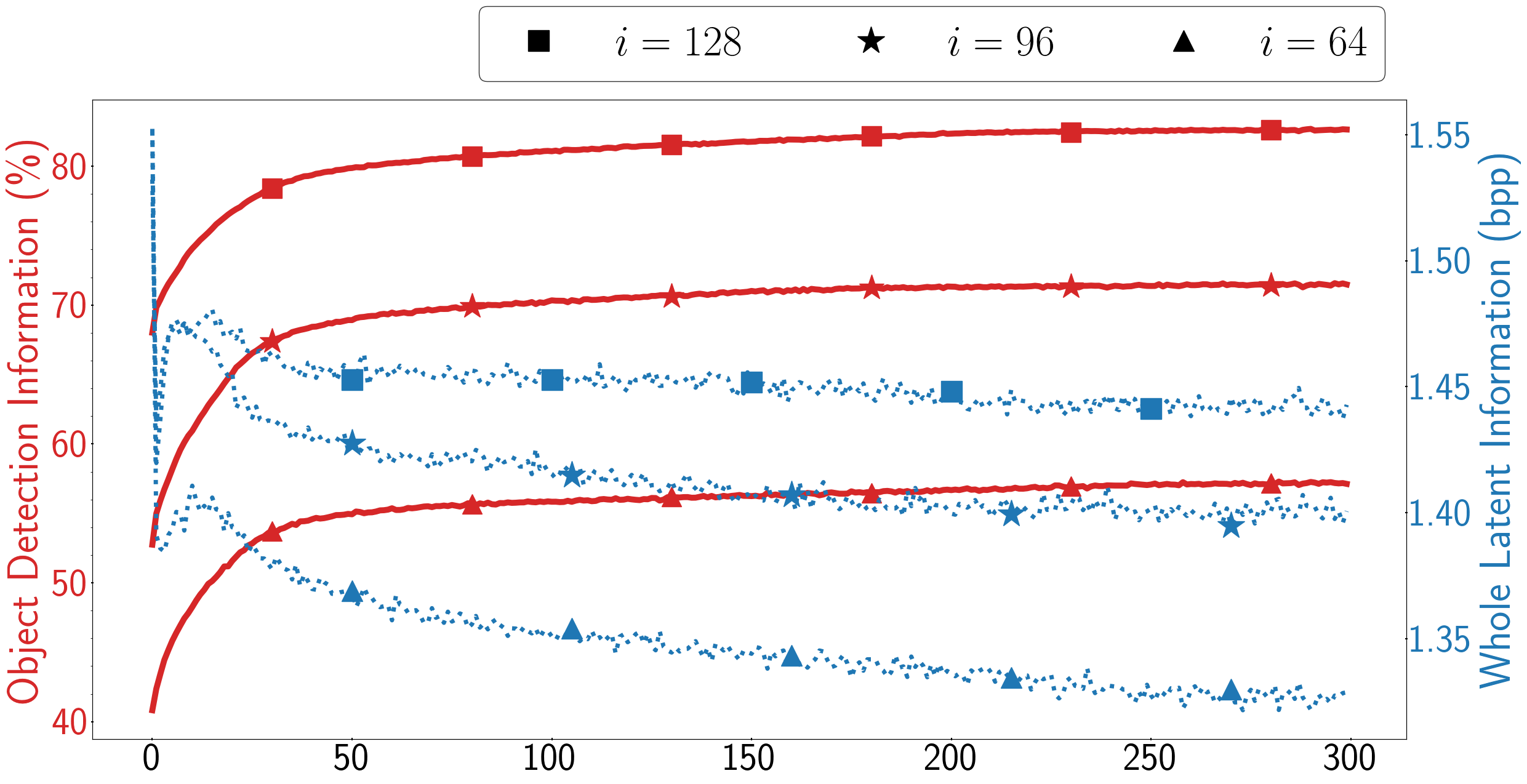}
    \centerline{(a)}\medskip
    \end{minipage}
    \centering
    \begin{minipage}[b]{1\linewidth}
    \centering
    \includegraphics[width=\textwidth]{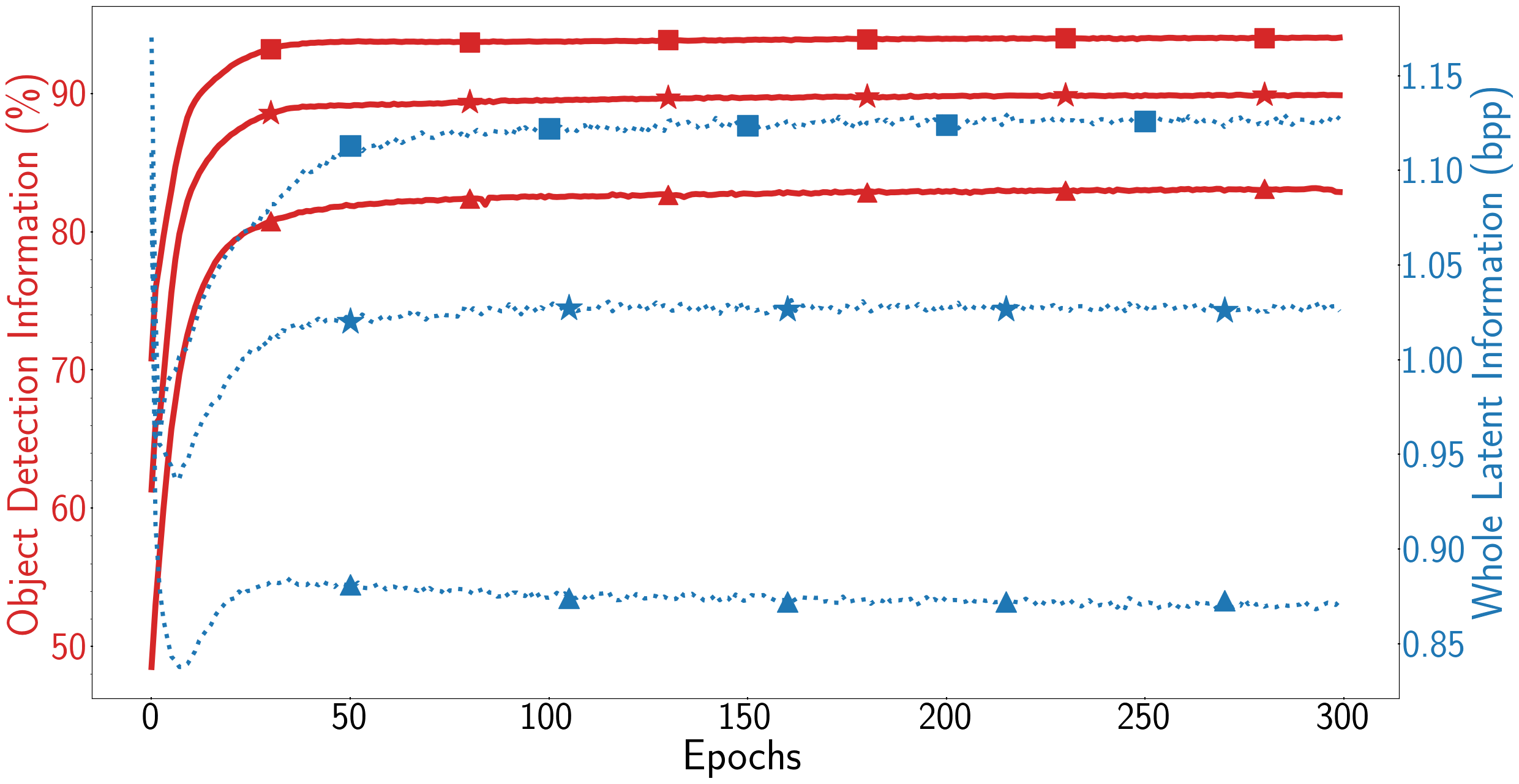}
    \centerline{(b)}\medskip
    \end{minipage}
\vspace{-0.5cm}    
\caption{Evolution of object detection information and total rate through training, when the number of latent-space channels dedicated to object detection is $i \in \{64, 96, 128\}$. Blue curves show the total rate $H(\mathbfcal{Y})$, with the scale shown on the right vertical axis. Red curves show the percentage of object detection information~(\ref{eq:obj_information}) out of the total rate, with the scale shown on the left vertical axis.  Sub-figure (a) corresponds to the model trained with $\lambda=0.0483$ in~(\ref{eq:RD_loss}), and (b) to the model trained with $\lambda=0.0035$ in~(\ref{eq:RD_loss}).}
\label{fig:latent_entropy}
\end{figure}

Entropy estimates of the base features $\mathbfcal{Y}_1$ and the entire latent representation $\mathbfcal{Y}$ are obtained using rate estimates~\cite{minnen2018joint}

\begin{equation}
\begin{split}
H(\mathbfcal{Y}_1) & \approx \, \mathbb{E}_{x\sim p_x}\left [ -\textup{log}_{2}p_{\hat{y}_1}(\hat{y}_1) \right ],\\
H(\mathbfcal{Y}) & \approx \, \mathbb{E}_{x\sim p_x}\left [ -\textup{log}_{2}p_{\hat{y}}(\hat{y}) \right ].
\end{split}
\label{eq:entropy_y}
\end{equation}

Fig.~\ref{fig:latent_entropy} shows the results. The right-hand side Y-axis shows $H(\mathbfcal{Y})$ in bits per pixel (bpp), computed as total bits divided by input image resolution.
The left-hand side Y-axis shows the percentage of object detection information out of the total rate, that is
\begin{equation}
    \frac{H(\mathbfcal{Y}_1)}{H(\mathbfcal{Y})} \cdot 100\%.
\label{eq:obj_information}
\end{equation}
Fig.~\ref{fig:latent_entropy}(a) shows the results for $\lambda=0.0483$ in~(\ref{eq:RD_loss}), while Fig.~\ref{fig:latent_entropy}(b) shows the results for $\lambda=0.0035$. In both cases, the number of base channels $i\in\{64, 96, 128\}$. 

From Fig.~\ref{fig:latent_entropy}, it can be seen that as $\lambda$ decreases from $0.0483$ in Fig.~\ref{fig:latent_entropy}(a) to $0.0035$ in Fig.~\ref{fig:latent_entropy}(b), the total rate reduces from the range $[1.45, 1.55]$ bpp down to $[1.10, 1.17]$ bpp. This is clear from~(\ref{eq:RD_loss}) because, as $\lambda$ decreases, the importance of $R$ in the Lagrangian cost increases, so there is more pressure to reduce the rate. The same happens with the object detection information rate. With $i=128$ feature channels dedicated to object detection, in Fig.~\ref{fig:latent_entropy}(a) the object detection information rate is approximately $0.8\cdot1.45=1.16$ bpp, and this falls down to approximately $0.92\cdot1.13=1.04$ bpp in Fig.~\ref{fig:latent_entropy}(a), after 300 epochs. The same trend holds for other $i$'s. 

We can also notice that the fraction of the total rate devoted to the base layer does not scale proportionally to the fraction of the latent space occupied by the base layer. For example, when the number of base layer channels halves from $i=128$ to $i=64$, the percentage of the total rate it occupies only reduces from $80\%$ to $55\%$ in Fig.~\ref{fig:latent_entropy}(a), and from from $92\%$ to $82\%$ in Fig.~\ref{fig:latent_entropy}(b). 
This is due to the fact that the base layer serves two purposes - it is responsible for object detection, and also plays a role in input reconstruction. Hence, through RD optimization, it receives more rate than would be expected from the fraction of the latent space it occupies.

It is well known that scalable extensions of conventional codecs, such as H.264/AVC~\cite{avc_std} and H.265/HEVC~\cite{hevc_std_2019}, suffer a rate 
increase of about 15-25\% per enhancement layer compared to 
single-layer coding
~\cite{shvc_overview}  in terms of the rate-distortion performance. This is due to information redundancy between layers. 
It is natural to ask whether such redundancy exists in our scalable coding framework? 
Ideally,  $I(\widehat{\mathbfcal{Y}}_1; \widehat{\mathbfcal{Y}}_2) = 0$, meaning that base and enhancement layer are independent, and they can be coded separately without loss of coding efficiency. If this were the case, then because $\widehat{\mathbfcal{Y}}_1 \to \widetilde{\mathbfcal{F}}_{1}^{(13)}$ through LST, we would have $I(\widehat{\mathbfcal{Y}}_2; \widetilde{\mathbfcal{F}}_{1}^{(13)})=0$. To test this, we measure the mutual information between $\widehat{\mathbfcal{Y}}_j, j\in\{1,2\}$ and  $\widetilde{\mathbfcal{F}}_{1}^{(13)}$. Mutual information measurements are obtained using the Kernel Density Estimator (KDE) from~\cite{saxe}, as follows. 

\begin{figure}[t]
    \centering
    \begin{minipage}[b]{0.8\linewidth}
    \centering
    \includegraphics[width=\textwidth]{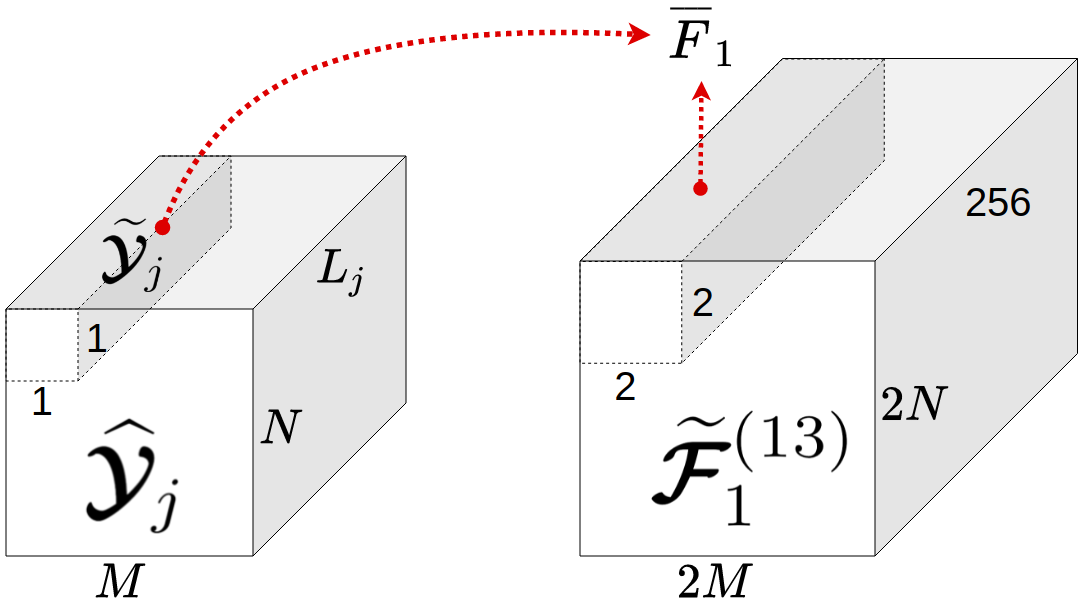}
    \end{minipage}
\caption{Mutual information estimation between $\widehat{\mathbfcal{Y}}_j$ and $\widetilde{\mathbfcal{F}}_1^{(13)}$}
\label{fig:mi_estimation}
\end{figure}

\begin{figure}[t]
    \centering
    \begin{minipage}[b]{1\linewidth}
    \centering
    \includegraphics[width=\textwidth]{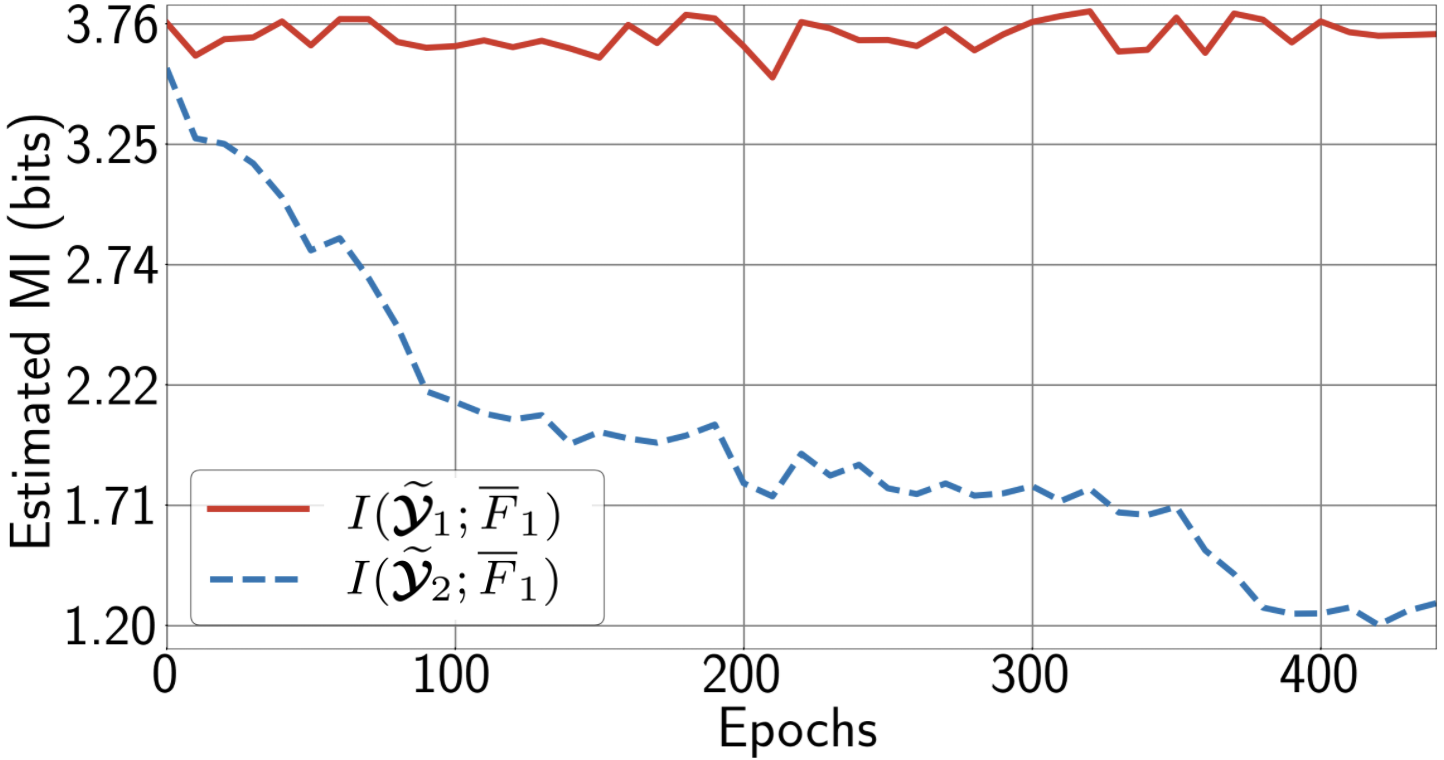}
    \end{minipage}
\caption{Estimated mutual information (MI)  $I(\widetilde{\mathbfcal{Y}}_1;\overline{F}_1)$ and $I(\widetilde{\mathbfcal{Y}}_2;\overline{F}_1)$} 
\label{fig:evolution_mi_estimation}
\end{figure}

We first divide sub-latent tensors $\widehat{\mathbfcal{Y}}_j$ into fibers $\widetilde{\mathbfcal{Y}}_j \in \mathbb{R}^{1 \times 1 \times L_j}$, where $L_1 = 128$ and $L_2 = 64$. Also  $\widetilde{\mathbfcal{F}}_{1}^{(13)}$ is divided into spatially corresponding fibers with the size of $2 \times2 \times 256$, as illustrated in Fig.~\ref{fig:mi_estimation}. Then, we cluster the estimated output feature fibers into $K=16$ clusters (similarly  to~\cite{Lee_ISBI2021}) with cluster index denoted $\overline{F}_1$, and then compute
\begin{equation}
\begin{split}
I(\widetilde{\mathbfcal{Y}}_j;\overline{F}_1) & = \, H(\widetilde{\mathbfcal{Y}}_j) - {H(\widetilde{\mathbfcal{Y}}_j|\overline{F}_1)} \\
& = \, H(\widetilde{\mathbfcal{Y}}_j) - \sum_{\bar{f}_1 \in \{1,...,K\}}p(\bar{f}_1)\cdot H(\widetilde{\mathbfcal{Y}}_j|\overline{F}_1=\bar{f}_1).
\end{split}
\label{eq:mi_estimation}
\end{equation}
Due to clustering, $I(\widetilde{\mathbfcal{Y}}_j;\overline{F}_1)$ is an underestimate of $I(\widetilde{\mathbfcal{Y}}_j;\widetilde{\mathbfcal{F}}_{1}^{(13)})$, but the measurements still provide useful insight into the behavior of the system. Fig.~\ref{fig:evolution_mi_estimation} shows the evolution of the estimated mutual information over 400 epochs for $\lambda=0.0483$ and $i=128$. 
Since $\widehat{\mathbfcal{Y}}_1 \to \widetilde{\mathbfcal{F}}_{1}^{(13)}$, we see that $I(\widetilde{\mathbfcal{Y}}_1;\overline{F}_1)$ remains relatively high, around 3.76 bits per fiber, throughout the 400 epochs. Meanwhile, $I(\widetilde{\mathbfcal{Y}}_2;\overline{F}_1)$ reduces down to 1.20 bits per fibre, showing that the enhancement layer carries less and less information about object detection as the training goes on. While $I(\widetilde{\mathbfcal{Y}}_2;\overline{F}_1)$ does not reach the ideal value of zero, the decreasing trend is evident, which confirms that the loss function~(\ref{eq:RD_loss})-(\ref{eq:distortion_term}) provides appropriate information steering to various portions of the latent space. At the same time, since the base and enhancement latents are coded independently, a non-zero mutual information between them is an indication of inefficiency of scalable coding. Indeed, as will be seen in Section~\ref{ssec:evaluation_human_vision}, in terms of input reconstruction, our 2-layer system is less efficient than~\cite{cheng2020image}, while our 3-layer system is less efficient than our 2-layer system, even though all three systems share the same encoder architecture. This loss of efficiency due to scalability has also been observed in earlier studies on scalable coding~\cite{shvc_overview}, so it is not a new phenomenon. We believe our approach of looking at the problem through the lens of mutual information will contribute to further advances in this area.


\subsection{Extensions to new tasks}
\label{ssec:new_tasks}
One may also ask what happens if a new task needs to be added to the system post-hoc, after the initial system has already been trained? This is a challenge for all coding-for-machine approaches, because the features trained for initial tasks may not be appropriate for the new task. In general, some modifications need to be made, but there are several possibilities, depending on how efficient the new system needs to be and whether or not it is possible to retrain the encoder. Four our proposed systems, the possibilities are as follows:

\begin{enumerate}
    \item If the new task happens to be a subset of an existing task,\footnote{For example, an existing task is detection of 10 different breads of dogs and 5 breeds of cats, while the new task is just detecting `dogs' and `cats'.} then one can simply train a new LST from the portion of the latent space dedicated to the previous task into the latent space of the new back-end, because all the information is already available there. In this case, encoder does not need to be retrained, only the LST.
    \item If the new task is not a subset of any of the previous tasks, one could retrain the encoder to include the new task and allocate a portion of the latent space to it.
    \item As shown in Section~\ref{subsec:motivation}, any task $T$ is a subset of the input reconstruction task. Therefore, one can simply train a LST from the full latent space into the latent space of the new back-end. This would avoid retraining the encoder, but the new task would likely operate at a higher bitrate compared to approach 2) above, because the full latent space needs to be decoded. 
\end{enumerate}

\section{Experiments}
\label{sec:experimental_results}
Here we describe the experiments to assess the performance of 2- and 3-layer networks presented earlier. Our choice of machine vision tasks (object detection, object segmentation) and the corresponding back-end models (YOLOv3~\cite{Redmon2018_yolov3} and FPN~\cite{FPN}) is motivated by the MPEG VCM activities~\cite{vcm_call_for_evidence}, where these are some of the recommended tasks and models. Individual tasks of each network are evaluated on the relevant datasets and the results are compared with the relevant benchmarks. Both 2- and 3-layer networks are trained on the same datasets using a two-stage training strategy described below. 

\subsection{Training setup}
\label{ssec:experiments_training}

\begin{figure*}[t]
    \centering
    \begin{minipage}[b]{0.45\linewidth}
    \centering
    \includegraphics[width=\textwidth]{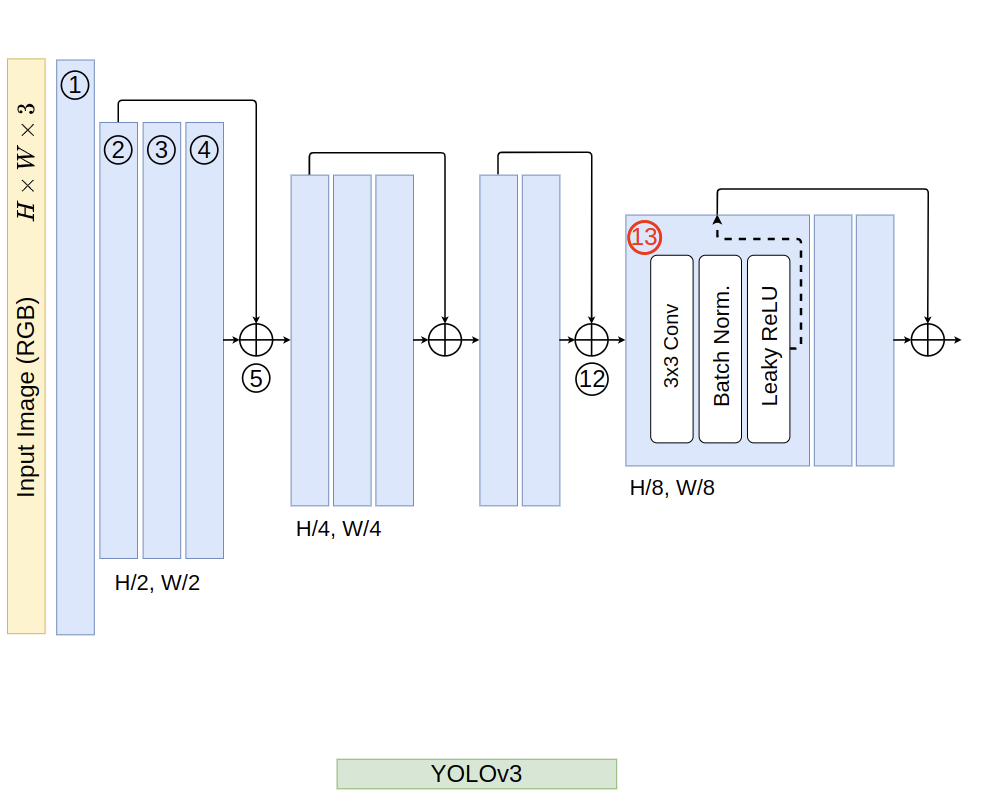}
    \centerline{(a)}\medskip
    \end{minipage}
    \centering
    \begin{minipage}[b]{0.45\linewidth}
    \centering
    \includegraphics[width=\textwidth]{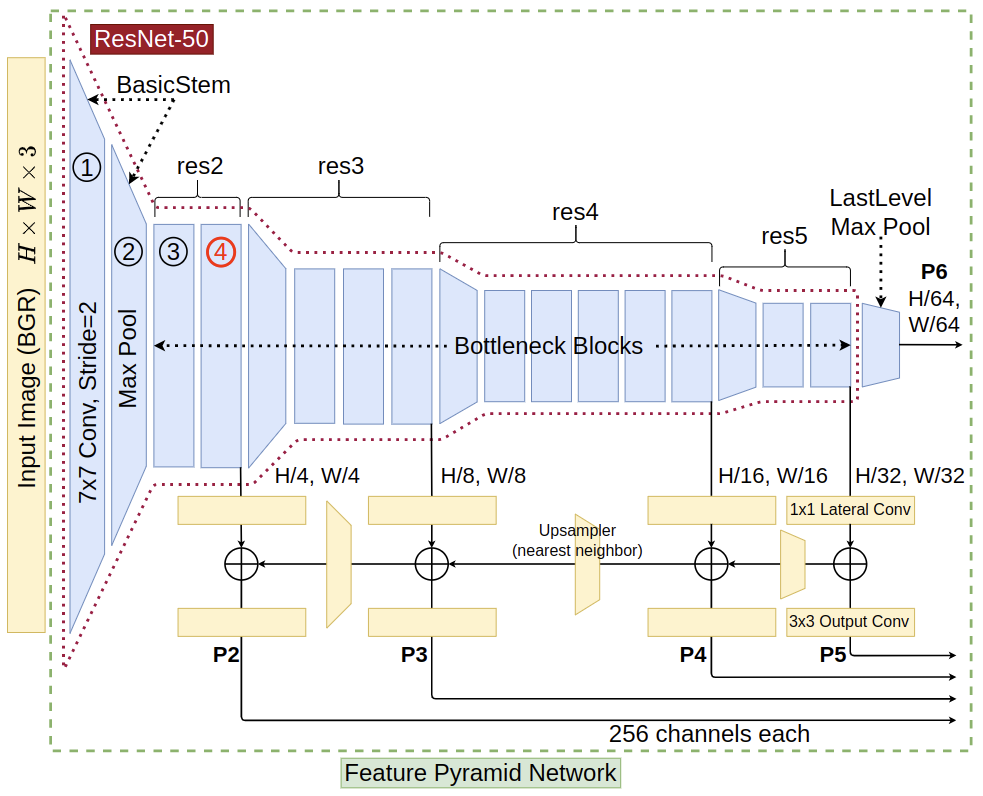}
    \centerline{(b)}\medskip
    \end{minipage}
\vspace{-0.5cm}
\caption{Part of network architectures for targeted vision tasks: (a) YOLOv3~\cite{Redmon2018_yolov3} and (b) Feature Pyramid Network used for Faster~\cite{ren2016faster} and Mask~\cite{he2017mask} R-CNN }
\label{fig:vision_task_networks}
\end{figure*}

Our multi-task networks are trained in two stages. In the first stage, the networks are trained on CLIC~\cite{clic_dataset} and JPEG-AI~\cite{jpeg_ai_dataset} datasets. Randomly cropped patches with size of $256 \times 256$ from both datasets are used as input. We also set the mini-batch size to 16. Adam optimizer with a fixed learning rate of $10^{-4}$ is used for 400 epochs on a GeForce RTX 2080 GPU with 11 GB RAM. Then, we change the dataset to VIMEO-90K~\cite{xue2019video_vimeo} to continue the training for another 300-400 epochs in the second stage. Likewise, randomly cropped patches are drawn from the dataset, but this time the learning rate decreases with polynomial decay for every 10 epochs. Six values of $\lambda$, shown in Table~\ref{tbl:lambda_values}, are used in~(\ref{eq:RD_loss}) to produce six versions of the trained networks, as in~\cite{begaint2020compressai}. The total number of latent-space channels is set to $C=192$ in each case, as in~\cite{balle2018variational, minnen2018joint, cheng2020image}. Specific details of the 2- and 3-layer networks are given below. 

\begin{table}[t]
\centering
\caption{$\lambda$ values for training models}
\label{tbl:lambda_values}
\smallskip\noindent
\resizebox{1.0\linewidth}{!}{%
\begin{tabular}{@{}ccccccc@{}}
\toprule
Quality Index & 1      & 2      & 3      & 4     & 5     & 6 \\ \midrule
$\lambda$   & 0.0018 & 0.0035 & 0.0067 & 0.013 & 0.025 & 0.0483 \\ \bottomrule
\end{tabular}}
\end{table}

\textbf{Two-layer network:} The 192 latent-space channels are split into $L_1=128$ base-layer channels and $L_2=64$  enhancement layer channels. A LST in the base layer maps the base features $\widehat{\mathbfcal{Y}}_1 \in \mathbb{R}^{N \times M \times 128}$ 
to the target features at the convolution 
output of layer 13 of YOLOv3, $\widetilde{\mathbfcal{F}}_1^{(13)} \in \mathbb{R}^{2N \times 2M \times 256}$. To match the resolution of the target feature tensor, the LST scaling factors $r_k$, $k\in\{1,2,3,4\}$, are set to 2, 1, 1, and 1, respectively. The processing at layer 13 of YOLOv3 includes a convolutional layer followed by batch normalization and Leaky ReLU activation, as shown in Fig~\ref{fig:vision_task_networks}(a). Since the 
pre-trained weights of YOLOv3~\cite{Redmon2018_yolov3} represent prior knowledge obtained over the training data, we keep and re-use them for the batch normalization followed by an activation function at layer 13, so the last layer of the LST simply adopts a linear activation. The estimate of the layer 13 convolution output, produced by our LST, is then used as input to the batch normalization with the learned weights at layer 13. To account for this, the distortion equation is slightly modified from~(\ref{eq:distortion_term}), to
\begin{equation}
\begin{split}
D = \, \textup{MSE}(\mathbf{X},\widehat{\mathbf{X}}) 
 + \gamma \cdot \textup{MSE}\left(\mathbfcal{F}^{(13)}_{1},V^{(13)}(\widetilde{\mathbfcal{F}}^{(13)}_{1},\rho^{*})\right),
\end{split}
\label{eq:m12_3_changed_distortion_term}
\end{equation}
\noindent where $\gamma=0.006$ and $V^{(13)}$ includes batch normalization followed by Leaky ReLU activation, with pre-trained weights $\rho^*$ from~\cite{Redmon2018_yolov3}.

\textbf{Three-layer network:} Here, the 192 latent-space channels are split into $L_1=96$ channels for the base layer (which will support object detection), ${L_2=32}$ channels for the first enhancement layer (which will support segmentation) and the remaining ${L_3=64}$ channels for the last enhancement layer, which will support input reconstruction. LSTs in the base and the first enhancement layer individually estimate intermediate tensors of Faster R-CNN~\cite{ren2016faster} for object detection and Mask R-CNN~\cite{he2017mask} for segmentation, respectively. Fig.~\ref{fig:vision_task_networks}(b) presents the ResNet-50~\cite{he2016deep} based Feature Pyramid Network (FPN)~\cite{FPN} used as a backbone network for both R-CNN networks. In particular, the LSTs estimate the outputs of layer 4 in the FPN, $\widetilde{\mathbfcal{F}}_j^{(4)} \in \mathbb{R}^{4N \times 4M \times 256}$, where $j\in\{1,2\}$. Hence, to generate the correct size of the feature tensors from the sub-latents $\widetilde{\mathbfcal{Y}}_1$ and $\{\widehat{\mathbfcal{Y}}_1, \widehat{\mathbfcal{Y}}_2\}$, both LSTs have the same configuration of scaling factors: $r_1=r_2=2$ and $r_3=r_4=1$. The activation at layer 4 in the FPN is ReLU, so we use the same function for the last activation layer for both LSTs. 
For distortion computation, the MSE is measured at various points \textbf{P}2--\textbf{P}6 in the FPN, which are shown in Fig.~\ref{fig:vision_task_networks}(b). 
To account for this, the distortion equation~(\ref{eq:distortion_term}) is slightly modified to
\begin{equation}
\begin{split}
D & = \, \textup{MSE}(\mathbf{X},\widehat{\mathbf{X}}) \\
& \, + \gamma \cdot \frac{1}{5} \cdot \sum_{j=1}^{2}\sum_{l=2}^{6} \textup{MSE}(\textbf{P}l_j, V^{\textup{back-end,\textbf{P}}l}_{\textup{FPN,j}}(\widetilde{\mathbfcal{F}}_j^{(4)}, \rho^{*}_j))
\end{split}
\label{eq:m12_3_changed_distortion_term_three_layer}
\end{equation}
\noindent where $\gamma = 0.0015$ and $V^{\textup{back-end,\textbf{P}}l}_{\textup{FPN,j}}$ represents the portion of the FPN back-end up to \textbf{P}$l$, with pre-trained weights $\rho^{*}_j$~\cite{wu2019detectron2}, using $\widetilde{\mathbfcal{F}}_j^{(4)}$ as input. 

\subsection{Evaluation on machine vision tasks}
\label{ssec:evaluation_machine_vision}

Our multi-task networks are evaluated on relevant datasets associated with targeted tasks, in terms of task accuracy vs. bitrate. 

\textbf{Benchmarks:} We compare our proposed networks against several benchmarks, including the latest standard codecs such as HEVC~\cite{hevc_std_2019} and VVC~\cite{vvc_std}, as well as five recent DNN-based image codecs~\cite{minnen2018joint, cheng2020image, hu2020coarse, lee2018context}. Specifically, the reference software versions for HEVC and VVC are HM-16.20~\cite{HM16.20} and VTM-12.3~\cite{VTM12.3}, and are used in all intra configuration~\cite{hevc_ctc,vvc_ctc} with quantization parameters QP $\in\{22, 25, 28, ..., 40\}$. To encode RGB images using the standard codecs, these images are first converted to YUV444 in accordance with ITU-R BT.709~\cite{itu_r_bt709}, then the chrominance channels are sub-sampled to create the YUV420 version of the image, which can then be fed to an HEVC or VVC encoder. After decoding, chrominance channels of the YUV420 output are interpolated to YUV444 using bilinear interpolation, then converted to RGB in accordance with ITU-R BT.709~\cite{itu_r_bt709}. DNN-based codecs are trained to compress RGB images directly, so no conversion to YUV is needed. Finally, decoded RGB images are used as input to the pre-trained computer vision models\footnote{In each case, the same model whose back-end is used in our multi-task network.} to examine task-specific accuracy.

\begin{figure}[t]
    \centering
    \begin{minipage}[b]{1.0\linewidth}
    \centering
    \includegraphics[width=\textwidth]{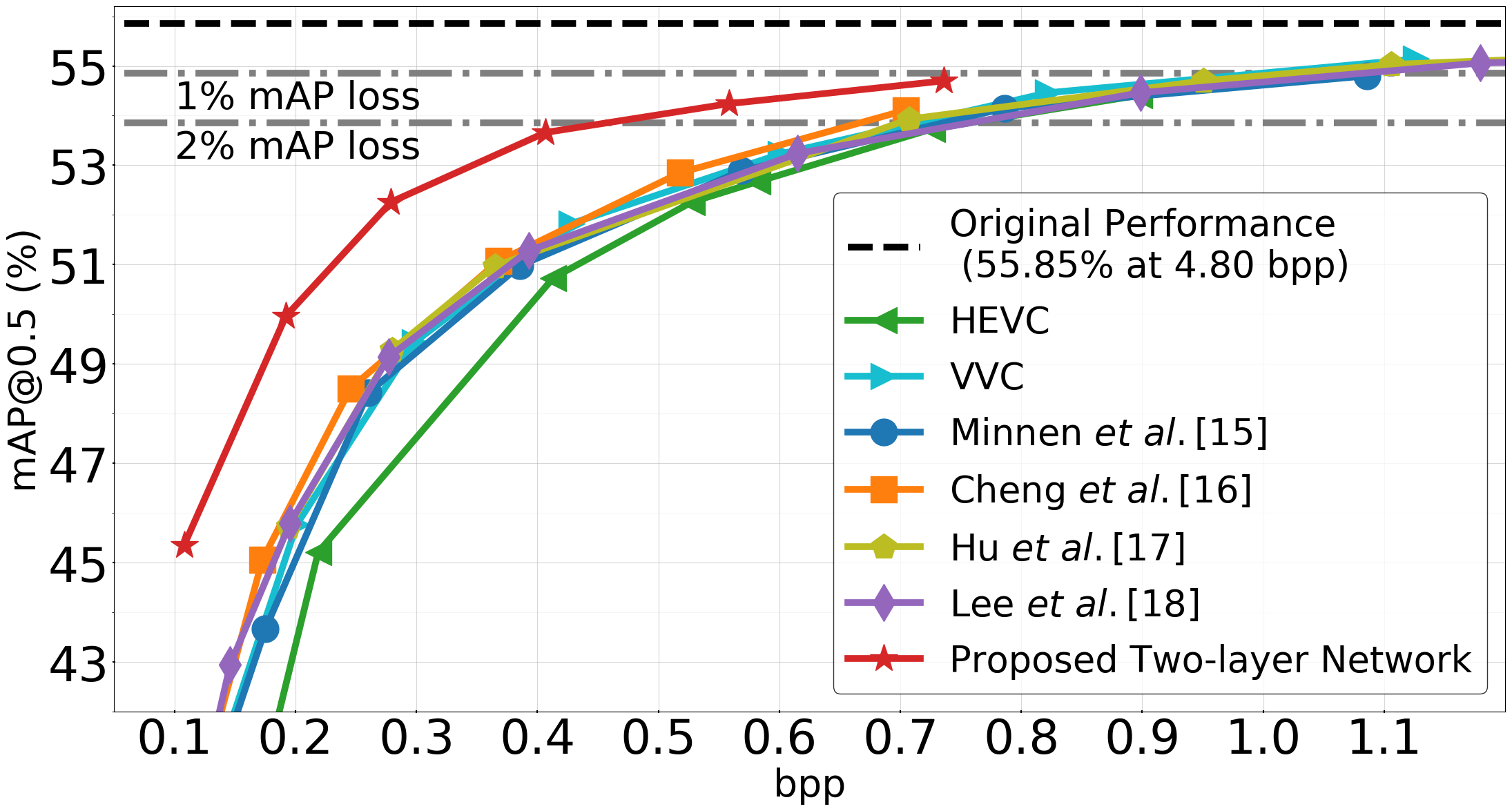}
    \end{minipage}
\vspace{-0.5cm}
\caption{Object detection performance of the two-layer network compared with benchmarks on 
the COCO2014 validation set}

\label{fig:two_task_vision}
\end{figure}

\textbf{Two-layer network:} Our two-layer network supports object detection in the base layer using the YOLOv3~\cite{Redmon2018_yolov3} back-end. We first evaluate the object detection performance on the COCO2014 validation set~\cite{COCO}, which includes about 5,000 images. Since most vision networks resize the input to a specific resolution before processing, we also resize input images to $512\times512$ 
using bilinear interpolation without letterboxing, in order to generate (via LST) a feature tensor $\widetilde{\mathbfcal{F}}^{(13)}_{1} \in \mathbb{R}^{64 \times 64 \times 256}$ that can be directly fed into the YOLOv3 back-end. The same resizing is done with benchmark image codecs. 

Fig.~\ref{fig:two_task_vision} presents a comparison of our object detection performance against various benchmarks in terms of bitrate in bits per pixel (bpp) vs. mean Average Precision (mAP), where bpp is computed by dividing the total number of coded bits (base and side bitstreams) by the number of input pixels. The mAP uses the Intersection of Union (IoU) threshold of 0.5. 
In the figure, the black dashed line shows the default mAP performance of 55.85\% when using the COCO2014 validation images as input to YOLOv3~\cite{Redmon2018_yolov3} with pre-trained weights. The average bitrate of the images in the COCO2014 validation set, coded with JPEG, is 4.80 bpp, which is indicated in the legend in the figure. Our object detection achieves the best rate-accuracy performance, with mAP loss of about 1\% at 0.74 bpp, where most benchmarks suffer a mAP loss of about 2\%.  Moreover, at 0.56 bpp, our method operates within a 2\% mAP loss margin, whereas most benchmarks have lost about 4\% mAP at this point. 
Cheng $\textit{et al.}$~\cite{cheng2020image} shows the best performance among the benchmarks, but there is still significant gap between it and our proposed method.

\begin{table}[t]
\centering
\caption{Summarized performance of vision tasks against various benchmarks with BD metrics}
\label{tbl:vision_task_bd_results}
\smallskip\noindent
\resizebox{1.0\linewidth}{!}{%
\begin{tabular}{@{}c|cccccc@{}}
\toprule
\multirow{2}{*}{\begin{tabular}[c]{@{}c@{}}Proposed \\ network\end{tabular}} & \multicolumn{2}{c|}{Two-layer}        & \multicolumn{4}{c}{Three-layer}                                          \\ \cmidrule(l){2-7} 
                                                                             & \multicolumn{4}{c|}{Object Detection}                                         & \multicolumn{2}{c}{Segmentation} \\ \midrule
Benchmark                                                                    & BD-rate & \multicolumn{1}{c|}{BD-mAP} & BD-rate & \multicolumn{1}{c|}{BD-mAP} & BD-rate         & BD-mAP         \\ \midrule
VVC                                                                          & -39.8 & \multicolumn{1}{c|}{2.79}   & -70.5 & \multicolumn{1}{c|}{2.33}   & -71.2         & 2.34           \\
HEVC                                                                         & -47.9 & \multicolumn{1}{c|}{4.55}   & -73.2 & \multicolumn{1}{c|}{3.05}   & -74.7         & 2.96           \\
{\cite{minnen2018joint}}                                                                     & -41.3 & \multicolumn{1}{c|}{3.26}   & -78.7 & \multicolumn{1}{c|}{3.73}   & -77.2         & 3.38           \\
{\cite{cheng2020image}}                                                                     & -37.4 & \multicolumn{1}{c|}{2.89}   & -76.6 & \multicolumn{1}{c|}{3.62}   & -75.4         & 3.49           \\
{\cite{hu2020coarse}}                                                                     & -38.1 & \multicolumn{1}{c|}{1.99}   & -76.3 & \multicolumn{1}{c|}{2.80}   & -75.6         & 2.56           \\
{\cite{lee2018context}}                                                                     & -39.6 & \multicolumn{1}{c|}{2.93}   & -77.5 & \multicolumn{1}{c|}{3.66}   & -76.2         & 3.42           \\ \bottomrule
\end{tabular}}
\end{table}

Table~\ref{tbl:vision_task_bd_results} (first three columns) summarizes object detection vs. bitrate results using extended versions of Bj\o{}ntegaard Delta (BD) metric~\cite{Bjontegaard, vcm_template}. For the BD-mAP metric, positive numbers represent an average increase of mAP at the same bitrate. For BD-rate, negative numbers indicate average bit savings at the same accuracy. Our method shows a noticeable bit savings and increased accuracy compared to all benchmarks. For example, against HEVC, we achieve BD-rate savings of --47.9\%, and BD-mAP gain of 4.55\%. Against Cheng $\textit{et al.}$\cite{cheng2020image}, we achieve BD-rate savings of --37.4\%, and BD-mAP gain of 2.89\%. The greatest bit reduction of 41.3\% is accomplished against Minnen $\textit{et al.}$\cite{minnen2018joint}, also marginally less significant reductions are achieved against other DNN-based networks\cite{hu2020coarse, lee2018context}.

\begin{figure}[t]
    \centering
    \begin{minipage}[b]{1.0\linewidth}
    \centering
    \includegraphics[width=\textwidth]{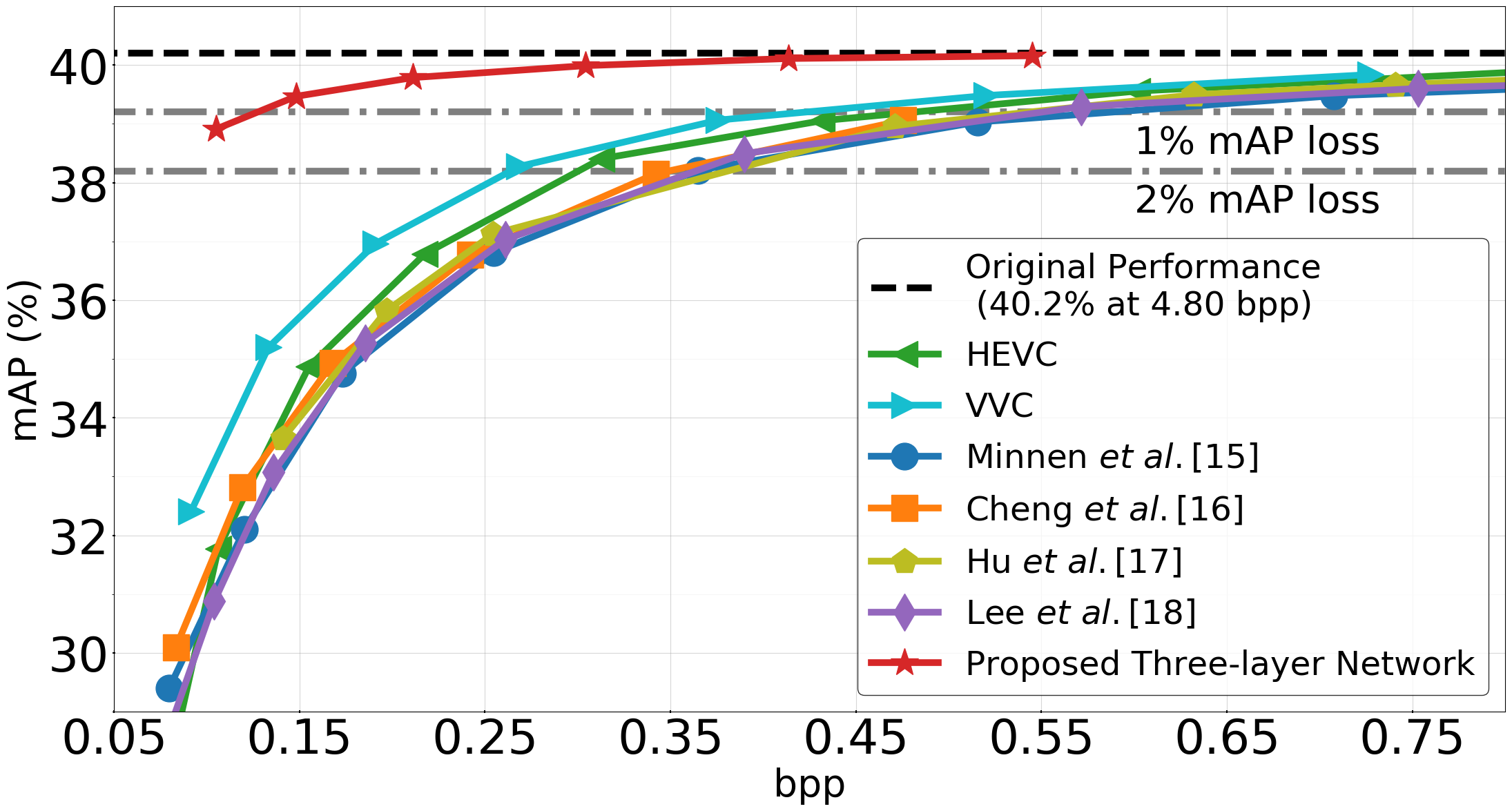}
    \centerline{(a)}\medskip
    \end{minipage}
    
    \begin{minipage}[b]{1.0\linewidth}
    \centering
    \includegraphics[width=\textwidth]{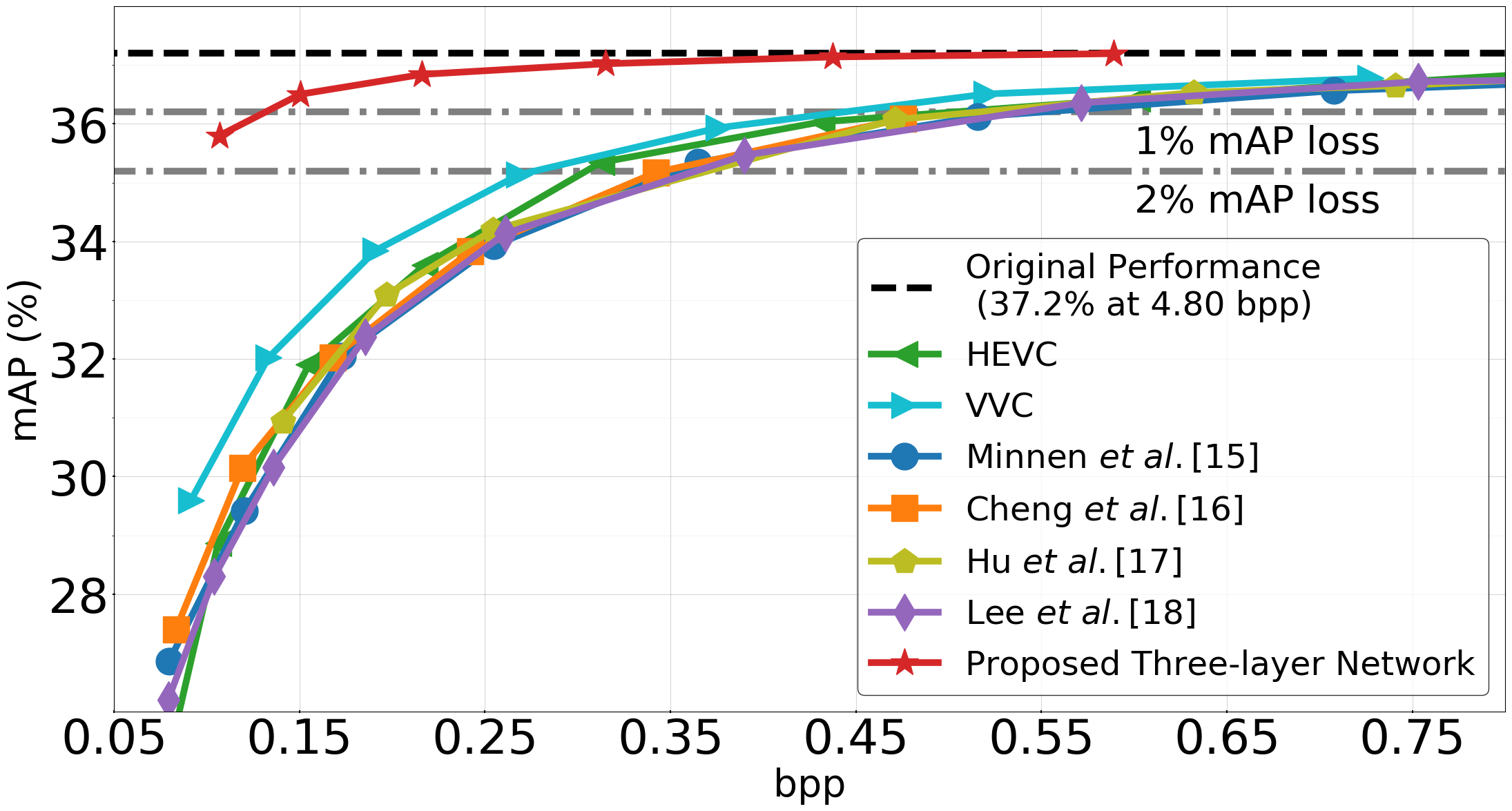}
    \centerline{(b)}\medskip
    \end{minipage}
    \hfill
\vspace{-0.2cm}
\caption{Performance of (a) object detection and (b) segmentation of the three-layer network compared with benchmarks on the COCO2017 validation set}
\label{fig:three_layer_vision}
\end{figure}

%
%

\textbf{Three-layer network:}
The three-layer network supports object detection in the base layer and object segmentation in the first enhancement layer. The corresponding back-ends use ResNet-50-based Faster~\cite{ren2016faster} and Mask~\cite{he2017mask} R-CNN with pre-trained weights~\cite{wu2019detectron2}, respectively. Since the benchmarks' performance of two tasks using those R-CNNs are evaluated on the COCO2017 validation set~\cite{COCO} and publicly reported via Detectron2~\cite{wu2019detectron2}, we use the same dataset for fair evaluation and comparison with our proposed network. 
Faster R-CNN and Mask R-CNN have a constraint on the input resolution that the shorter edge must be less than or equal to 800 pixels according to given default configuration. Hence, we resize the test images to meet the constraint using bilinear interpolation prior to the experiment, then use the resized images as input to our three-layer network.
As a result, generated feature tensors $\widetilde{\mathbfcal{F}}_{1}^{(4)}$ and $\widetilde{\mathbfcal{F}}_{2}^{(4)}$ from the base and the first enhancement layer can be fed into the Faster and Mask R-CNN back-ends, respectively. The resized images are also used as input to benchmark image codecs. 

Fig.~\ref{fig:three_layer_vision} shows the performance of our three layer network and the benchmarks on both tasks. In Fig.~\ref{fig:three_layer_vision}(a) and (b), dashed black lines show the default mAP performance of 40.2\% and 37.2\% at 4.80 bpp on average\footnote{Although the COCO2014 and COCO2017 validation sets are not identical, the average bitrate of images in each set, coded by JPEG, is 4.80 bpp.} for object detection and segmentation, respectively. Here, mAP is obtained using the IoU threshold from 0.5 to 0.95 with a step size of 0.05. On both tasks, our 3-layer network shows excellent performance: less than 1\% mAP drop down to 0.15 bpp, and less than about 1.5\% mAP drop at 0.1 bpp. 
Meanwhile, all benchmarks lose 1\% mAP already at 0.4 bpp, while at 0.1 bpp, they have lost 7-8\% mAP. 


Table~\ref{tbl:vision_task_bd_results} (last four columns) shows object detection and segmentation performance vs. bitrate in terms of extended versions of BD metrics~\cite{vcm_template}. Here, the gains of our network against the benchmarks are even higher. Against VVC, which was the best-performing benchmark, our network achieves BD-rate savings of --73.2\% on object detection and --71.2\% on object segmentation. At the same time, the BD-mAP gains against VVC are 2.33\% on object detection and 2.34\% on object segmentation.

\begin{figure*}[t]
    \centering
    \begin{minipage}[b]{0.45\linewidth}
    \centering
    \includegraphics[width=\textwidth]{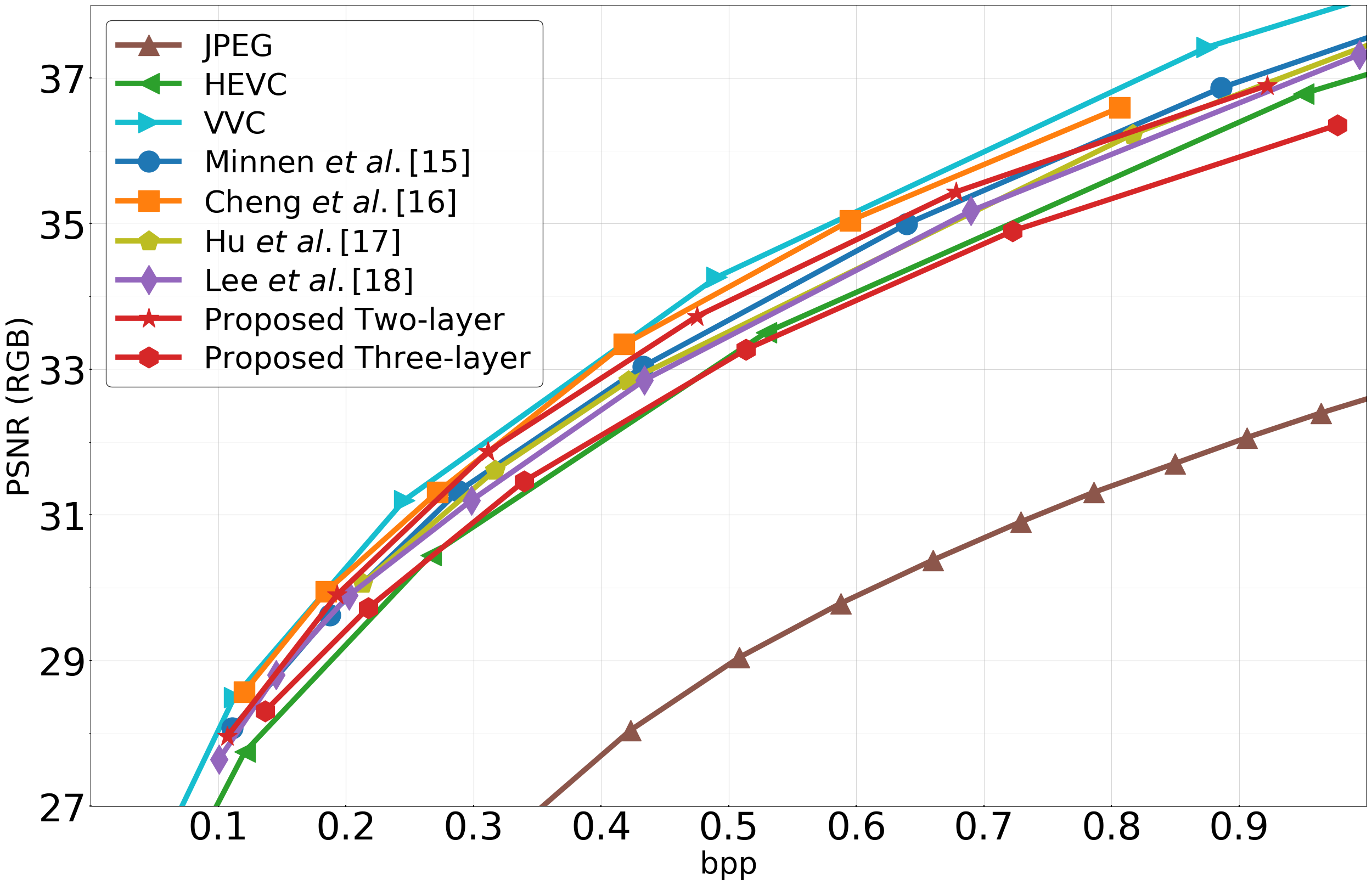}
    \centerline{(a)}\medskip
    \end{minipage}
    \hspace{0.3cm}
    \begin{minipage}[b]{0.466\linewidth}
    \centering
    \includegraphics[width=\textwidth]{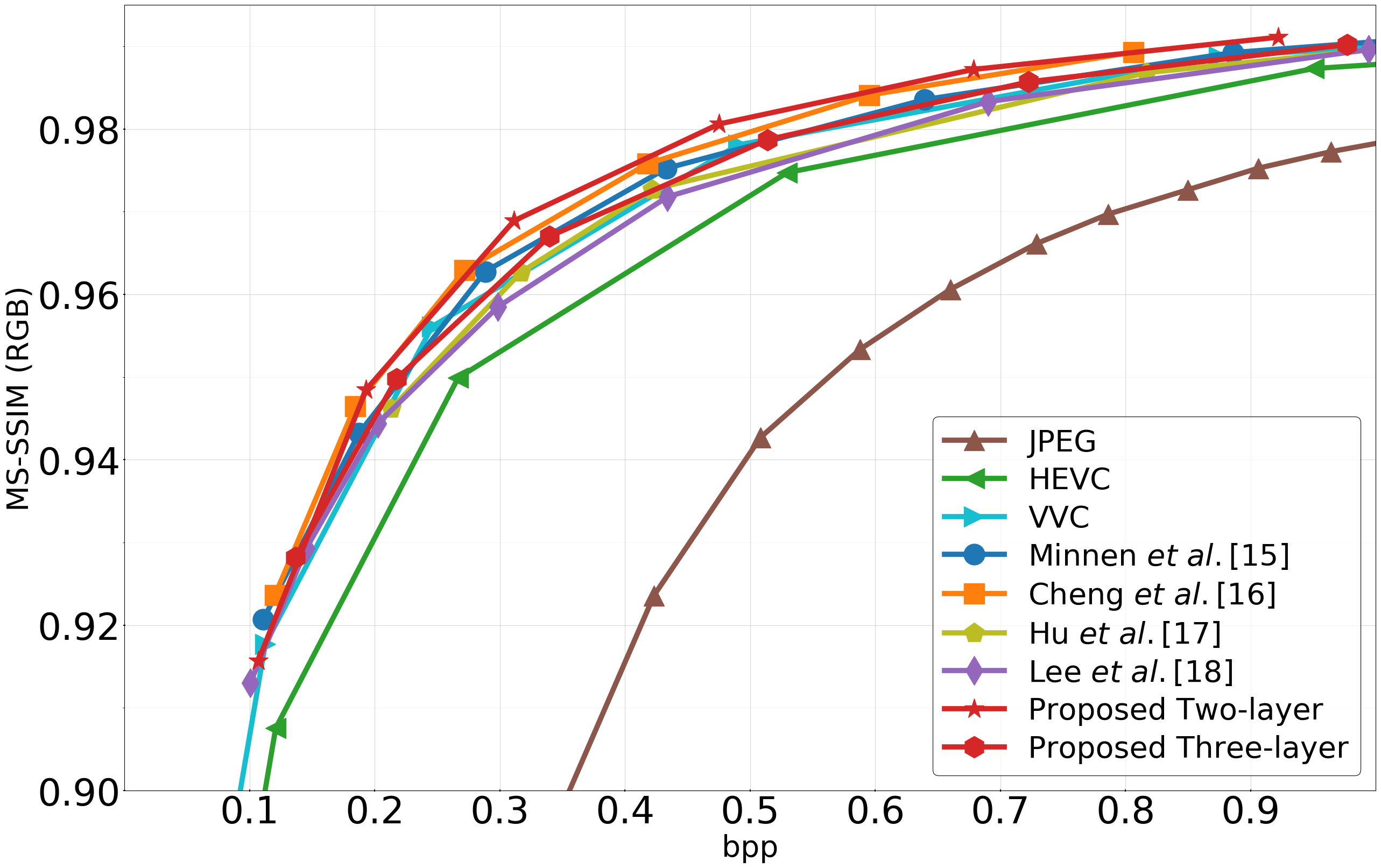}
    \centerline{(b)}\medskip
    \end{minipage}
\vspace{-0.2cm}
\caption{Comparison of input reconstruction performance on Kodak in terms of (a) PSNR vs. bpp and (b) MS-SSIM vs. bpp 
}
\label{fig:multi_task_input_recon}
\end{figure*}


\begin{table}[t]
\centering
\caption{BD-rate relative to various benchmarks on Kodak~\cite{kodak_dataset}}
\label{tbl:recon_bdbr}
\smallskip\noindent
\resizebox{1.0\linewidth}{!}{%
\begin{tabular}{@{}ccccc@{}}
\toprule
                                                 & \multicolumn{4}{c}{Proposed methods}                                                                                                                                                                                                                                             \\ \midrule
\multicolumn{1}{c|}{\multirow{2}{*}{Benchmarks}} & \multicolumn{2}{c|}{Two-layer Network}                                                                                                            & \multicolumn{2}{c}{Three-layer Network}                                                                                      \\ \cmidrule(l){2-5} 
\multicolumn{1}{c|}{}                            & \begin{tabular}[c]{@{}c@{}}BD-rate\\ (PSNR)\end{tabular} & \multicolumn{1}{c|}{\begin{tabular}[c]{@{}c@{}}BD-rate\\ (MS-SSIM)\end{tabular}} & \begin{tabular}[c]{@{}c@{}}BD-rate\\ (PSNR)\end{tabular} & \begin{tabular}[c]{@{}c@{}}BD-rate\\ (MS-SSIM)\end{tabular} \\ \midrule
\multicolumn{1}{c|}{VVC}                         & 10.17\%                                                     & \multicolumn{1}{c|}{-7.83\%}                                                        & 30.43\%                                                     & 2.14\%                                                         \\
\multicolumn{1}{c|}{HEVC}                        & -14.27\%                                                    & \multicolumn{1}{c|}{-26.15\%}                                                       & 1.38\%                                                      & -17.96\%                                                       \\
\multicolumn{1}{c|}{JPEG}                        & -64.52\%                                                    & \multicolumn{1}{c|}{-64.06\%}                                                       & -57.28\%                                                    & -57.84\%                                                       \\ \midrule
\multicolumn{1}{c|}{\cite{minnen2018joint}}            & -3.58\%                                                     & \multicolumn{1}{c|}{-7.83\%}                                                        & 14.02\%                                                     & 2.06\%                                                         \\
\multicolumn{1}{c|}{\cite{cheng2020image}}          & 4.49\%                                                      & \multicolumn{1}{c|}{-1.90\%}                                                        & 24.24\%                                                     & 9.55\%                                                         \\
\multicolumn{1}{c|}{\cite{hu2020coarse}}        & -5.58\%                                                     & \multicolumn{1}{c|}{-16.84\%}                                                       & 12.19\%                                                     & -4.92\%                                                        \\
\multicolumn{1}{c|}{\cite{lee2018context}}    & -7.01\%                                                     & \multicolumn{1}{c|}{-15.23\%}                                                       & 9.94\%                                                      & -7.37\%                                                        \\
\multicolumn{1}{c|}{\begin{tabular}[c]{@{}c@{}}Two-layer\\ Network\end{tabular}}         & \multicolumn{2}{c|}{-}                                                                                                                            & 18.84\%                                                     & 11.95\%                                                        \\ \bottomrule
\end{tabular}}
\end{table}

\subsection{Evaluation on input reconstruction}
\label{ssec:evaluation_human_vision}

The highest enhancement layer of our 2- and 3-layer networks supports input reconstruction for human viewing. 
The performance here is examined on the Kodak dataset~\cite{kodak_dataset}, which includes 24 uncompressed RGB images. 

\textbf{Benchmarks:} Our network performance is compared with the benchmarks, which include standard codecs~\cite{hevc_std_2019, vvc_std} and DNN-based codecs~\cite{minnen2018joint, cheng2020image, hu2020coarse, lee2018context}, all optimized for MSE. The results reported here for HEVC and VVC are borrowed from CompressAI~\cite{begaint2020compressai}. For HEVC~\cite{hevc_std_2019} and VVC~\cite{vvc_std}, HM-16.20~\cite{HM16.20} with range extension profile and VTM-9.1~\cite{VTM9.1} were used, respectively. RGB images were first converted to YUV444 in accordance with BT.709~\cite{itu_r_bt709}, then directly input to the codecs. Decoded images in the YUV444 format were converted back to RGB in accordance with BT.709~\cite{itu_r_bt709}. Since each input image is coded separately, all intra configurations following the common test conditions~\cite{hevc_ctc,vvc_ctc} was used. 
Images were coded using various QPs from 12 to 47 with a step size of 5, and we show the RD points that cover the range of bitrates acheived by DNN-based codecs. We also borrowed JPEG results within this bitrate range from CompressAI~\cite{begaint2020compressai}. 

Fig.~\ref{fig:multi_task_input_recon}(a) shows PSNR (RGB) vs. bitrate curves. VVC achieves the best performance among all methods in this figure. The method from which our backbone is derived, Cheng $\textit{et al.}$~\cite{cheng2020image}, achieves the second-best performance. Our 2-layer network shows competitive performance compared to~\cite{cheng2020image} at lower bitrates, but the reconstruction quality slightly degrades (by about 0.3dB) compared to~\cite{cheng2020image} at higher bitrates. This is the price to pay for scalability and supporting object detection in the base layer. Compared to~\cite{hu2020coarse, lee2018context} and other conventional codecs, our network still outperforms, while simultaneously supporting  scalability. Table~\ref{tbl:recon_bdbr} shows BD-rate~\cite{Bjontegaard, vcm_template} comparisons among various codecs. Overall, our 2-layer network has a loss of 10.17\% against VVC and 4.49\% against~\cite{cheng2020image}, but saves --3.58\%, --5.58\%, --7.01\%, --14.27\%, and --63.99\% of bits compared to Minnen $\textit{et al.}$~\cite{minnen2018joint}, Hu $\textit{et al.}$~\cite{hu2020coarse}, Lee $\textit{et al.}$~\cite{lee2018context}, HEVC, and JPEG, respectively, while providing  scalability.

Our 3-layer network is less efficient in terms of rate-distortion performance on input reconstruction. As shown in the last row of Table~\ref{tbl:recon_bdbr}, the 3-layer network suffers an increase of 18.84\% BD-rate compared to the 2-layer network. Nonetheless, it is worthwhile to remark that earlier work on conventional scalable codecs suggests that adding one scalability layer costs about 15-25\% in terms of BD-rate~\cite{shvc_overview}, so the performance of our 3-layer network seems reasonable in this context. It is still comparable with HEVC (1.38\% increase in BD-rate) and much better than JPEG, while providing 3-layer scalability and superior efficiency on computer vision tasks. Overall, Fig.~\ref{fig:multi_task_input_recon}(a) shows that both our 2- and 3-layer networks are comparable with state of the art codecs in terms of PSNR at lower bitrates, but their relative PSNR performance degrades at higher bitrates. 

In addition to PSNR results, we also provide performance comparisons in terms of MS-SSIM\cite{wang2003multiscale} vs. bitrate in Fig.~\ref{fig:multi_task_input_recon}(b). It is known that DNN-based codecs perform very well on MS-SSIM (even if they are optimized for MSE), and this is also evident in Fig.~\ref{fig:multi_task_input_recon}(b) and Table~\ref{tbl:recon_bdbr}. In particular, our 2-layer network achieves the best results in this comparison, showing coding gains against all benchmarks, with --1.9\% savings compared to the best benchmark~\cite{cheng2020image}. Even our 3-layer network now shows a gain of almost --18\% relative to HEVC, and comparable performance (with a gap around 2\%) relative to VVC and~\cite{minnen2018joint}.  
We believe this improved performance of our networks in terms of MS-SSIM is due to better latent representations related to objects features associated with machine vision tasks in the lower layers, which encourages higher structural quality of the reconstructed input. 

Finally, we consider a comparison with~\cite{liu2021semantics}. While direct comparison is not possible since the tasks (and the corresponding image datasets) are different, an indirect comparison via BPG (HEVC Intra) is possible based on presented results. According to Fig.~9 in~\cite{liu2021semantics}, when it comes to input reconstruction, the approach in~\cite{liu2021semantics} is less efficient than HEVC in terms of PSNR, and comparable to HEVC on MS-SSIM. On the other hand, from Table~\ref{tbl:recon_bdbr}, our 2-layer network is 14.27\% more efficient than HEVC on PSNR, while our 3-layer network is slightly (about 1.3\%) less efficient than HEVC in terms of PSNR. Meanwhile, both our networks are significantly more efficient than HEVC Intra on MS-SSIM. Hence, it appears that our networks are more efficient than~\cite{liu2021semantics} in terms of input reconstruction.

\subsection{Run-time complexity of DNN-based methods}

We compare the run time complexity of our methods with DNN-based benchmarks~\cite{minnen2018joint, cheng2020image, hu2020coarse, lee2018context} by measuring the average encoding and decoding time per image on a GeForce RTX 2080 GPU with 11 GiB RAM. We evaluate the DNN-based codecs on the Kodak dataset~\cite{kodak_dataset}.  
Our methods target the input reconstruction at the highest enhancement layer, but entail extra computation related to sub-latent coding in the lower layer(s) for machine vision task(s). Meanwhile, the benchmarks target only input reconstruction. Three benchmarks\cite{minnen2018joint, cheng2020image, hu2020coarse} and ours are implemented based on CompressAI~\cite{begaint2020compressai}, while~\cite{lee2018context} is developed using Tensorflow-Compression~\cite{tf_compression}. All models run on a single GPU, but some of the computation for some models~\cite{cheng2020image, lee2018context}, including ours, has to run on a CPU (e.g., auto-regressive context modelling for entropy coding), which slows the system down.

Table~\ref{tbl:run_time} presents run-time complexity of the models, and ours has the second-largest encoding and decoding time, after~\cite{lee2018context}. Compared to~\cite{cheng2020image}, which is the baseline network we used as the backbone for our models, run-time complexity increases in accordance (roughly proportionally) with the number of layers. This seems reasonable. Nonetheless, complexity is an issue for all DNN-based codecs, including ours.

\begin{table}[t]
\centering
\caption{Average execution time (in seconds) for DNN-based models}
\label{tbl:run_time}
\smallskip\noindent
\resizebox{1.0\linewidth}{!}{%
\begin{tabular}{@{}c|cccccc@{}}
\toprule
\multirow{2}{*}{\begin{tabular}[c]{@{}c@{}}Avg. \\ Time (s)\end{tabular}} & \multicolumn{4}{c}{DNN-based benchmarks}                                              & \multicolumn{2}{c}{Proposed methods} \\ \cmidrule(l){2-7} 
                               & \cite{minnen2018joint} & \cite{cheng2020image} & \cite{hu2020coarse} & \cite{lee2018context} & Two-layer        & Three-layer       \\ \midrule
Encode                         & 4.12            & 4.05              & 5.60                & 20.52        & 9.36                 & 12.96                  \\
Decode                         & 8.29            & 8.30              & 6.25                & 70.88             & 17.26                  &  24.94                 \\ \bottomrule
\end{tabular}}
\end{table}

\subsection{Ablation study for LST}

\begin{figure}[t]
    \centering
    \begin{minipage}[b]{1\linewidth}
    \centering
    \includegraphics[width=\textwidth]{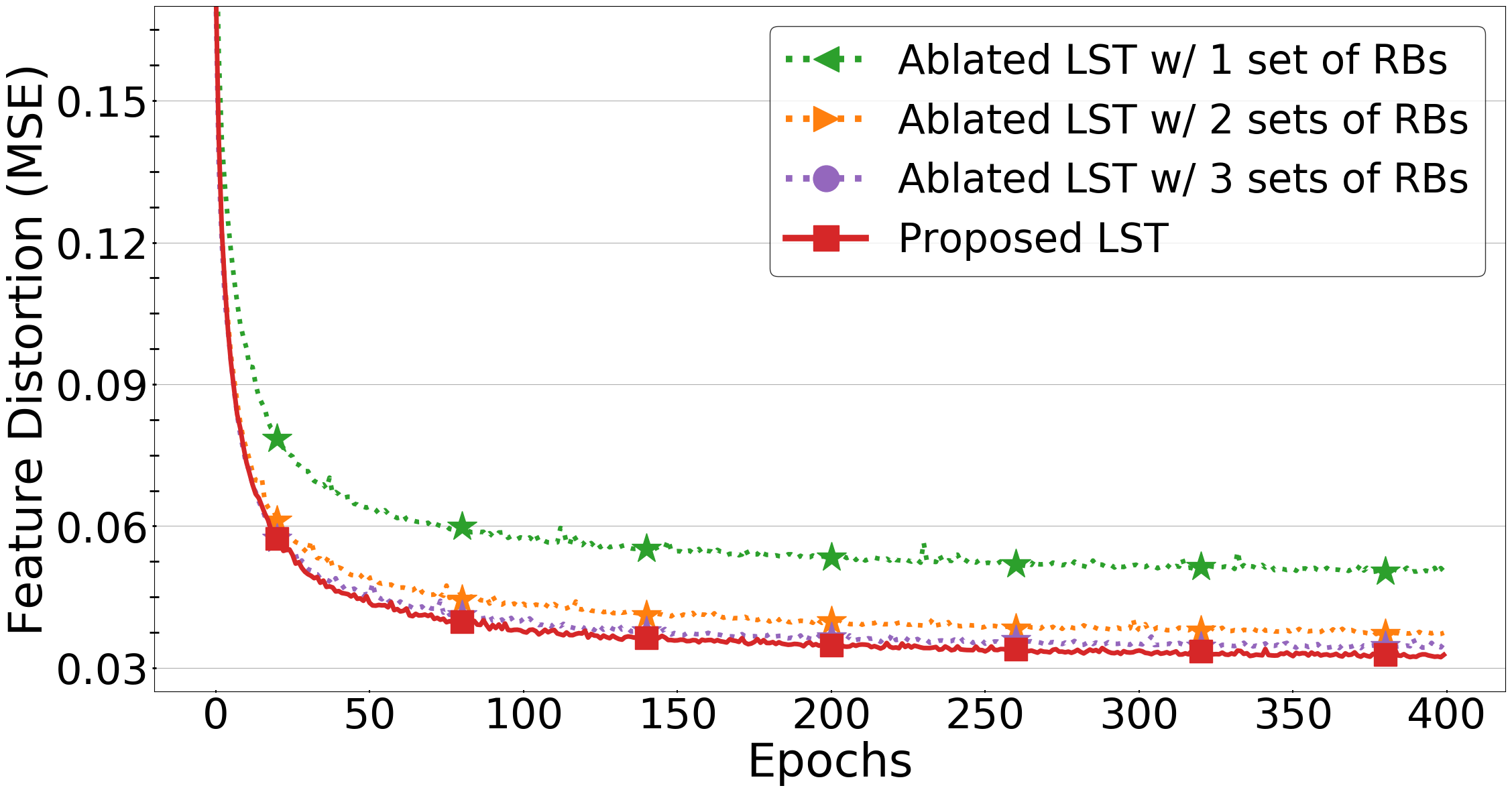}
    \centerline{(a)}\medskip
    \end{minipage}
    \centering
    \begin{minipage}[b]{1\linewidth}
    \centering
    \includegraphics[width=\textwidth]{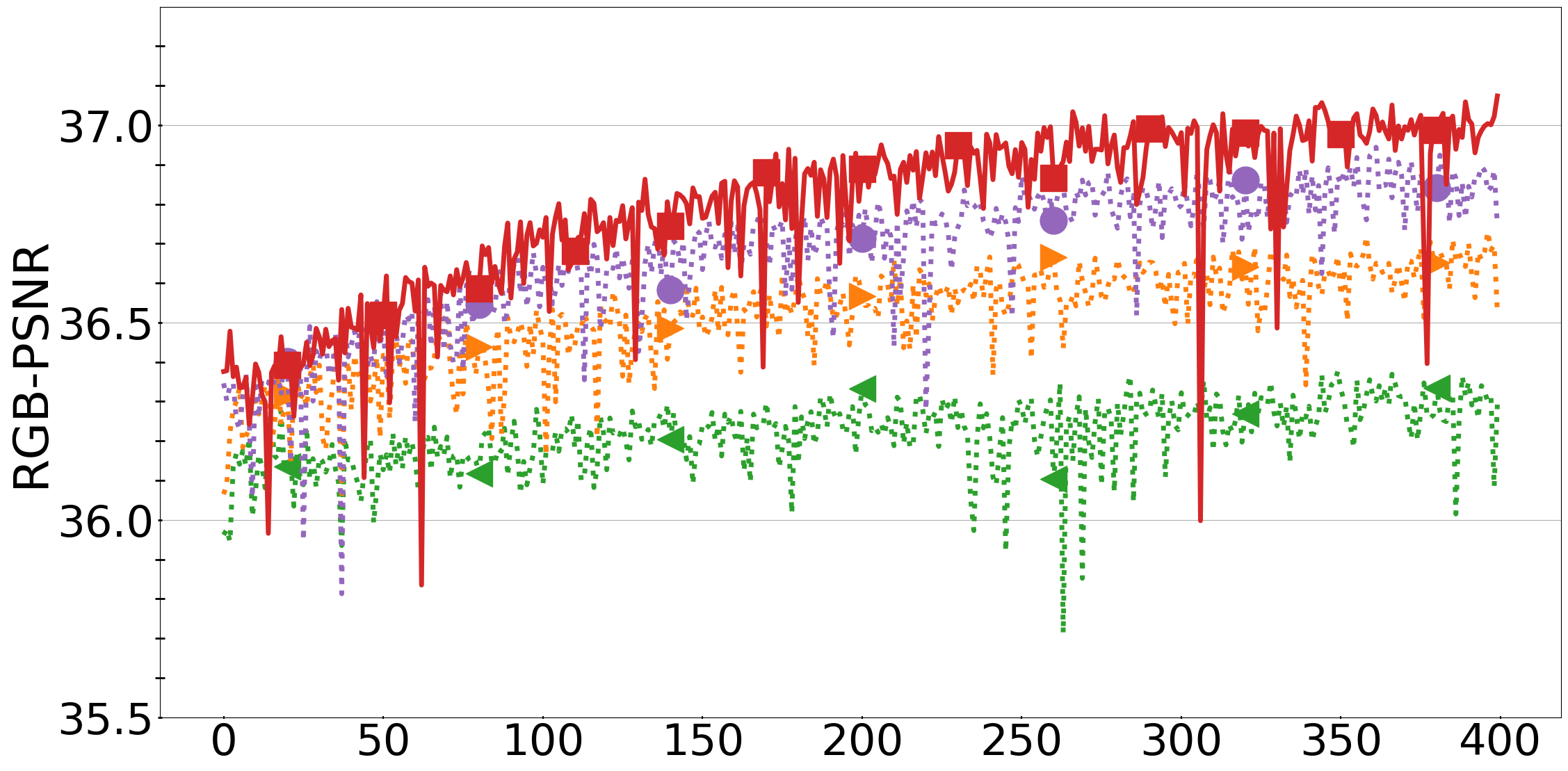}
    \centerline{(b)}\medskip
    \end{minipage}
\vspace{-0.5cm}    
\caption{LST ablation results on the 2-layer network: (a) feature distortion in the base layer, (b) input reconstruction in the enhancement layer.  
}
\label{fig:ablation}
\end{figure}

In the proposed approach to latent space scalability, the LST plays a crucial role of transforming a portion of the encoder's latent space into another latent space containing task-relevant features. In this section, we perform an ablation study on the LST in our 2-layer network, by gradually removing residual blocks (RBs, shown in Fig.~\ref{fig:latent_transform_module}) and examining the impact of such removal on both outputs of the network. The study is performed on the highest-rate 2-layer model (${\lambda}=0.0483$) for the first 400 epochs of training. The results are shown in Fig.~\ref{fig:ablation}. Fig.~\ref{fig:ablation}(a) shows the feature distortion in the target latent space (which is the latent space of YOLOv3) for four cases: LST with one set of RBs,\footnote{One ``set of RBs'' consists of one RB and one RB w/ Upsample, illustrated in Fig.~\ref{fig:latent_transform_module}.} LST with two sets of RBs, LST with three sets of RBS, and the proposed LS, which has four sets of RBs. As seen in the figure, increasing the number of RBs improves the ability of the LST to match the target features of YOLOv3 and reduces the distortion in the target latent space. With four sets of RBs (the proposed LST), we seem to have reached diminishing returns in terms of performance vs. complexity, so four sets of RBs seems to be a reasonable choice for the LST.

Fig.~\ref{fig:ablation}(b) shows the impact of LST ablation on input reconstruction. Even though the LST is in the object detection processing pipeline, since the network is trained end-to-end, a change in the LST has an impact on input reconstruction as well. We see that the performance of input reconstruction improves as the number of sets of RBs increases to four, which is the proposed LST. In essence, the ability of LST to better extract object detection-relevant features from the base layer helps the system better distribute the information between the base and enhancement layers, thereby improving both tasks. 

\subsection{Bitstream analysis}
\begin{figure}[t]
    \centering
    \begin{minipage}[b]{0.95\linewidth}
    \centering
    \includegraphics[width=\textwidth]{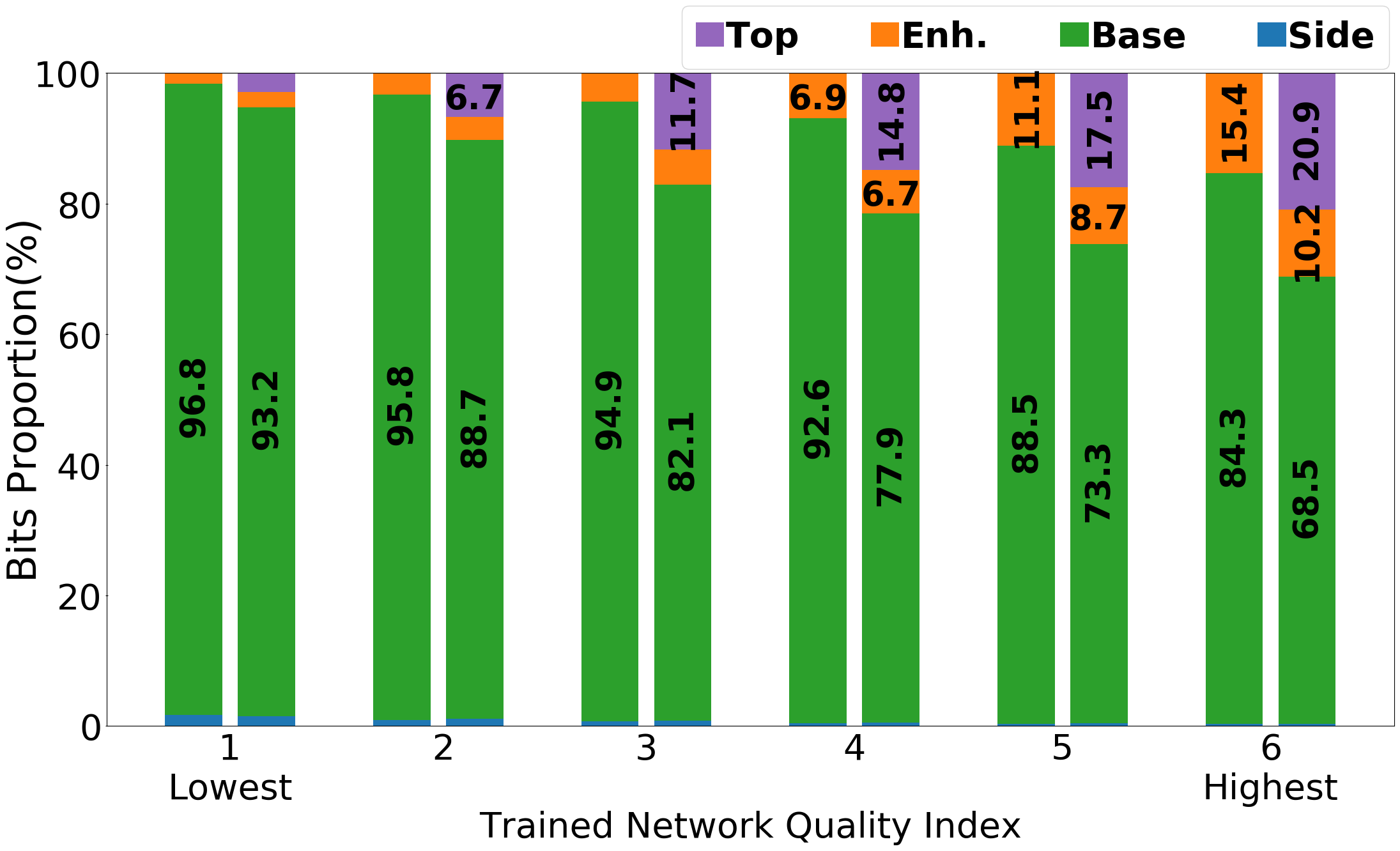}
    \end{minipage}
\caption{Proportion of bits in various layers of the bitstream. There are six quality indices, corresponding to six values of $\lambda$ shown in Table~\ref{tbl:lambda_values}. For each quality index, the left bar corresponds to the 2-layer network, and the right bar to the 3-layer network.}
\label{fig:bits_proportion_analysis}
\end{figure}

Fig.~\ref{fig:bits_proportion_analysis} shows the bitsteam composition of six versions of our networks, from those trained for the lowest bitrate (quality index 1), to those trained for the highest bitrate (quality index 6). For each index, left bar corresponds to the 2-layer network, and the right bar to the 3-layer network. There are relatively few bits in the side bitstream (blue), which accounts for less than 2\% of the total bitstream. The base layer (green) accounts for most bits, and its fraction drops from over 93\% at lowest bitrates to less than 85\% at highest bitrates for the 2-layer network, and less than 70\% for the 3-layer network. The first enhancement layer (orange, denoted ``Enh.'' in the figure) and the second enhancement layer in the 3-layer network (purple, denoted ``Top'') take up progressively more bits at higher bitrates. 
Considering the fact that our networks show more competitive performance of input reconstruction at lower bitrates (Fig.~\ref{fig:multi_task_input_recon}), these results seem to indicate that the base layer  conveys significant information for input reconstruction as well, in addition to enabling object detection.

\subsection{Visual examples}

\begin{figure*}[!th]
    \centering
    \begin{minipage}[b]{0.161\linewidth}
    \centering
    \centerline{\small Original}
    \vspace{0.08cm}
    \includegraphics[width=\textwidth]{Figures/visual_examples/original/000000396863.jpg}
    \vspace{0.215cm}
    \end{minipage}
    \begin{minipage}[b]{0.161\linewidth}
    \centering
    \centerline{\small \textcolor{white}{g}VVC\textcolor{white}{g}}
    \vspace{0.08cm}
    \includegraphics[width=\textwidth]{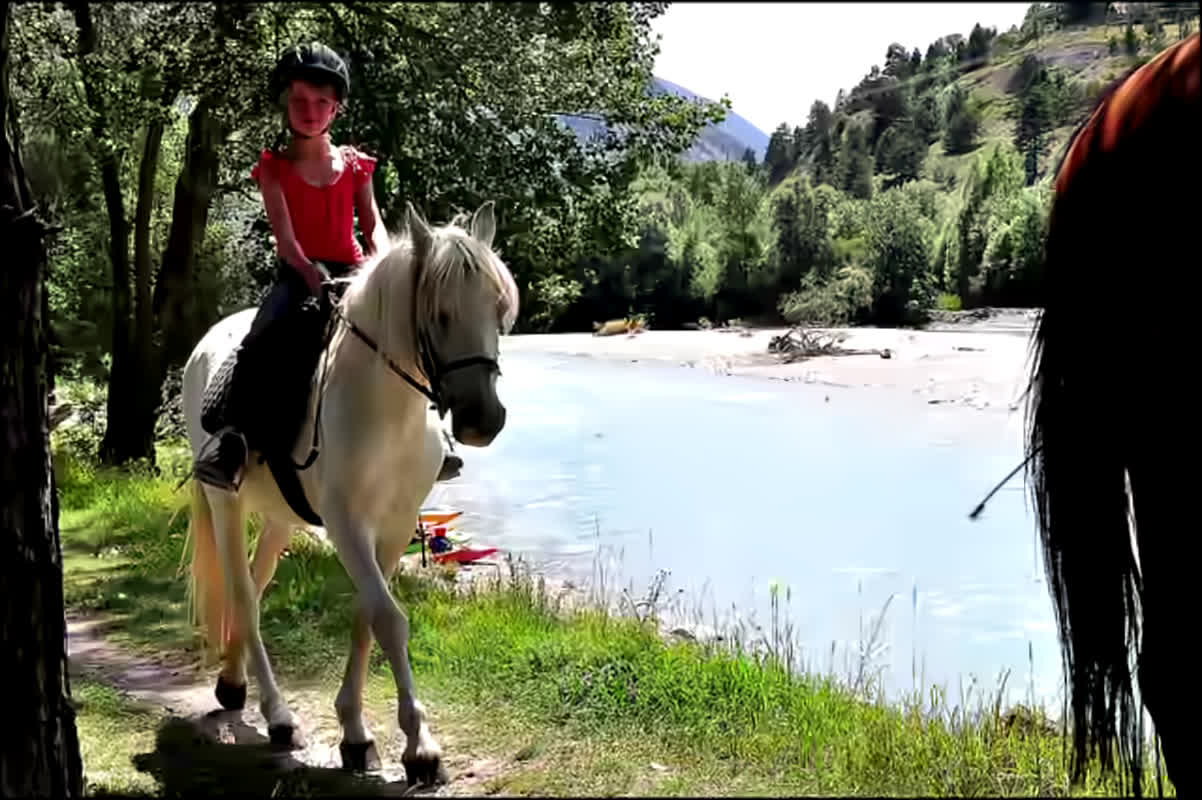}
    \centerline{0.195/23.83}\medskip
    \end{minipage}
    \begin{minipage}[b]{0.161\linewidth}
    \centering
    \centerline{\small \textcolor{white}{g}HEVC\textcolor{white}{g}}
    \vspace{0.08cm}
    \includegraphics[width=\textwidth]{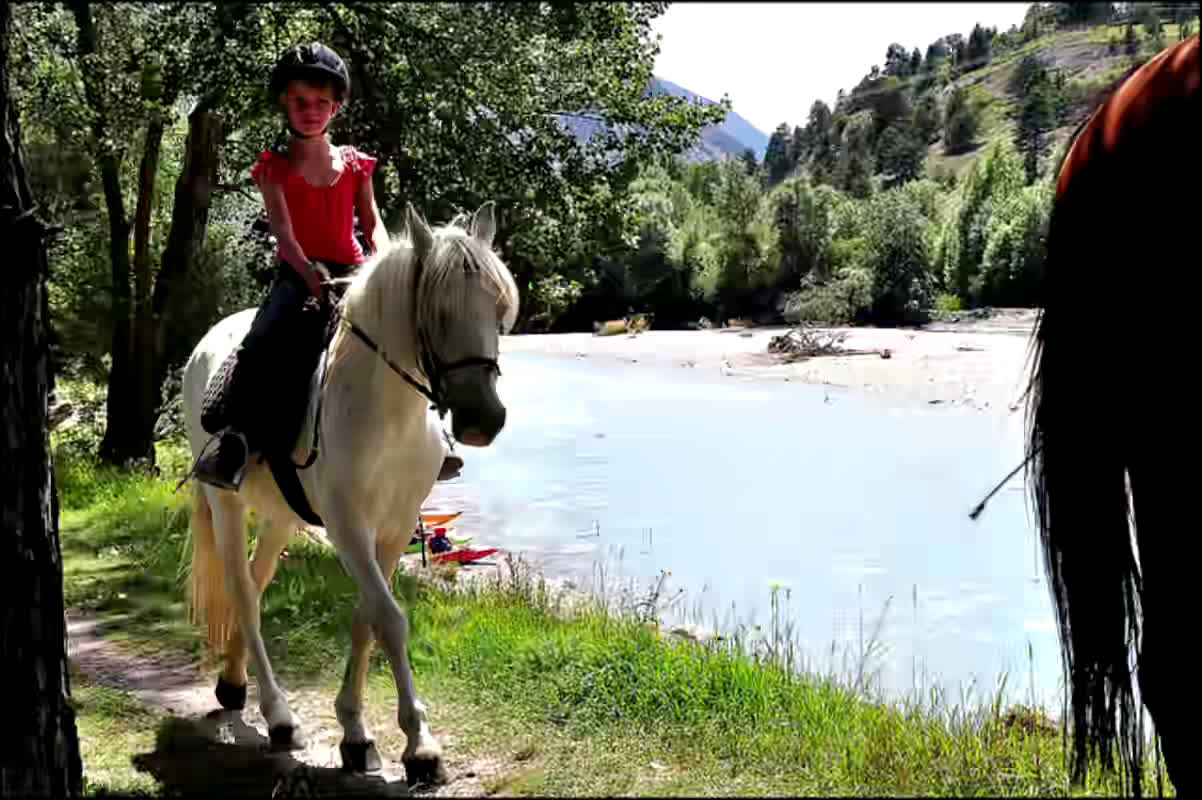}
    \centerline{0.211/23.64}\medskip
    \end{minipage}
    \begin{minipage}[b]{0.161\linewidth}
    \centering
    \centerline{\small Cheng $\textit{et al.}$~\cite{cheng2020image}}
    \vspace{0.08cm}
    \includegraphics[width=\textwidth]{Figures/visual_examples/cheng/dec_cheng_q1_img_396863.jpg}
    \centerline{0.206/26.59}\medskip
    \end{minipage}
    \begin{minipage}[b]{0.161\linewidth}
    \centering
    \centerline{\small \textcolor{white}{g}Minnen $\textit{et al.}$~\cite{minnen2018joint}\textcolor{white}{g}}
    \vspace{0.08cm}
    \includegraphics[width=\textwidth]{Figures/visual_examples/mbt/dec_mbt_q1_img_396863.jpg}
    \centerline{0.199/26.44}\medskip
    \end{minipage}
    \begin{minipage}[b]{0.161\linewidth}
    \centering
    \centerline{\small \textcolor{white}{g}Proposed\textcolor{white}{g}}
    \vspace{0.08cm}
    \includegraphics[width=\textwidth]{Figures/visual_examples/proposed/dec_000000396863.jpg}
    \centerline{0.212/26.04}\medskip
    \end{minipage}
    
    \centering
    \begin{minipage}[b]{0.161\linewidth}
    \centering
    \includegraphics[width=\textwidth]{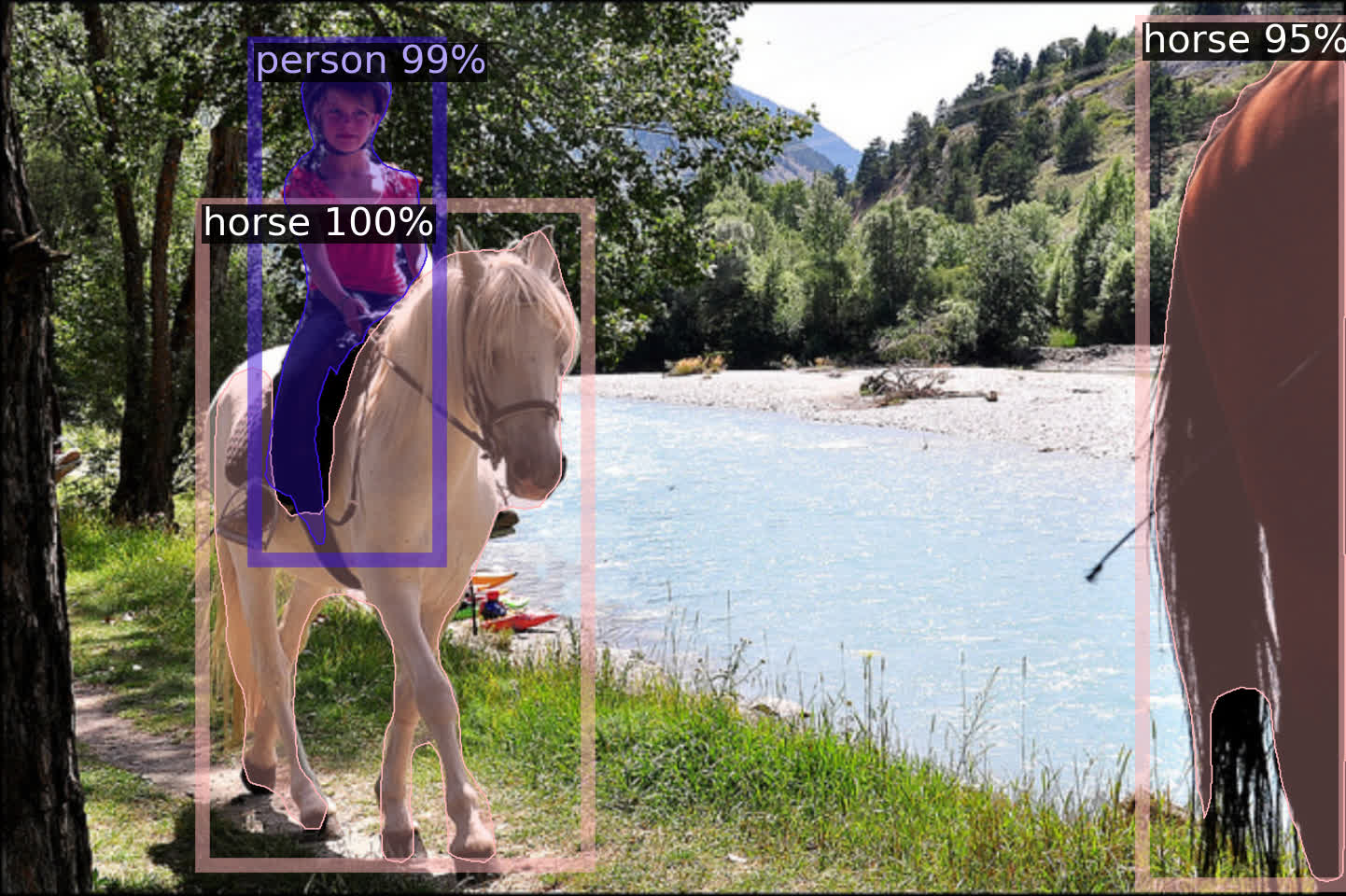}
    \centerline{Segmentation}\medskip
    \end{minipage}
    \begin{minipage}[b]{0.161\linewidth}
    \centering
    \includegraphics[width=\textwidth]{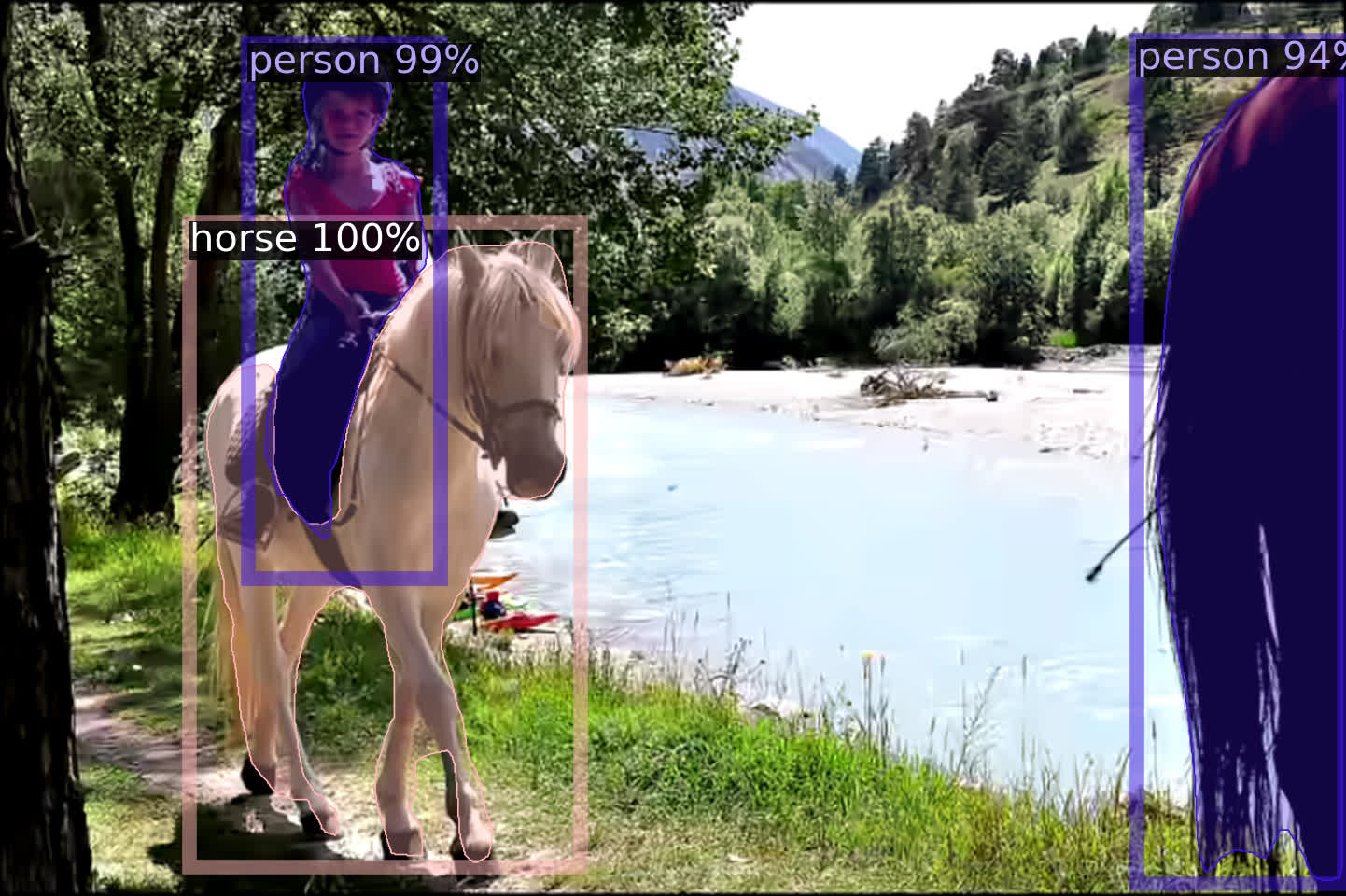}
    \vspace{0.29cm}
    \end{minipage}
    \begin{minipage}[b]{0.161\linewidth}
    \centering
    \includegraphics[width=\textwidth]{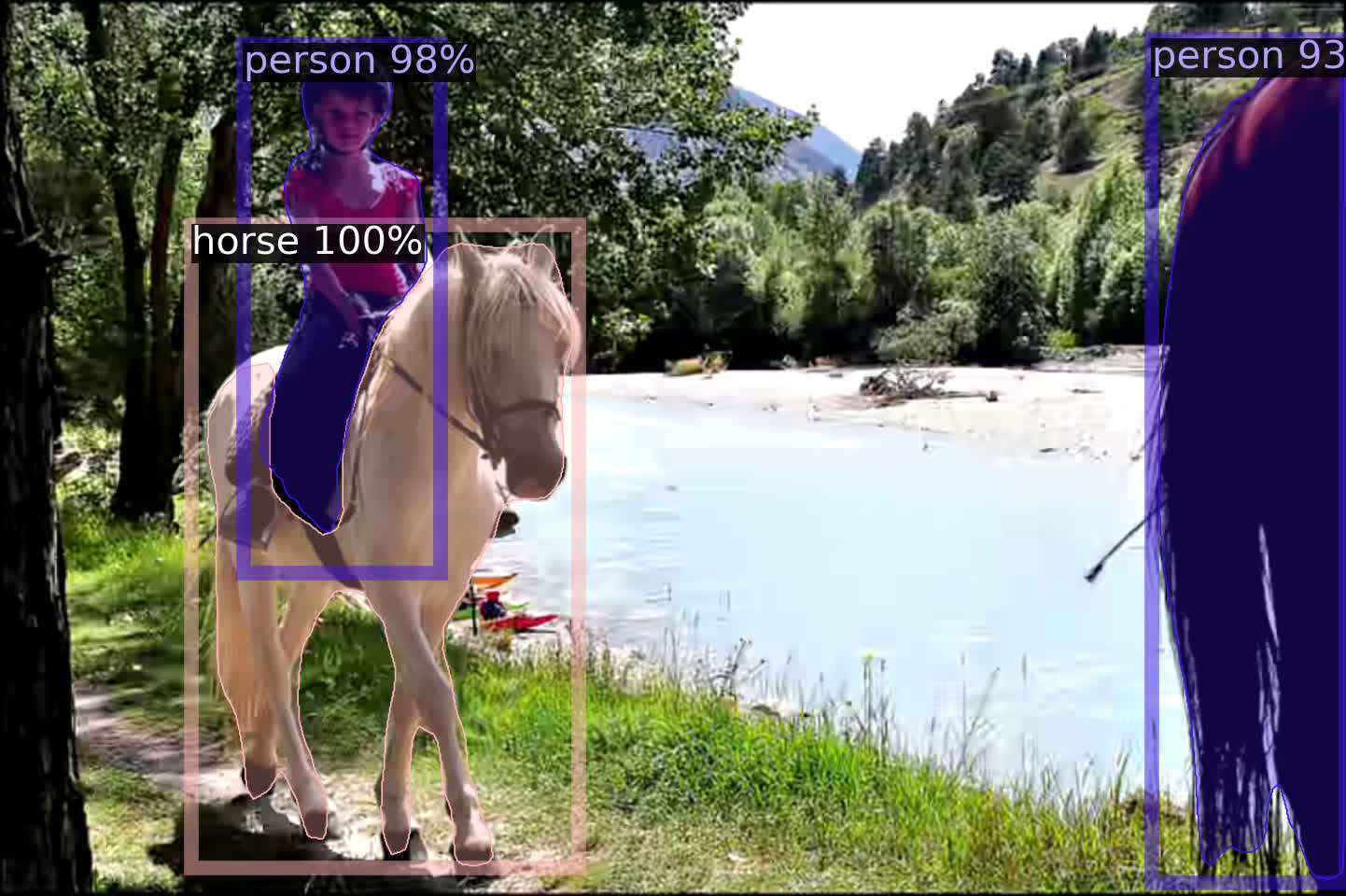}
    \vspace{0.29cm}
    \end{minipage}
    \begin{minipage}[b]{0.161\linewidth}
    \centering
    \includegraphics[width=\textwidth]{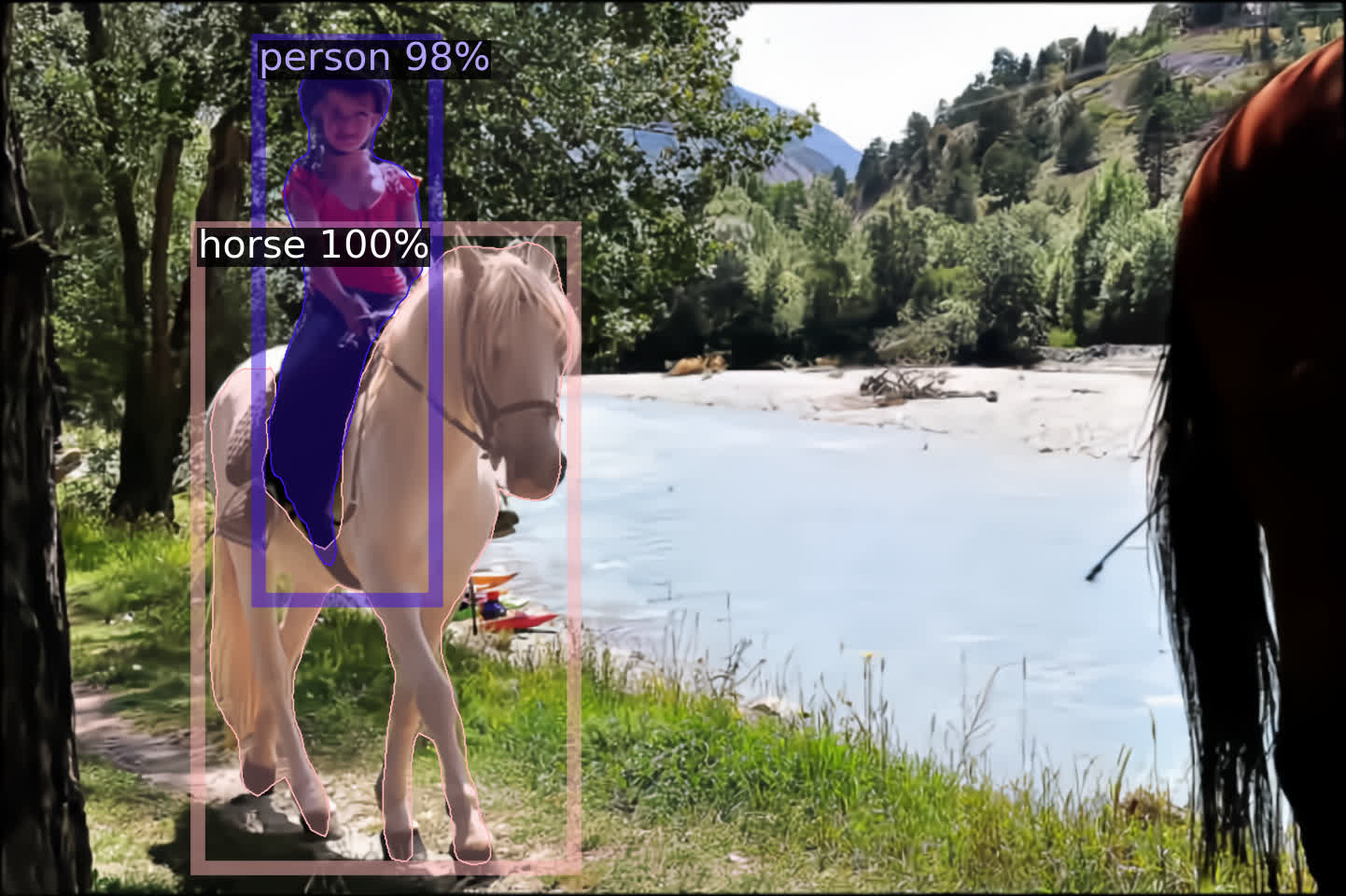}
    \vspace{0.29cm}
    \end{minipage}
    \begin{minipage}[b]{0.161\linewidth}
    \centering
    \includegraphics[width=\textwidth]{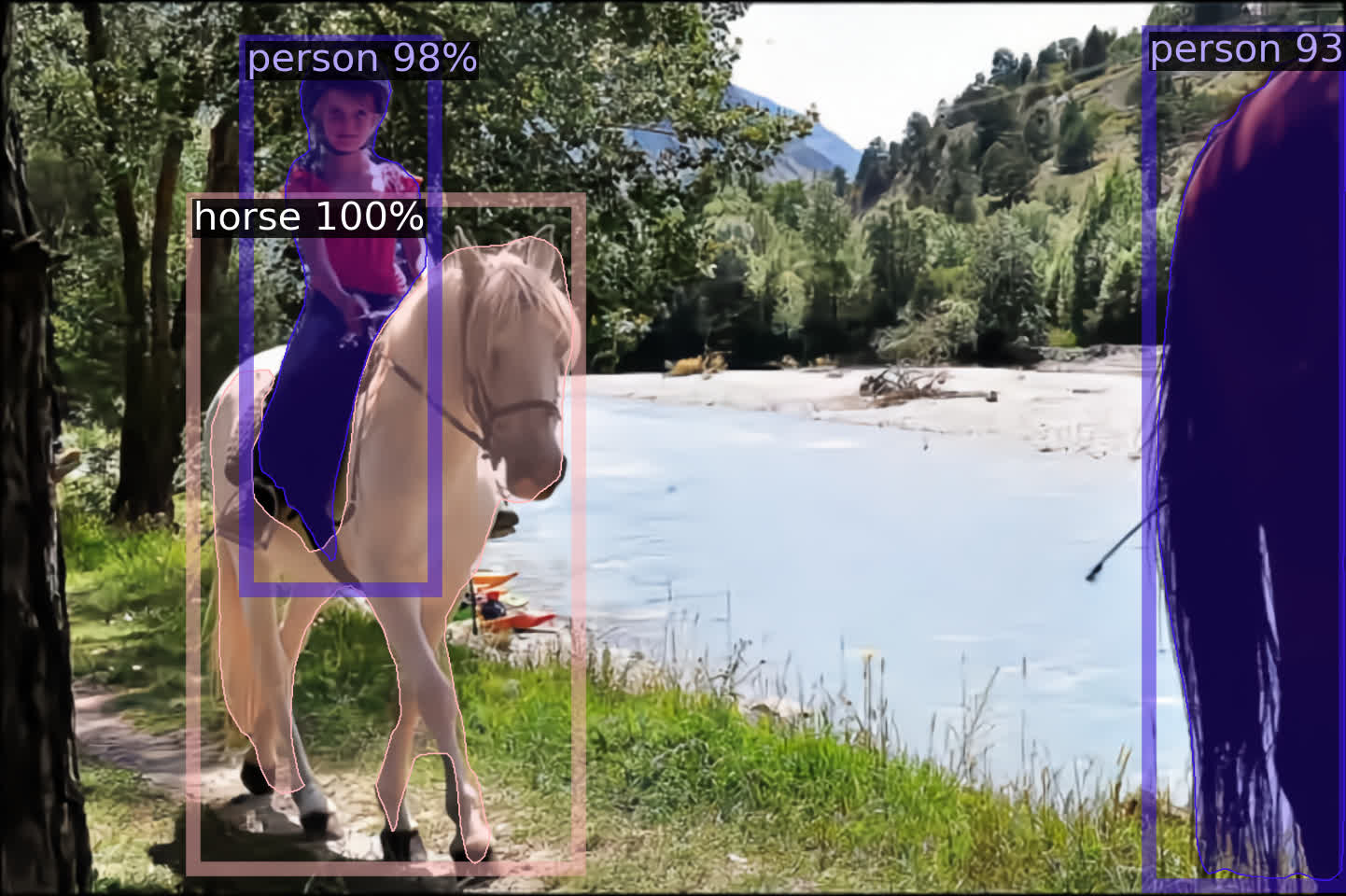}
    \vspace{0.29cm}
    \end{minipage}
    \begin{minipage}[b]{0.161\linewidth}
    \centering
    \includegraphics[width=\textwidth]{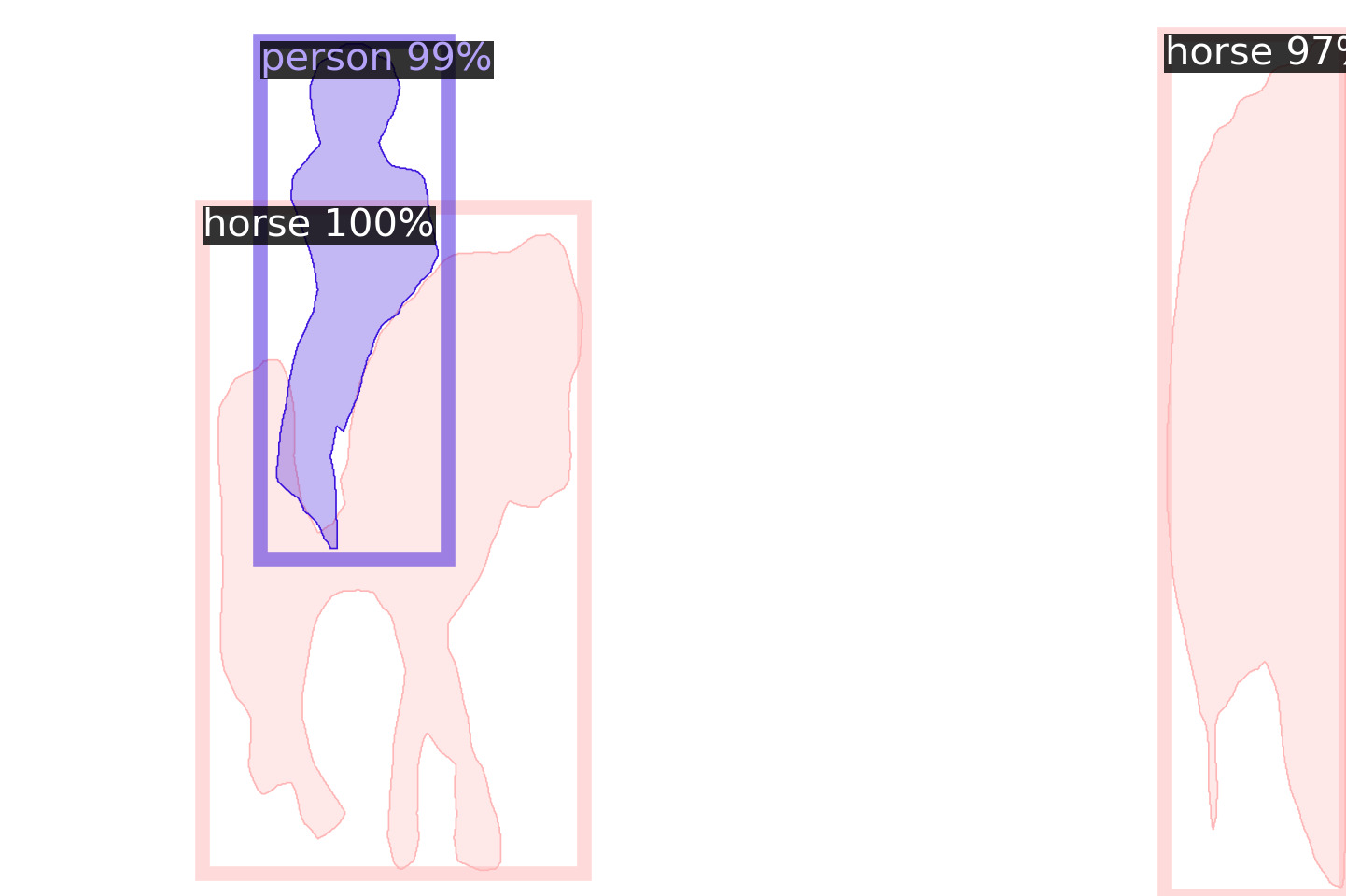}
    \vspace{0.01cm}
    \centerline{0.205 bpp}\medskip
    \end{minipage}
    
    \centering
    \begin{minipage}[b]{0.161\linewidth}
    \centering
    \includegraphics[width=\textwidth]{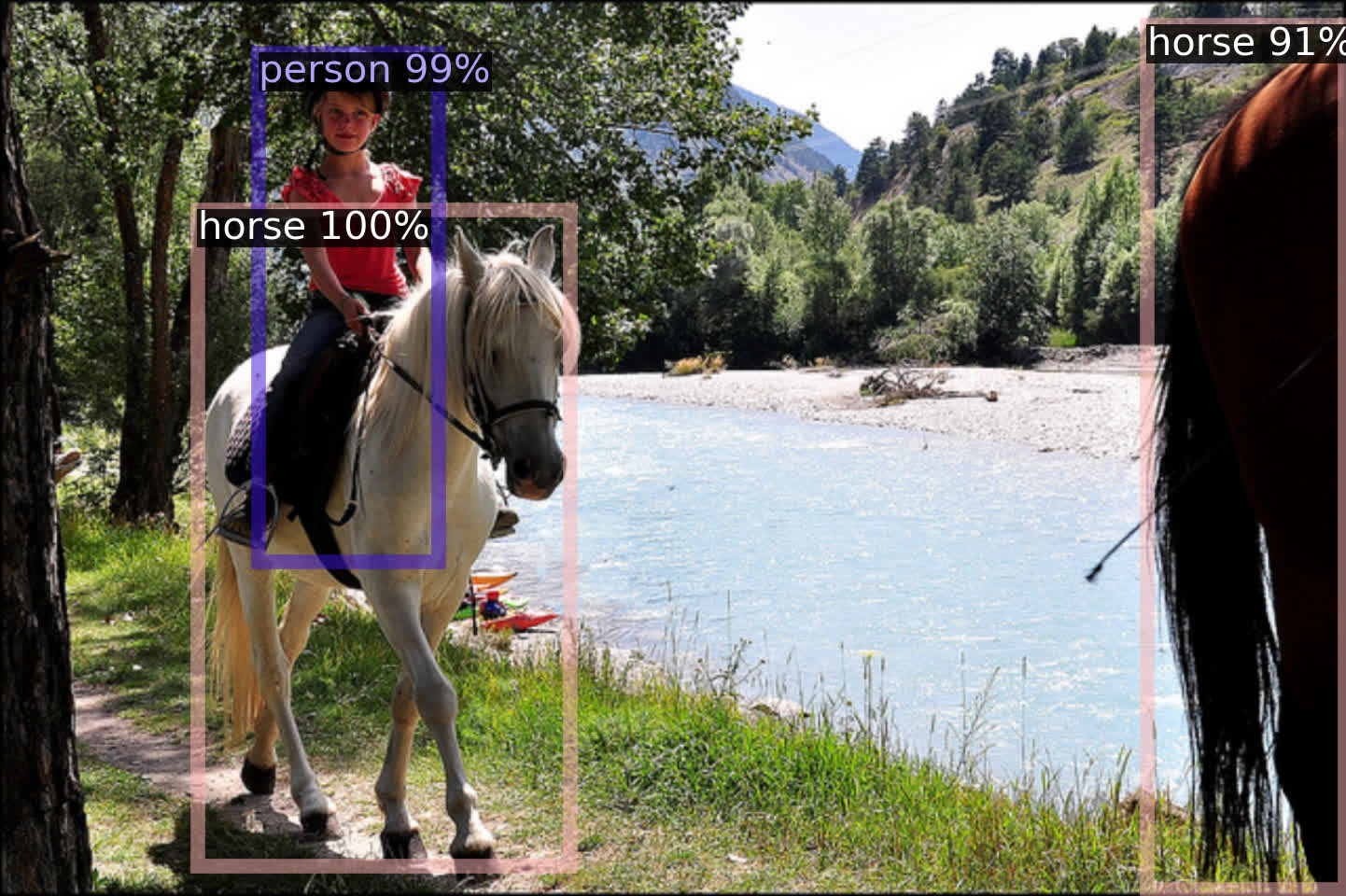}
    \centerline{Object Detection}\medskip
    \end{minipage}
    \begin{minipage}[b]{0.161\linewidth}
    \centering
    \includegraphics[width=\textwidth]{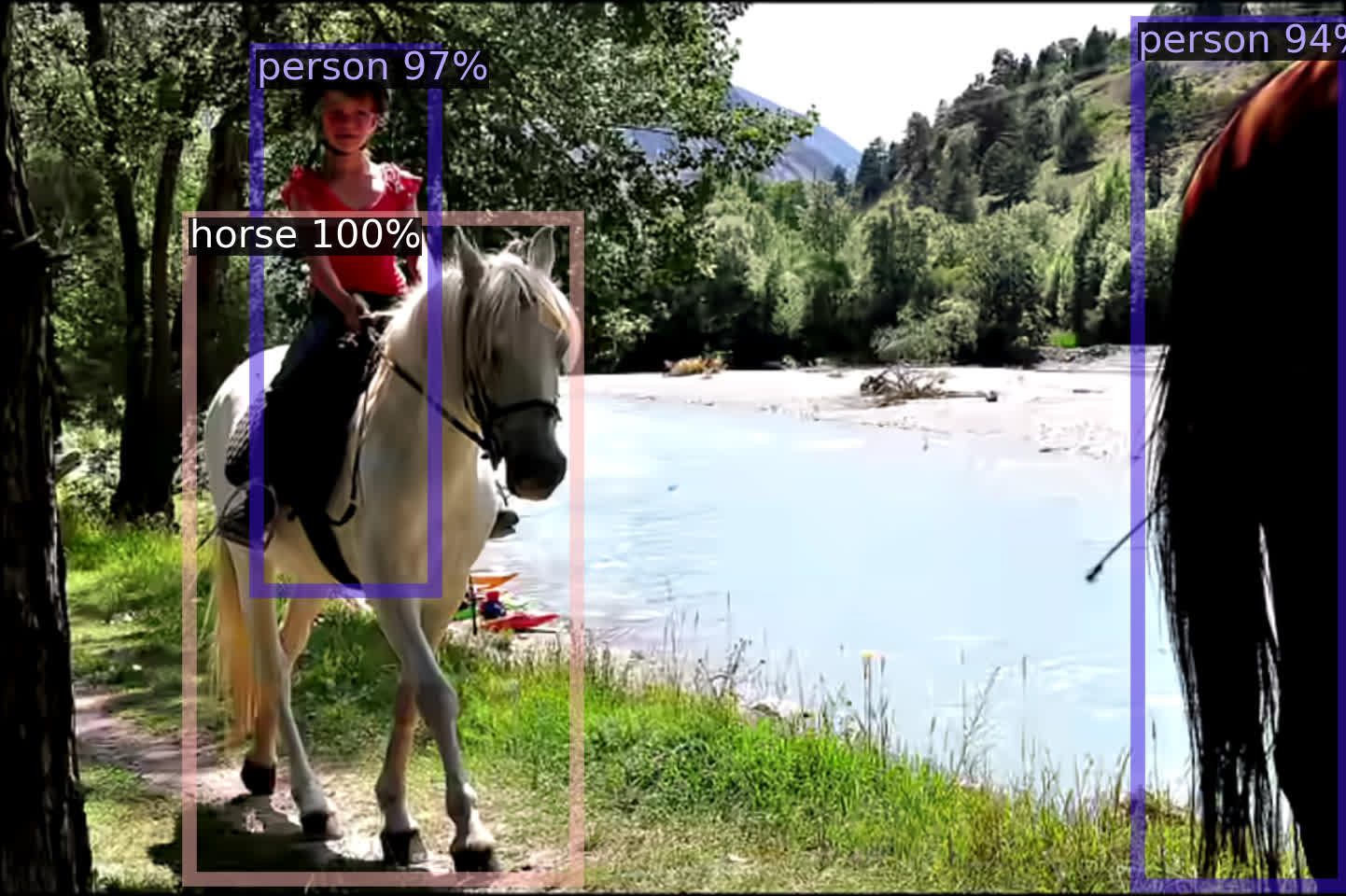}
    \vspace{0.29cm}
    \end{minipage}
    \begin{minipage}[b]{0.161\linewidth}
    \centering
    \includegraphics[width=\textwidth]{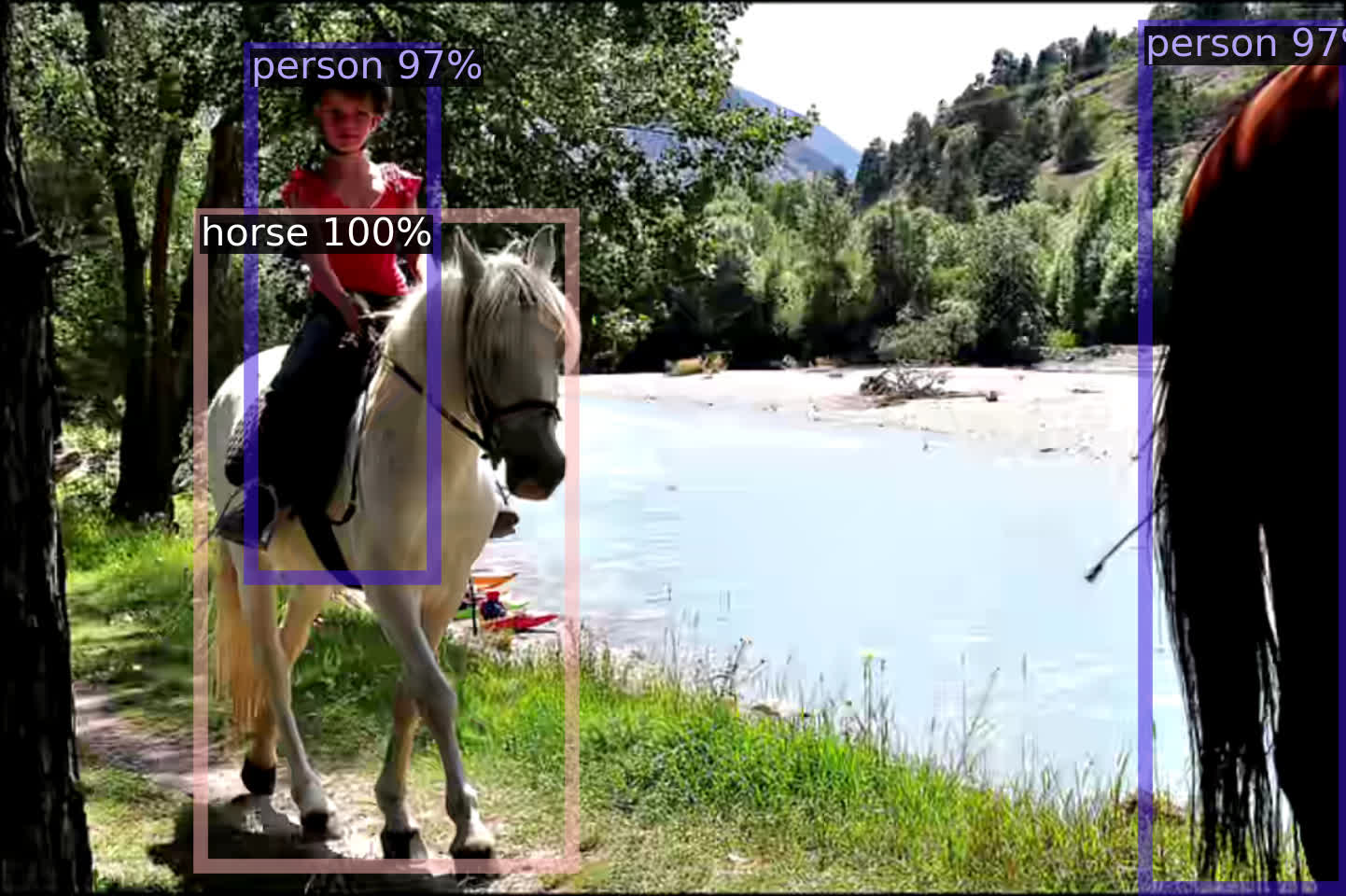}
    \vspace{0.29cm}
    \end{minipage}
    \begin{minipage}[b]{0.161\linewidth}
    \centering
    \includegraphics[width=\textwidth]{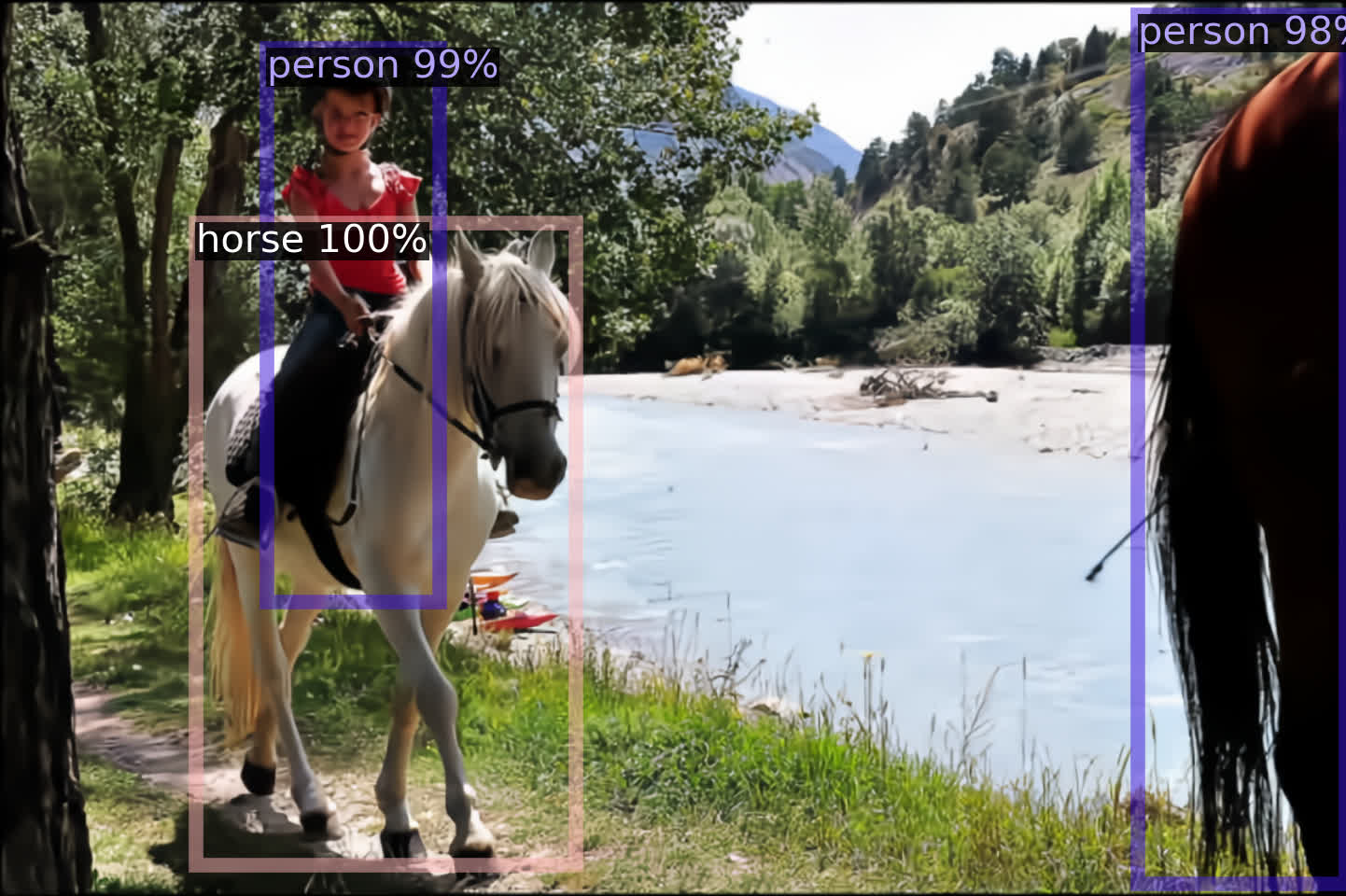}
    \vspace{0.29cm}
    \end{minipage}
    \begin{minipage}[b]{0.161\linewidth}
    \centering
    \includegraphics[width=\textwidth]{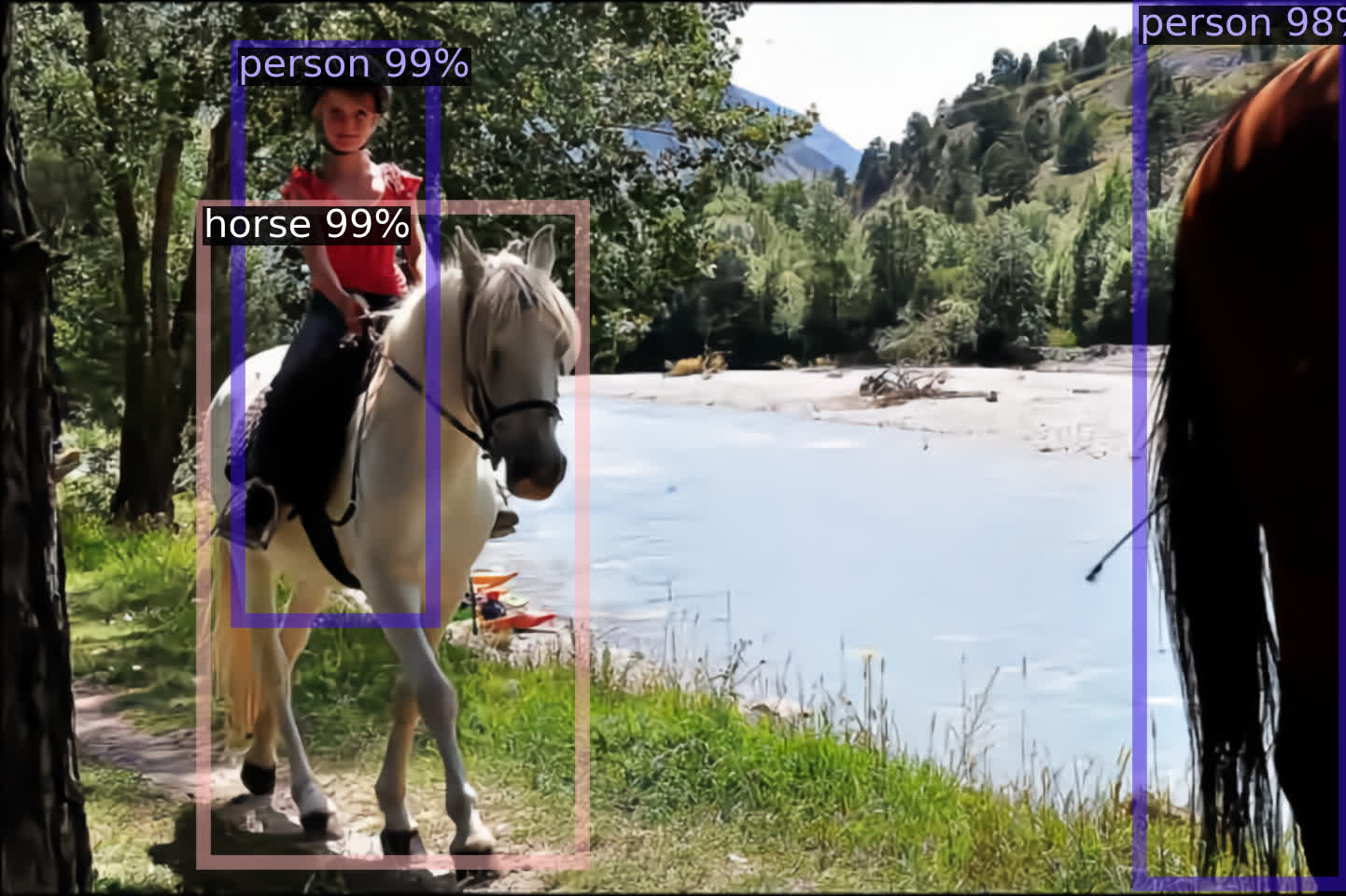}
    \vspace{0.29cm}
    \end{minipage}
    \begin{minipage}[b]{0.161\linewidth}
    \centering
    \includegraphics[width=\textwidth]{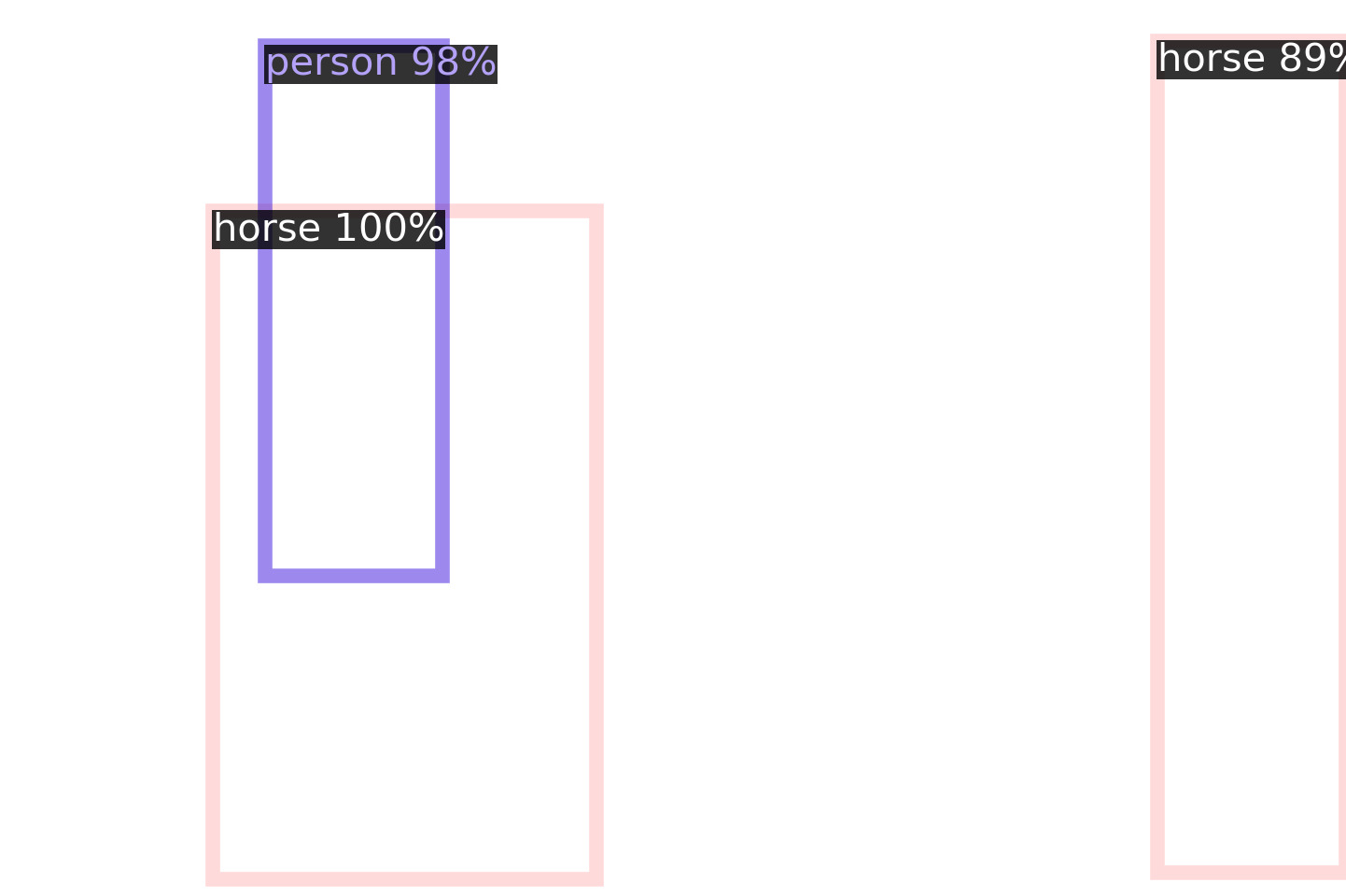}
    \vspace{0.01cm}
    \centerline{0.195 bpp}\medskip
    \end{minipage}
    
    \vspace{0.3cm}
    
    \centering
    \begin{minipage}[b]{0.161\linewidth}
    \centering
    \centerline{\small Original}
    \vspace{0.08cm}
    \includegraphics[width=\textwidth]{Figures/visual_examples/original/000000464089.jpg}
    \vspace{0.215cm}
    \end{minipage}
    \begin{minipage}[b]{0.161\linewidth}
    \centering
    \centerline{\small \textcolor{white}{g}VVC\textcolor{white}{g}}
    \vspace{0.08cm}
    \includegraphics[width=\textwidth]{Figures/visual_examples/vvc/converted_img_464089.jpg}
    \centerline{0.219/27.93}\medskip
    \end{minipage}
    \begin{minipage}[b]{0.161\linewidth}
    \centering
    \centerline{\small \textcolor{white}{g}HEVC\textcolor{white}{g}}
    \vspace{0.08cm}
    \includegraphics[width=\textwidth]{Figures/visual_examples/hevc/converted_img_464089.jpg}
    \centerline{0.214/27.48}\medskip
    \end{minipage}
    \begin{minipage}[b]{0.161\linewidth}
    \centering
    \centerline{\small Cheng $\textit{et al.}$~\cite{cheng2020image}}
    \vspace{0.08cm}
    \includegraphics[width=\textwidth]{Figures/visual_examples/cheng/dec_cheng_q2_img_464089.jpg}
    \centerline{0.203/30.01}\medskip
    \end{minipage}
    \begin{minipage}[b]{0.161\linewidth}
    \centering
    \centerline{\small \textcolor{white}{g}Minnen $\textit{et al.}$~\cite{minnen2018joint}\textcolor{white}{g}}
    \vspace{0.08cm}
    \includegraphics[width=\textwidth]{Figures/visual_examples/mbt/dec_mbt_q2_img_464089.jpg}
    \centerline{0.218/29.94}\medskip
    \end{minipage}
    \begin{minipage}[b]{0.161\linewidth}
    \centering
    \centerline{\small \textcolor{white}{g}Proposed\textcolor{white}{g}}
    \vspace{0.08cm}
    \includegraphics[width=\textwidth]{Figures/visual_examples/proposed/dec_000000464089.jpg}
    \centerline{0.222/28.89}\medskip
    \end{minipage}
    
    \centering
    \begin{minipage}[b]{0.161\linewidth}
    \centering
    \includegraphics[width=\textwidth]{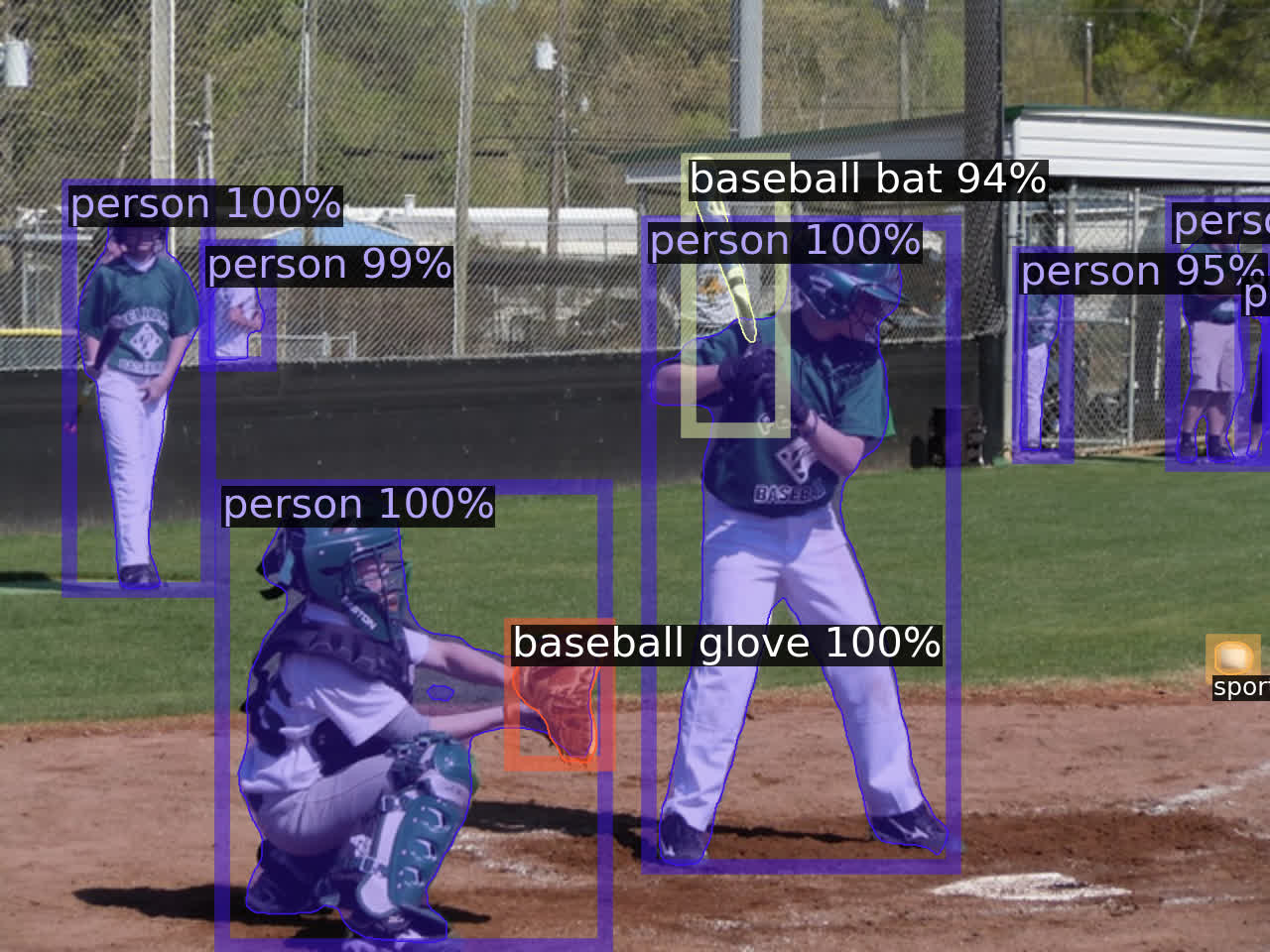}
    \centerline{Segmentation}\medskip
    \end{minipage}
    \begin{minipage}[b]{0.161\linewidth}
    \centering
    \includegraphics[width=\textwidth]{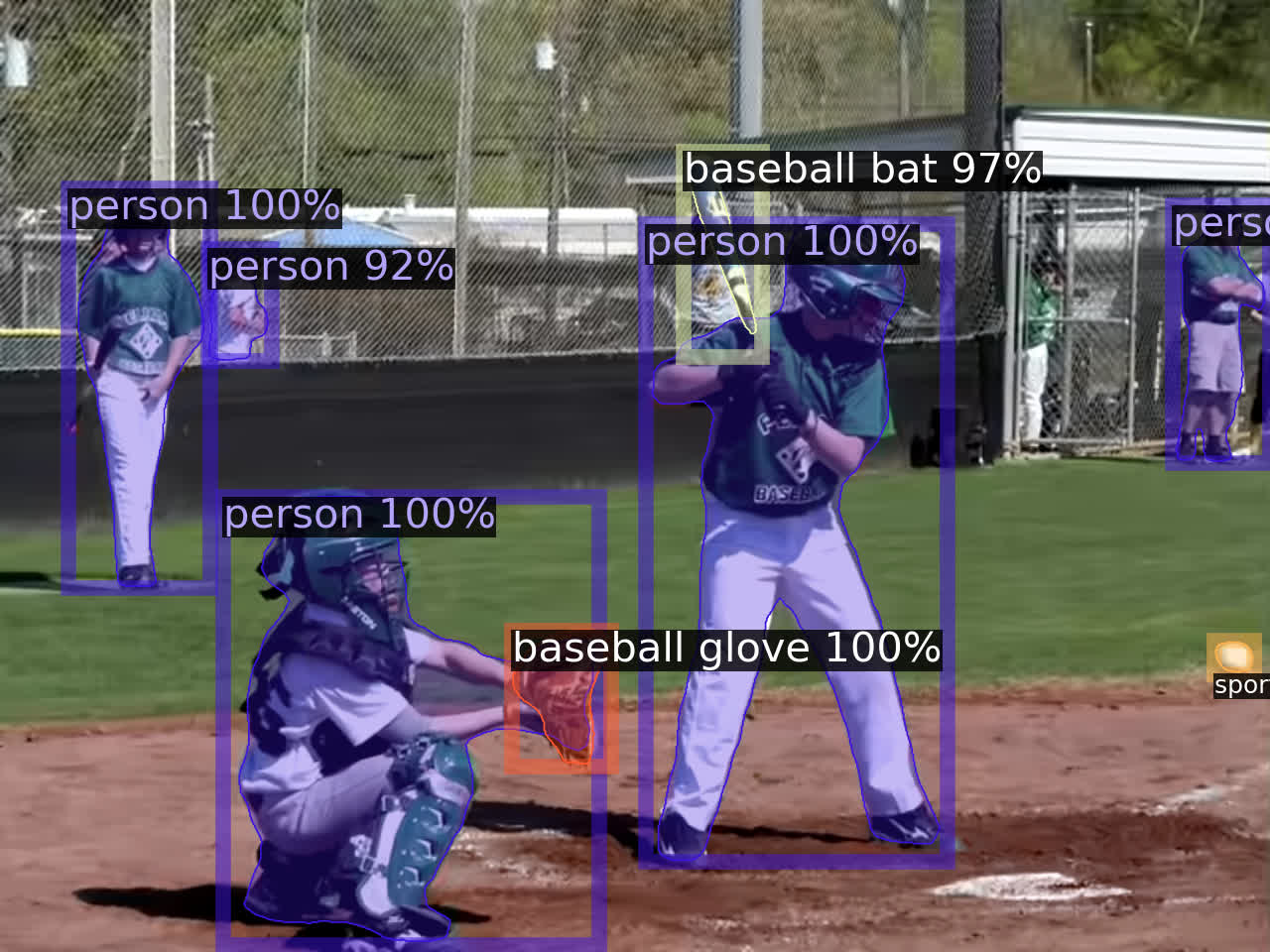}
    \vspace{0.29cm}
    \end{minipage}
    \begin{minipage}[b]{0.161\linewidth}
    \centering
    \includegraphics[width=\textwidth]{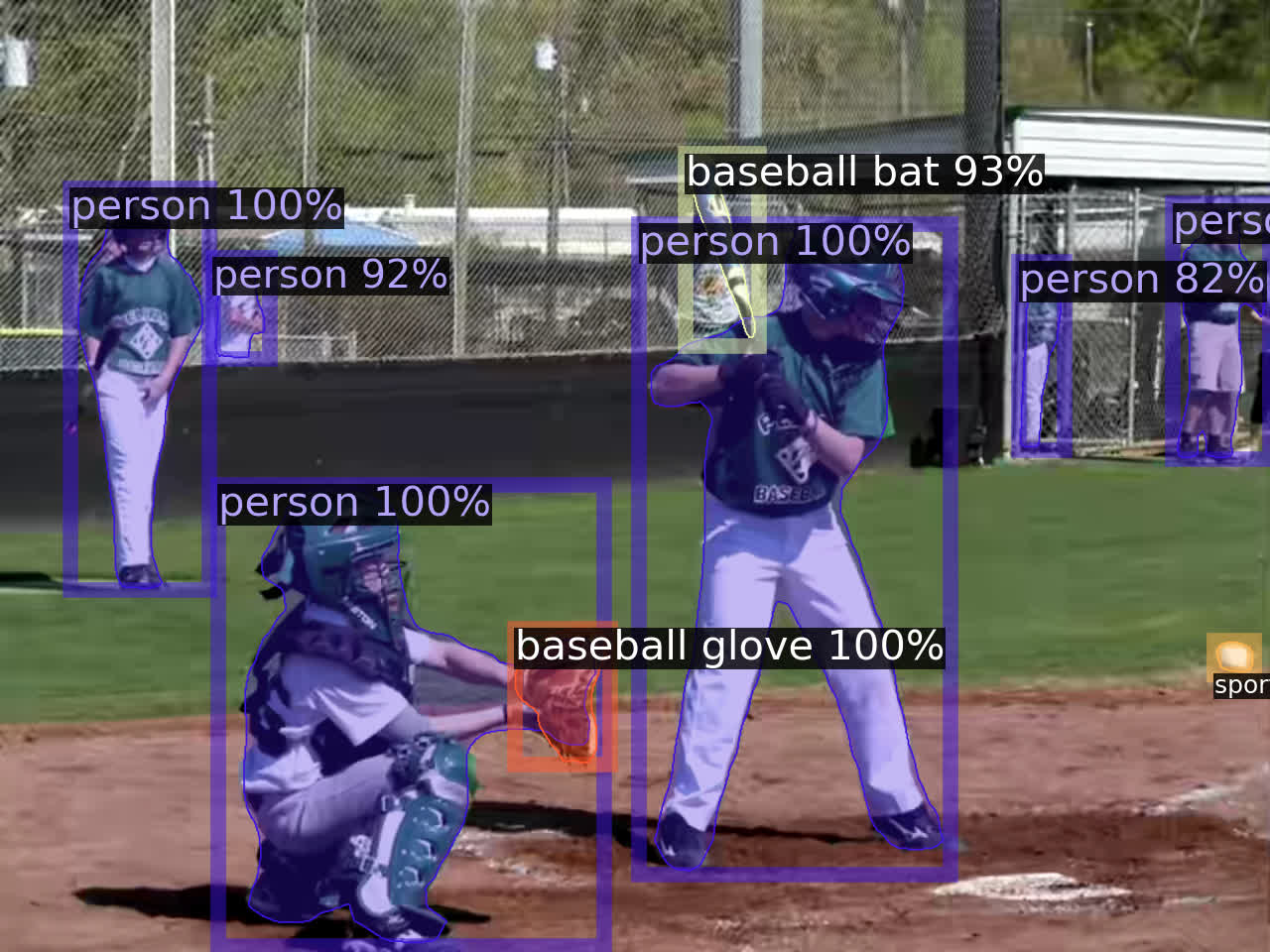}
    \vspace{0.29cm}
    \end{minipage}
    \begin{minipage}[b]{0.161\linewidth}
    \centering
    \includegraphics[width=\textwidth]{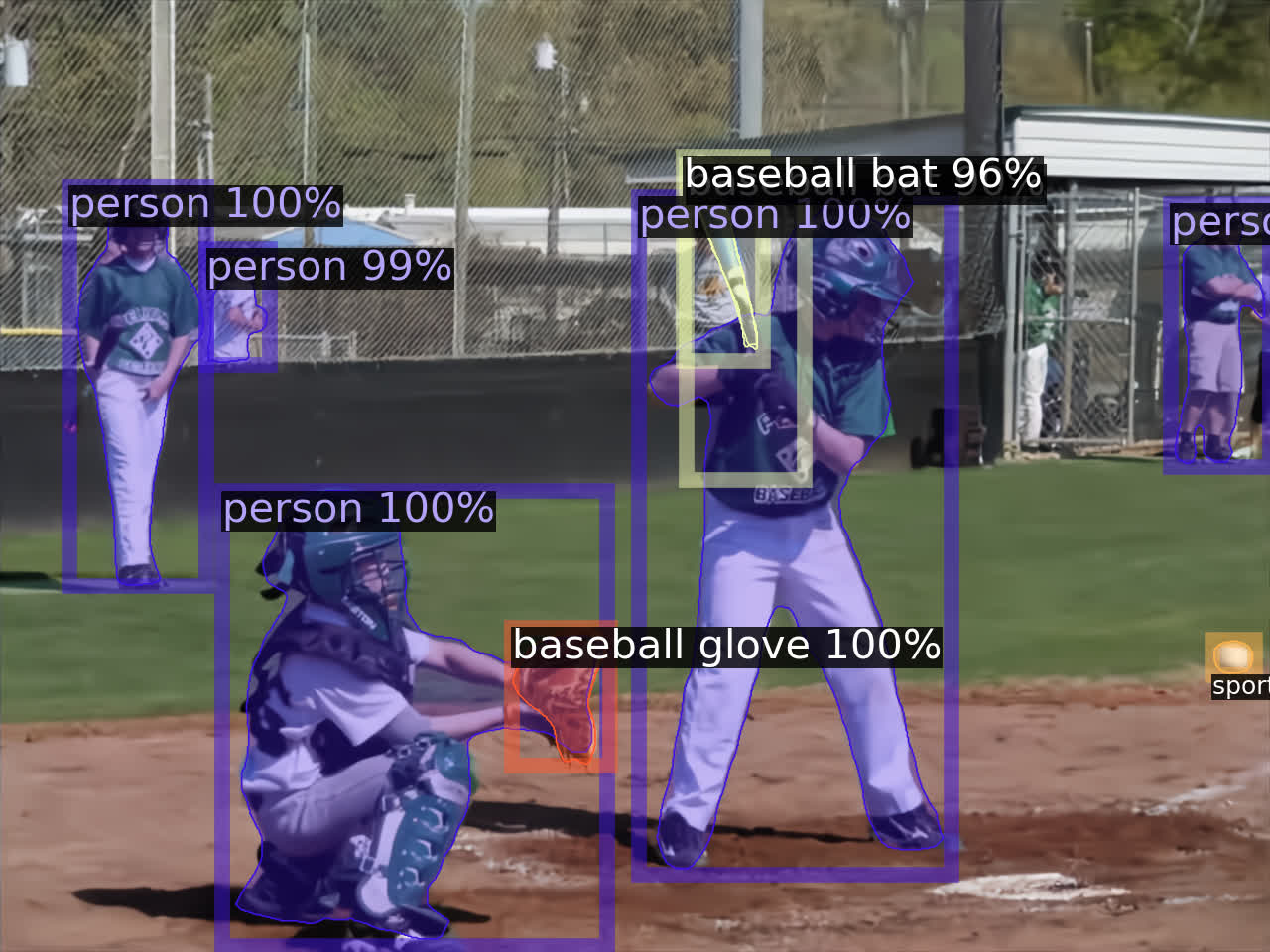}
    \vspace{0.29cm}
    \end{minipage}
    \begin{minipage}[b]{0.161\linewidth}
    \centering
    \includegraphics[width=\textwidth]{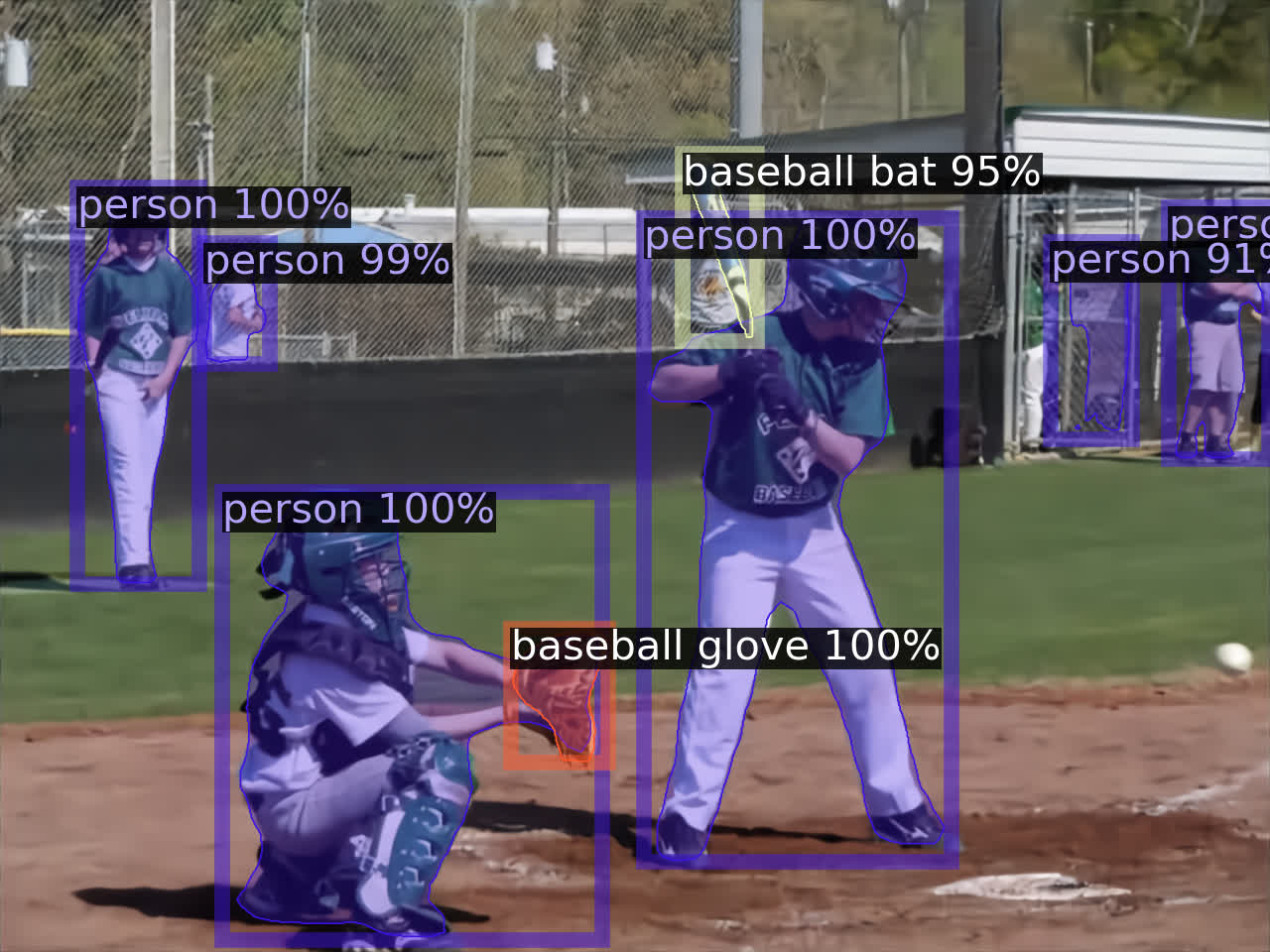}
    \vspace{0.29cm}
    \end{minipage}
    \begin{minipage}[b]{0.161\linewidth}
    \centering
    \includegraphics[width=\textwidth]{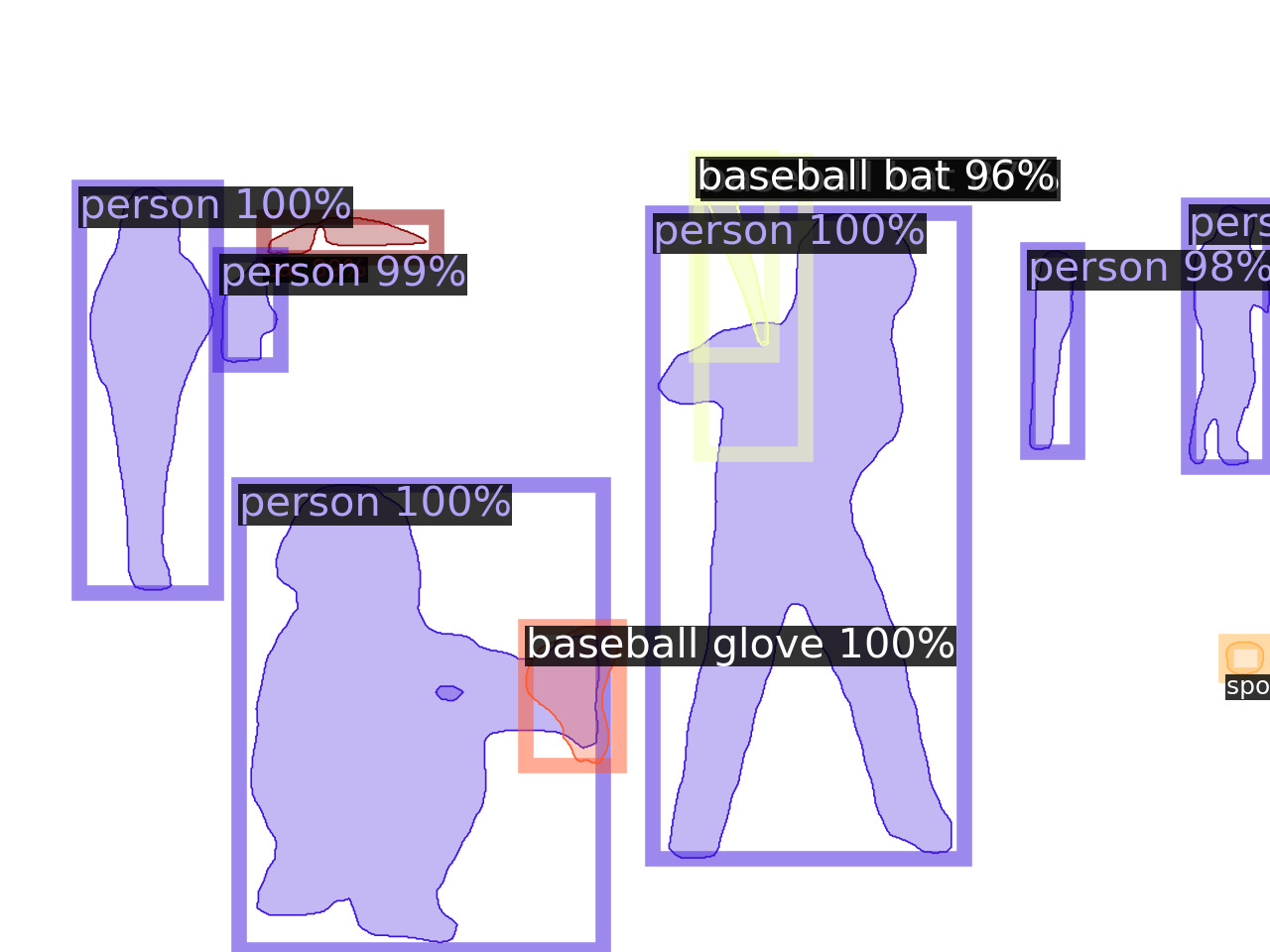}
    \vspace{0.01cm}        
    \centerline{0.215 bpp}\medskip
    \end{minipage}

    \centering
    \begin{minipage}[b]{0.161\linewidth}
    \centering
    \includegraphics[width=\textwidth]{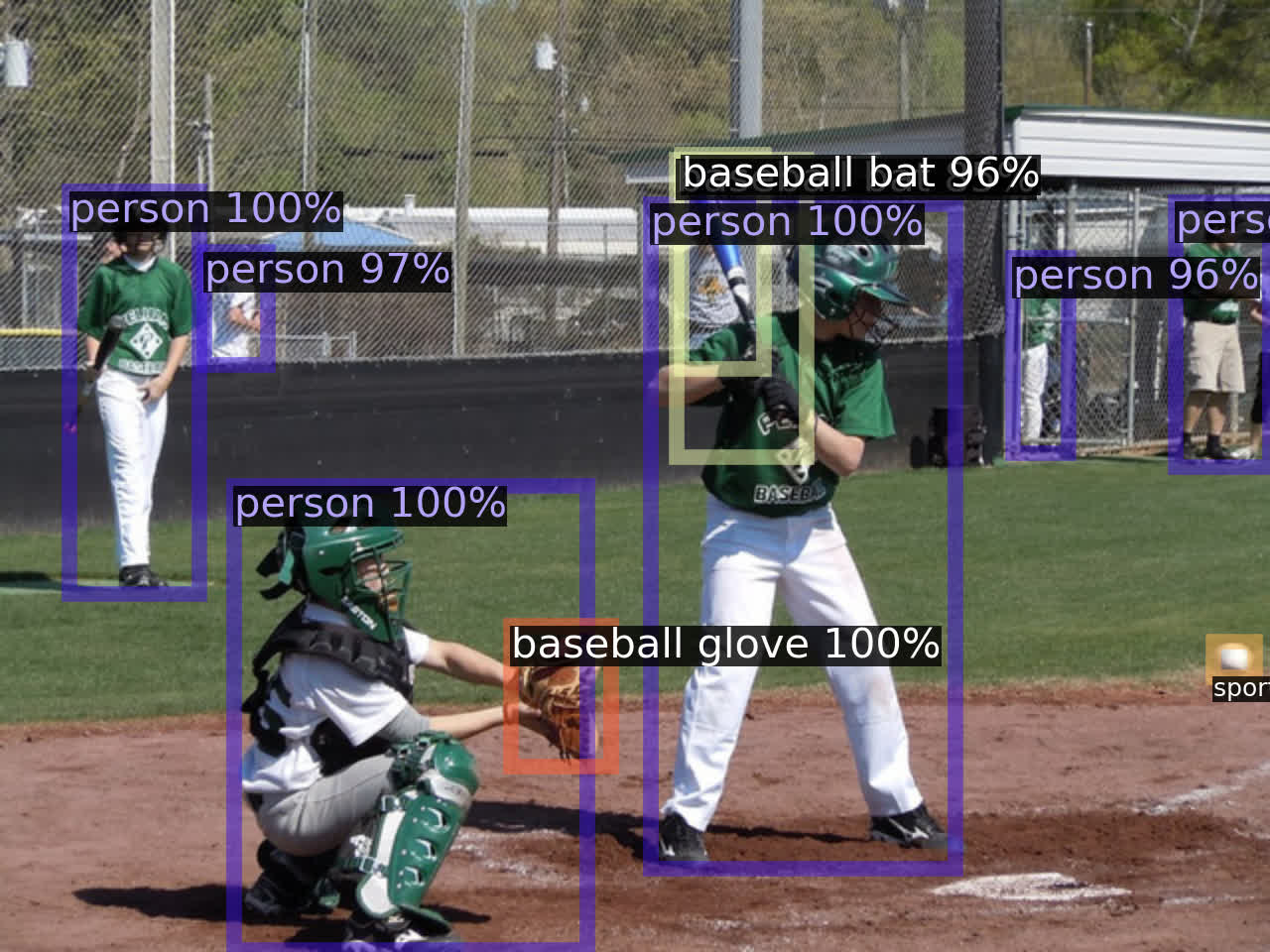}
    \centerline{Object Detection}\medskip
    \end{minipage}
    \begin{minipage}[b]{0.161\linewidth}
    \centering
    \includegraphics[width=\textwidth]{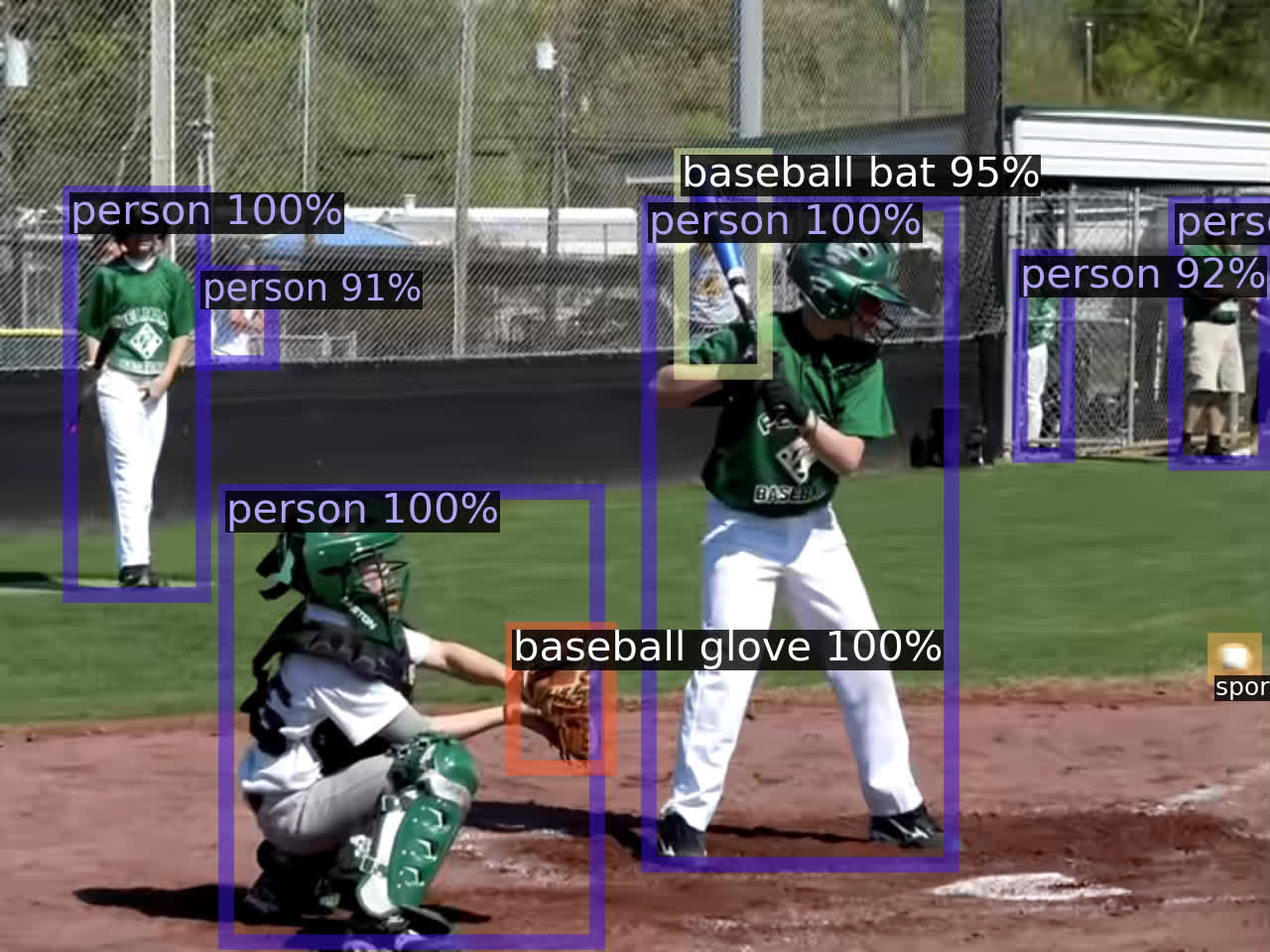}
    \vspace{0.29cm}
    \end{minipage}
    \begin{minipage}[b]{0.161\linewidth}
    \centering
    \includegraphics[width=\textwidth]{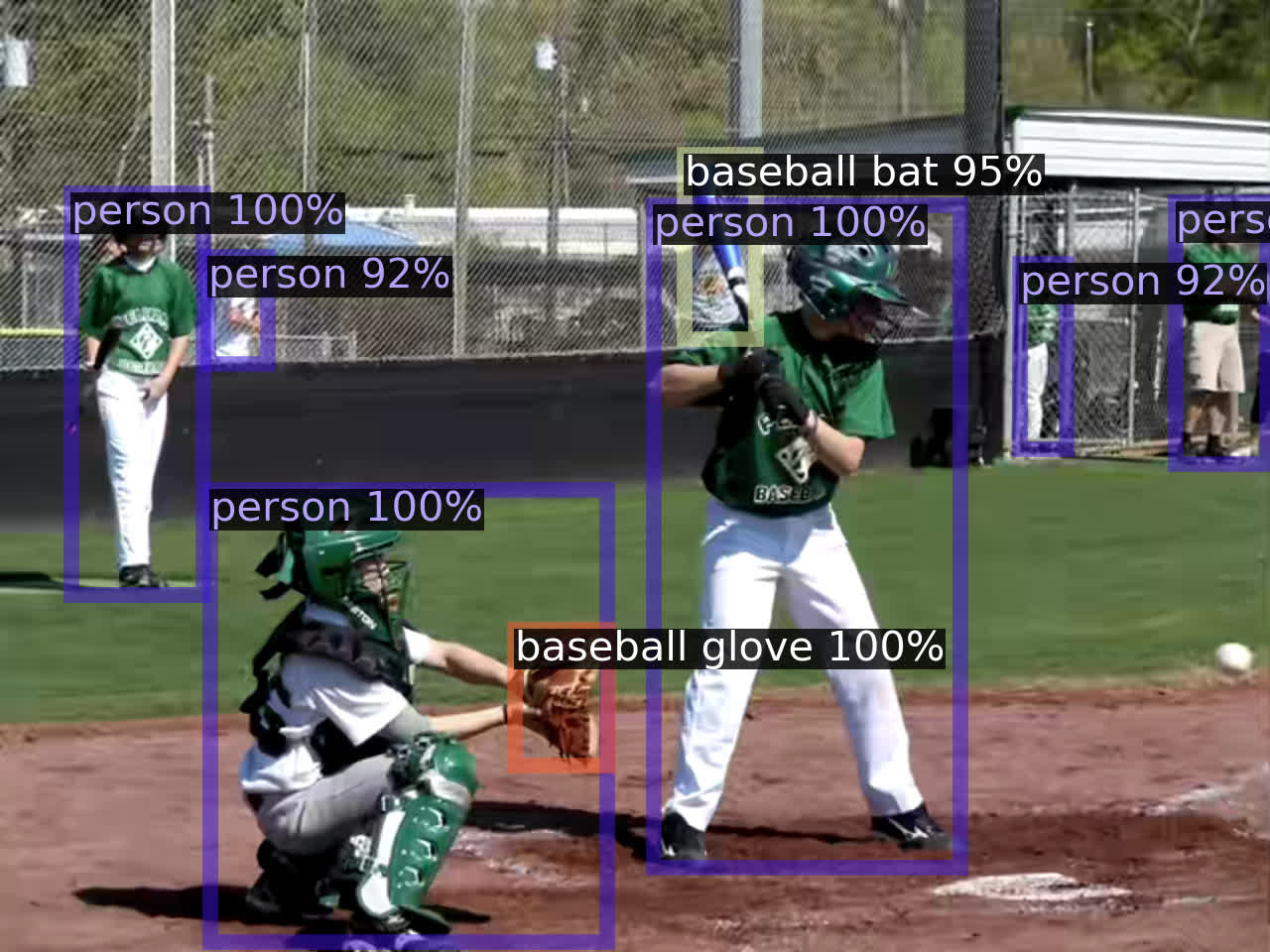}
    \vspace{0.29cm}
    \end{minipage}
    \begin{minipage}[b]{0.161\linewidth}
    \centering
    \includegraphics[width=\textwidth]{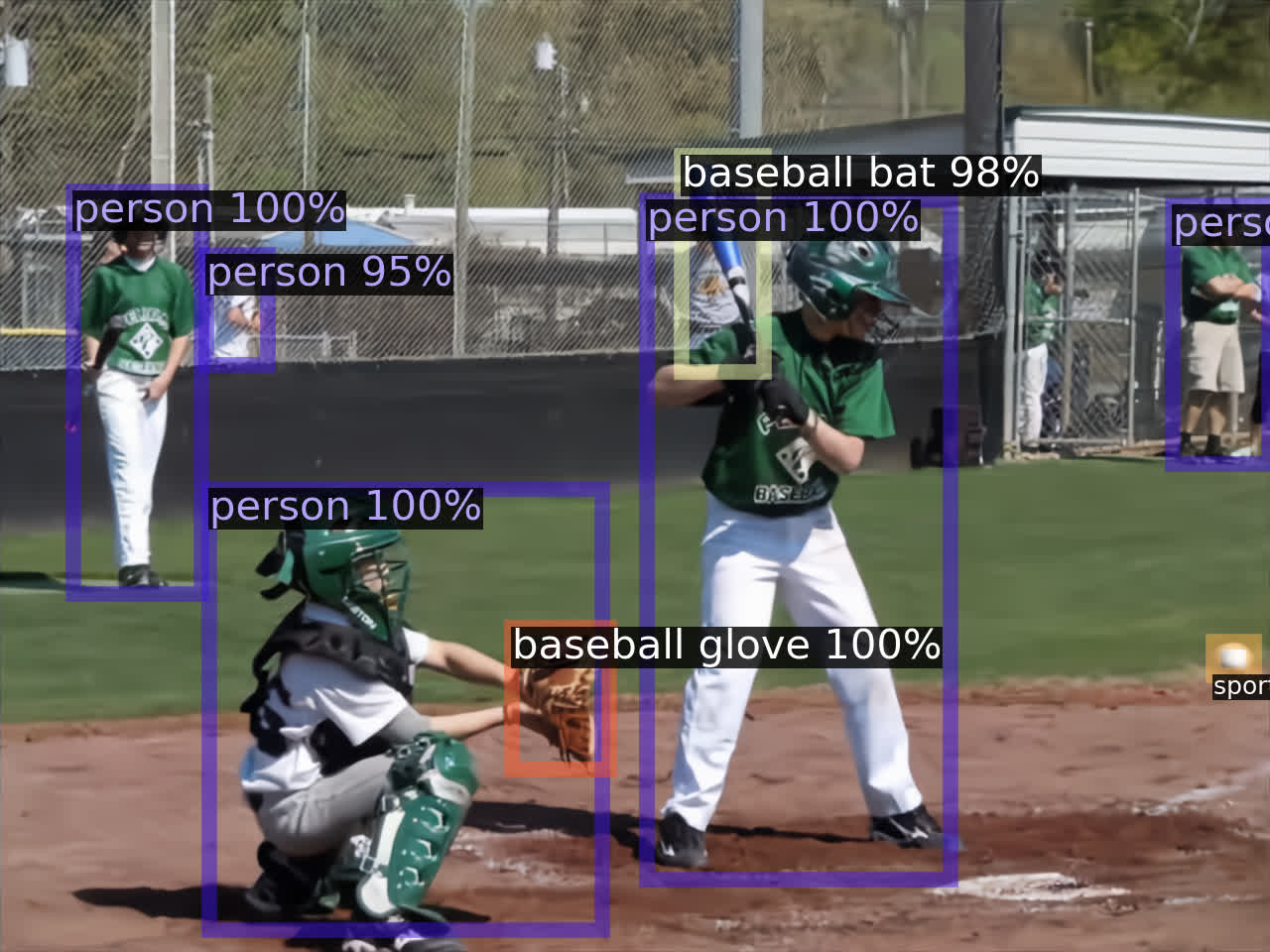}
    \vspace{0.29cm}
    \end{minipage}
    \begin{minipage}[b]{0.161\linewidth}
    \centering
    \includegraphics[width=\textwidth]{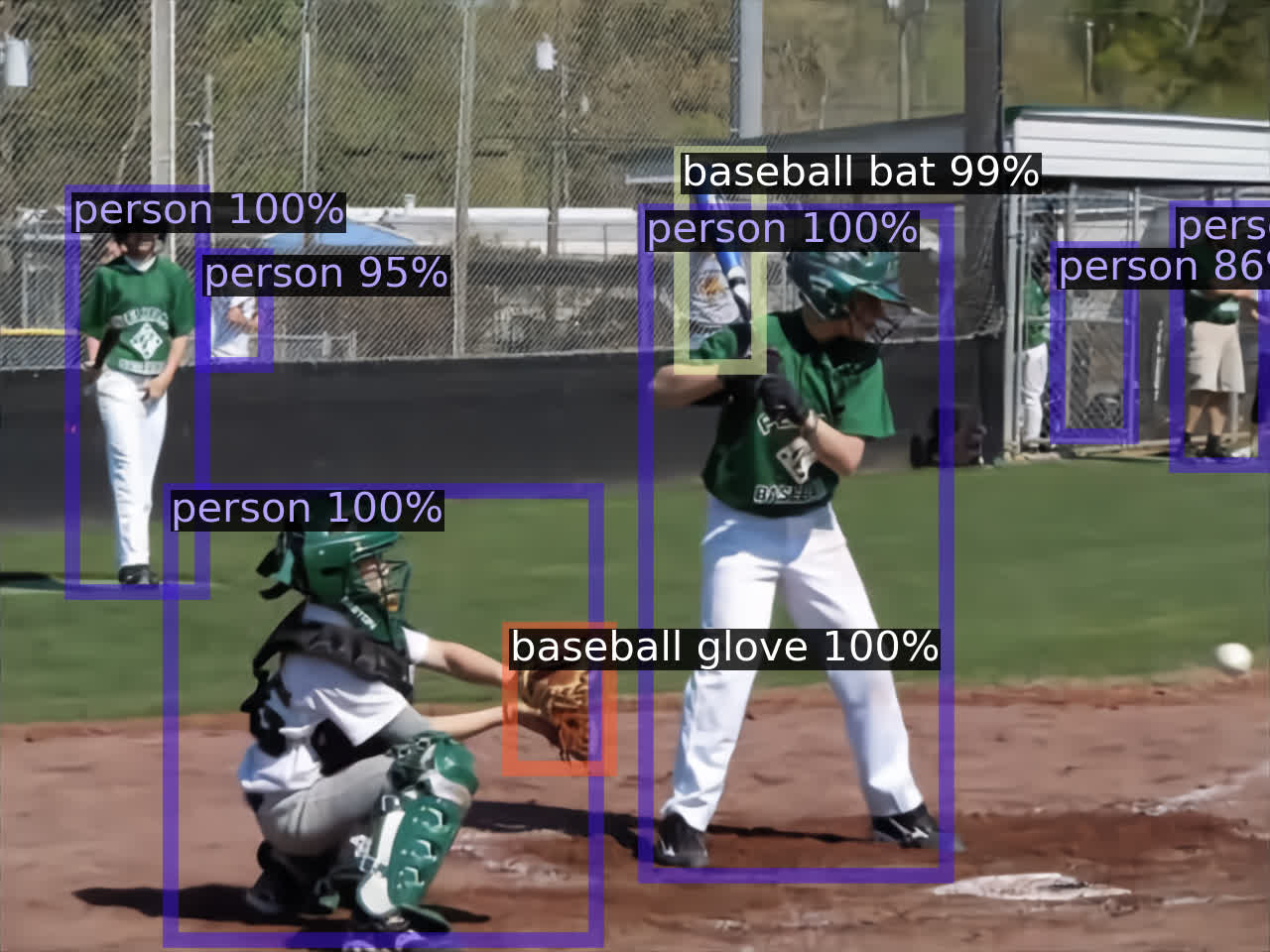}
    \vspace{0.29cm}
    \end{minipage}
    \begin{minipage}[b]{0.161\linewidth}
    \centering
    \includegraphics[width=\textwidth]{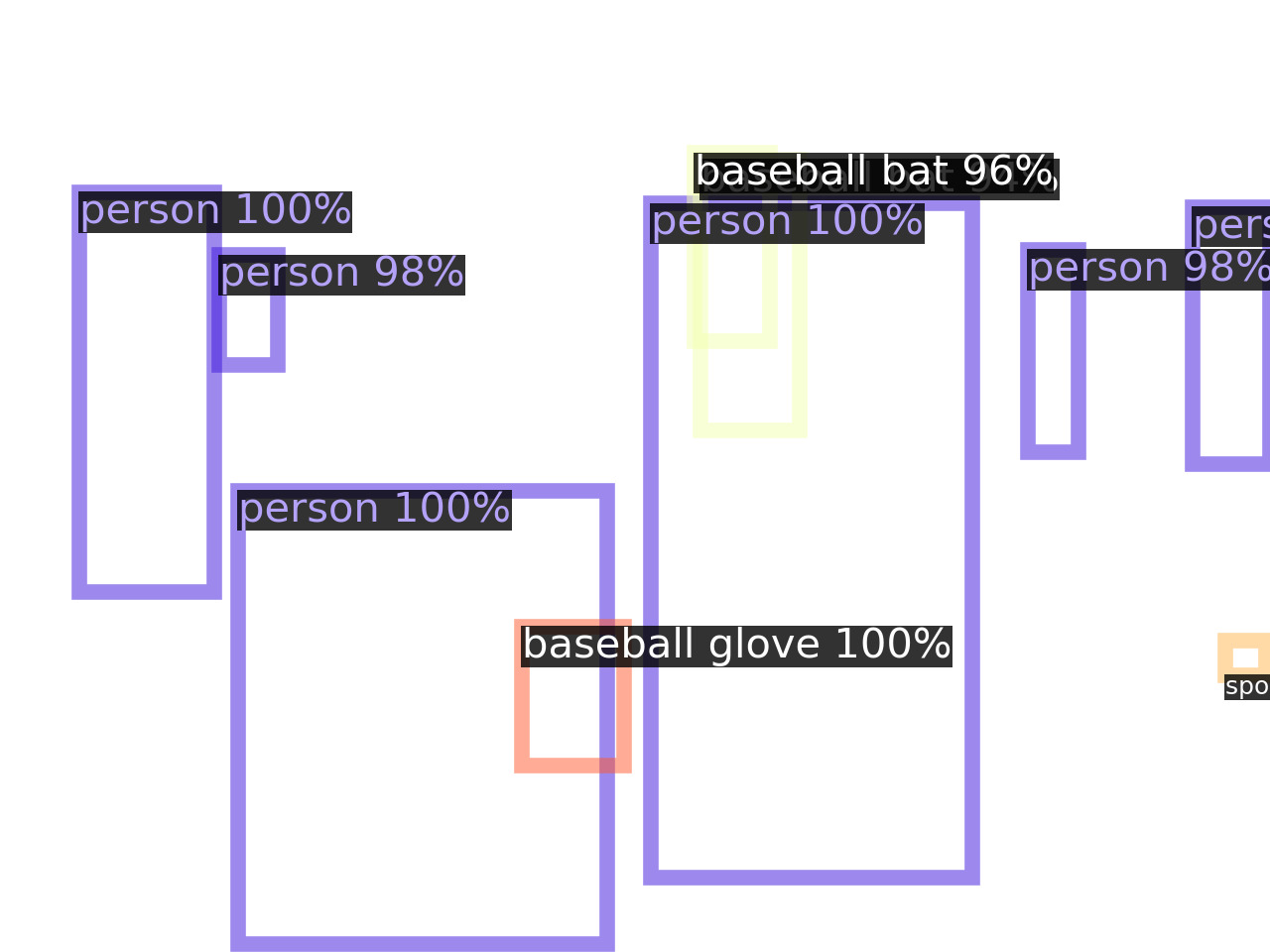}
    \vspace{0.01cm}    
    \centerline{0.211 bpp}\medskip
    \end{minipage}
    
\caption{Illustration of object detection, segmentation, and input reconstruction results for several benchmark methods and our 3-layer network on two images: the top three rows correspond to the first image, and the bottom three rows to the second image. For each input image, the top row shows input reconstruction, the next row shows object segmentation, and the last row shows object detection. The first column shows the original images and the results of object detection and segmentation produced by the FPN-based R-CNNs applied to those images. The detection and segmentation results are overlaid on the original images. The next four columns show the images encoded and decoded by various benchmarks, and the FPN-based R-CNN results produced from decoded images, again overlaid on those images. Two numbers are shown beneath each image, indicating the bitrate and RGB PSNR, respectively, in the format bpp/PSNR(dB). The last column shows the outputs of our 3-layer network. Object detection and segmentation results are shown on a blank background, because our system does not need to reconstruct the input image in order to perform object detection or segmentation. For these cases, only the bitrate is shown beneath each result. To reduce the clutter, in all cases we only show objects with a confidence score of over 80\%.}
\label{fig:visual_outcomes}
\end{figure*}

One of the key contributions of the present paper is that our scalable DNN-based image coding approach is able to support birate-efficient high-quality machine vision without resorting to input reconstruction.  Fig.~\ref{fig:visual_outcomes} shows two examples of the outputs of our 3-layer network, along with the results obtained by the benchmarks. For each example, the first row shows the input image and the reconstructed images with the corresponding bitrate and RGB PSNR (bpp/dB). The next two rows show the results of object segmentation and object detection. For the benchmarks, reconstructed image is fed to the corresponding model (Faster R-CNN~\cite{ren2016faster} for detection,  Mask R-CNN~\cite{he2017mask} for segmentation) with pre-trained weights to obtain the results. For our 3-layer network, only the corresponding part of the bitstream is decoded. Hence, since the input is not reconstructed in these cases, the results are shown 
on the empty background, and the corresponding rate is indicated below the image. 

In the first example, our network successfully detects and segments all three objects, with bitrates of 0.195 bpp for detection and 0.205 bpp for segmentation. In contrast, all benchmarks lead to mislabelling of a horse on the right as a person, even in the case of object detection, despite the fact they use more bits than our base layer. In the second example, benchmark-coded images lead to some missing objects, while our network correctly detect them all. For example, the image coded by~\cite{minnen2018joint} leads to missing the second person from the right in the background, as well as the baseball. Also, image coded by~\cite{cheng2020image} leads to missing the person in the background. With conventional codecs (HEVC and VVC), either the person or the baseball are missed. 
These examples illustrate why our 3-layer network provides superior performance in terms of object detection and segmentation in Fig.~\ref{fig:three_layer_vision}.

\section{Conclusion}
\label{sec:conclusion}
We presented a DNN-based image compression framework with latent-space scalability for human and machine vision. Latent image representation is coded into multiple layers, which can be separately decoded to enable the required task. We embodied the proposed ideas into 2- and 3-layer multi-task networks supporting object detection, segmentation, and input reconstruction. 
Mutual information estimates show that the  proposed loss function facilitates steering of relevant task-specific information into the corresponding portions of the latent space during training. The experiments show that our multi-task networks provide 37\% - 80\% bitrate savings on machine vision tasks compared to relevant benchmarks, while being comparable to state of the art image codecs in terms of input reconstruction quality. 



%




\ifCLASSOPTIONcaptionsoff
  \newpage
\fi



%

\bibliographystyle{IEEEbib-abbrev}
\bibliography{ref}

\appendices
\section{Deep feature compressibility}
\label{sec:feature_compressibility}
In this appendix, we prove that intermediate latent representations in a non-generative DNN are at least as compressible as it's input, both in terms of lossless compression and lossy compression. This provides some theoretical support for DNN-based image codecs, as well as collaborative intelligence applications~\cite{BLT_CI_ICASSP2021}, where intermediate DNN features computed from sensed signals are sent from the edge to the cloud for inference. The source of randomness in our analysis is the selection of DNN input data, so unless otherwise stated, all expectations are over the input data distribution. 

The proofs presented here are consequences of the Data Processing Inequality (DPI)~\cite{Cover_Thomas_2006,Yeung2002}, which we review briefly here. Random variables $X_1$, $X_2$, ..., $X_n$ form a Markov chain, indicated as $X_1 \to X_2 \to ... \to X_n$, if their joint distribution can be factored as $p(x_1,x_2,...,x_n)=p(x_1)p(x_2|x_1) ... p(x_n|x_{n-1})$. In that case, the chain also satisfies the Markov property in reverse, that is, $X_n \to ... \to X_2 \to X_1$ is also a Markov chain~\cite{Yeung2002}. Note that $X \to X \to Y$ is trivially a Markov chain. 
It has been noted that successive layers in non-generative feedforward networks form a Markov chain~\cite{tishby2015deep}. It is easy to see that if $f(\cdot)$ is a function, then $X \to Y \to f(Y)$ is a Markov chain.  A crucial result related to Markov chains that we will make repeated use of is the \emph{data processing inequality} (DPI)~\cite{Yeung2002}, which says that if $U \to X \to Y \to V$ is a Markov chain, then $I(U;V) \le I(X;Y)$. A special case of DPI~\cite{Cover_Thomas_2006} states that for $X \to Y \to V$, $I(X;V) \le I(X;Y)$.

For the DNN-related notation, input is denoted $\mathbf{X}$, $l$-th layer's output $\mathbfcal{Y}^{(l)}$, and the DNN inference output is $T$. We will focus on \emph{non-generative} feedforward networks, which do not introduce any randomness in their processing. Examples of such networks include most well-known DNNs, e.g., ResNet~\cite{he2016deep}, YOLO~\cite{Redmon2018_yolov3}, etc. In such networks, for fixed network weights, the output of any layer is uniquely determined by the input through feedforward processing. 

First we focus on the lossless compression case. Consider a deep feedforward non-generative network and let $\{\mathbfcal{Y}^{(1)}, \mathbfcal{Y}^{(2)},  ..., \mathbfcal{Y}^{(n)}\}$ be the outputs of $n$ of it's layers. 

\begin{theorem}
\label{thm:lossless}
The joint entropy of $\{\mathbfcal{Y}^{(1)}, \mathbfcal{Y}^{(2)},  ..., \mathbfcal{Y}^{(n)}\}$ is upper bounded by the entropy of the input $\mathbf{X}$, that is,
\begin{equation}
    H(\mathbfcal{Y}^{(1)}, \mathbfcal{Y}^{(2)},  ..., \mathbfcal{Y}^{(n)}) \leq H(\mathbf{X}),
    \label{eq:entropy_bound}
\end{equation}
with equality if and only if the input $\mathbf{X}$ can be reconstructed exactly from $\{\mathbfcal{Y}^{(1)}, \mathbfcal{Y}^{(2)},  ..., \mathbfcal{Y}^{(n)}\}$. 
\end{theorem}

\begin{proof}
Let $\mathbfcal{Y} = (\mathbfcal{Y}^{(1)}, \mathbfcal{Y}^{(2)},  ..., \mathbfcal{Y}^{(n)})$. Note that $\mathbfcal{Y}$ is a deterministic function of the input $\mathbf{X}$: since the model is non-generative, when $\mathbf{X}$ is input to the model, $\mathbfcal{Y}$
is uniquely determined by passing the signal forward through to all the $n$ layers of interest. Hence, when  $\mathbf{X}$ is given, there is no uncertainty about $\mathbfcal{Y}$, so  $H(\mathbfcal{Y}|\mathbf{X}) = 0$. By considering the Markov chain
$\mathbf{X} \to \mathbf{X} \to \mathbfcal{Y}$ and using the DPI, we can write
\begin{equation}
  \begin{split}
    H(\mathbf{X}) = I(\mathbf{X}; \mathbf{X})
    &\geq I(\mathbf{X}; \mathbfcal{Y}) \\
    &= H(\mathbfcal{Y}) - H(\mathbfcal{Y}|\mathbf{X}) \\
    &= H(\mathbfcal{Y}) \\
    &= H(\mathbfcal{Y}^{(1)}, \mathbfcal{Y}^{(2)}, \ldots, \mathbfcal{Y}^{(n)}).
  \end{split}
  \label{eq:entropy_multi-stream}
\end{equation}

This proves the claimed inequality. To see when equality holds in~(\ref{eq:entropy_multi-stream}), we start with $H(\mathbfcal{Y})=I(\mathbf{X}; \mathbfcal{Y})$, which is clear from~(\ref{eq:entropy_multi-stream}), and expand mutual information as follows:
\begin{equation}
    H(\mathbfcal{Y}) = I(\mathbf{X};\mathbfcal{Y}) = H(\mathbf{X}) - H(\mathbf{X}|\mathbfcal{Y}).
\label{eq:entropy_equality}
\end{equation}
So, in order to have $H(\mathbf{X})=H(\mathbfcal{Y})$, we must have $H(\mathbf{X}|\mathbfcal{Y})=0$. That is to say, the input $\mathbf{X}$ must be a deterministic function of $\mathbfcal{Y}$, which means that one must be able to reconstruct the input $\mathbf{X}$ from the layers' outputs $\{\mathbfcal{Y}^{(1)}, \mathbfcal{Y}^{(2)},  ..., \mathbfcal{Y}^{(n)}\}$ exactly. 
\end{proof}

Special cases of~(\ref{eq:entropy_multi-stream}) have been presented in the literature on information bottleneck in~\cite{saxe,shwartz-ziv_opening_2017}, but for a single layer rather than multiple intermediate layers. However, considering a collection of layers $\{\mathbfcal{Y}^{(1)}, \mathbfcal{Y}^{(2)},  ..., \mathbfcal{Y}^{(n)}\}$ rather than just a single layer $\mathbfcal{Y}^{(l)}$ is a practical necessity, because many useful models
have skip/parallel connections across layers, so compression of multiple intermediate layers may be needed. The above theorem  shows that the joint entropy of the outputs of any collection of DNN layers is no larger than the entropy of the input. According to this result, in the limit, we should be able to compress (losslessly, for now) outputs from any number of internal DNN layers at least as efficiently as the input, so long as we compress them jointly.

We next turn our attention to lossy compression, which is somewhat more involved. The central concept in lossy compression the \emph{rate-distortion function} $R(D)$, which, for a source $\mathbf{X}$, is defined as~\cite{Cover_Thomas_2006}:
\begin{equation}
    R(D) = \min_{p(\hat{\mathbf{x}}|\mathbf{x})~:~\sum_{(\mathbf{x},\hat{\mathbf{x}})} \left[p(\mathbf{x})p(\hat{\mathbf{x}}|\mathbf{x})d(\mathbf{x},\hat{\mathbf{x}})\right]~\leq~D} I(\mathbf{X};\widehat{\mathbf{X}}).
    \label{eq:RD_function}
\end{equation}
In the above equation, $\widehat{\mathbf{X}}$ is the quantized version of $\mathbf{X}$ -- this process of quantization, where values of $\mathbf{X}$ are represented by another, generally smaller set of values $\widehat{\mathbf{X}}$, is what causes ``loss'' in lossy compression. For a given pair  $(\mathbf{x},\hat{\mathbf{x}})$, the discrepancy is measured by a distortion metric $d(\mathbf{x},\hat{\mathbf{x}})$. 
The summation below the $\min$ operator computes the expected distortion $\mathbb{E}[d(\mathbf{X},\widehat{\mathbf{X}})]$ with respect to the joint distribution $p(\mathbf{x},\hat{\mathbf{x}})$. This expected distortion is required to be no larger than some value $D$. Within the joint distribution $p(\mathbf{x},\hat{\mathbf{x}})$, the  distribution of $\mathbf{X}$, $p(\mathbf{x})$, is fixed, and the minimization is carried out over the conditional (quantizer) distribution $p(\hat{\mathbf{x}}|\mathbf{x})$. The quantizer can be deterministic, in which case $p(\hat{\mathbf{x}}|\mathbf{x})$ is a delta function for any given $\mathbf{x}$, or random, in which case $p(\hat{\mathbf{x}}|\mathbf{x})$ is a proper, non-degenerate distribution. The theory is general enough to accommodate both cases. The source coding theorem and its converse~\cite{Cover_Thomas_2006} show that $R(D)$, as defined above, is the minimum achievable rate\footnote{Rate is in \emph{bits} if $\log_2$ is used in the computation of $I(\mathbf{X};\widehat{\mathbf{X}})$.} per source symbol that results in expected distortion of at most $D$. Hence, $R(D)$ represents a fundamental bound on lossy compression, just like entropy represents a fundamental bound on lossless compression.  

In order to use rate-distortion theory to analyze lossy feature compression, we will need several additional concepts. First, let the function implemented by the DNN from input $\mathbf{X}$ to output $T$ be $f(\cdot)$, so that the DNN's output is $T=f(\mathbf{X})$. Let the mapping from the input $\mathbf{X}$ to the set of $n$ layer outputs $\mathbfcal{Y} = (\mathbfcal{Y}^{(1)}, \mathbfcal{Y}^{(2)},  ..., \mathbfcal{Y}^{(n)})$, so that $\mathbfcal{Y} = g(\mathbf{X})$, and the mapping from $\mathbfcal{Y}$ to the output $T$ be $h(\cdot)$ so that $T=h(\mathbfcal{Y})$. Note that $T=h(g(\mathbf{X}))$, so $f=h\circ g$ is the composition of $g$ and $h$. These relationships are summarized below:
\begin{singlespace}
\begin{equation}
\begin{tikzpicture}[
  inner sep=0pt,
  outer sep=0pt,
  baseline=(x.base),
]
  \tikzstyle{arrow} = [thin,->,>=stealth]
  
  \path[every node/.append style={anchor=base west}]
    (0, 0)
    \foreach \name/\code in {
      x/ \mathbf{X}~,
      tmp/\,\qquad,
      t/ ~\mathbfcal{Y}~,
      tmp/\,\qquad,
      y/ ~T%
    } {
      node (\name) {$\code$}
      (\name.base east)
    }
  ;
  \path[
    every node/.append style={
      anchor=base,
      font=\scriptsize,
    },
  ]
    (x.base) -- node[above=2.0\baselineskip] (f) {$f$} (y)
    (x.base) -- node[below=0.2\baselineskip] (g) {$g$} (t)
    (t.base) -- node[below=0.2\baselineskip] (h) {$h$} (y)
  ;
  \draw [arrow] (x) -- (t);
  \draw [arrow] (t) -- (y);
    
  \begin{scope}[
    >={Stealth[length=5pt]},
    thin,
    rounded corners=2pt,
    shorten <=.3em,
    shorten >=.3em,
  ]
    
    \def\GebArrow#1#2#3{
      \draw[->]
        (#2.south) ++(0, -.3em) coordinate (tmp)
        (#1) |- (tmp) -| (#3)
      ;%
    }
    \GebArrow{x}{f}{y}
  \end{scope}
\end{tikzpicture}
\label{eq:mappings}
\end{equation}
\end{singlespace}

We will measure distortion at the output of the model. This is one difference with respect to the conventional rate-distortion formulation, where distortion $d(\mathbf{x},\hat{\mathbf{x}})$ measures how much the quantized $\hat{\mathbf{x}}$ deviates from $\mathbf{x}$. In our case, what is relevant is how the output of the model is affected by quantization, so our distortion is formulated using the mapping to the output, as $d(f(\mathbf{x}),f(\hat{\mathbf{x}}))$. The choice of the distortion metric itself depends on the model; for regression models, squared error distortion is the natural choice, while for classification models, cross-entropy seems appropriate. The key result below, however, is agnostic to the choice of the distortion metric. Using the distortion measured at the output, we can define the set of all  conditional distributions (quantizers) $p(\hat{\mathbf{x}}|\mathbf{x})$ that achieve output distortion of no more than $D$:
\begin{equation}
    \mathcal{P}_{\mathbf{X}}(D)=\left\{p\left(\hat{\mathbf{x}}|\mathbf{x}\right)~:~\mathbb{E}\left[d\left(f(\mathbf{X}),f(\widehat{\mathbf{X}})\right)\right]\leq D\right\}.
    \label{eq:quant_X}
\end{equation}
Using this set, we can formulate the rate-distortion function for the model's input $\mathbf{X}$ as:
\begin{equation}
    R_{\mathbf{X}}(D) = \min_{p(\hat{\mathbf{x}}|\mathbf{x})~\in~\mathcal{P}_{\mathbf{X}}(D)} I(\mathbf{X};\widehat{\mathbf{X}}).
    \label{eq:RD_function_X}
\end{equation}

Since we also need to consider quantization of intermediate layers $\mathbfcal{Y}$, we analogously define the set of all conditional distributions (quantizers) $p(\hat{\mathbf{y}}|\mathbf{y})$ that achieve output distortion of no more than $D$:
\begin{equation}
    \mathcal{P}_{\mathbfcal{Y}}(D)=\left\{p\left(\hat{\mathbf{y}}|\mathbf{y}\right)~:~\mathbb{E}\left[d\left(h(\mathbfcal{Y}),h(\widehat{\mathbfcal{Y}})\right)\right]\leq D\right\}.
    \label{eq:quant_T}
\end{equation}
Note that in~(\ref{eq:quant_T}), we use $h(\cdot)$, the mapping from $\mathbfcal{Y}$ to $T$, in the expected distortion. Using this set, we formulate the rate-distortion function for the model's intermediate layers $\mathbfcal{Y}$: 
\begin{equation}
    R_{\mathbfcal{Y}}(D) = \min_{p(\hat{\mathbf{y}}|\mathbf{y})~\in~\mathcal{P}_{\mathbfcal{Y}}(D)} I(\mathbfcal{Y};\widehat{\mathbfcal{Y}}).
    \label{eq:RD_function_T}
\end{equation}
We now state and prove the lossy compression result.

\begin{theorem}
\label{thm:lossy}
For any distortion level $D\geq 0$, the minimum achievable rate for compressing $\mathbfcal{Y}$ is upper bounded by the minimum achievable rate for compressing $\mathbf{X}$, that is, 
\begin{equation}
    R_{\mathbfcal{Y}}(D) \leq R_\mathbf{X}(D).
    \label{eq:RD_bound}
\end{equation}
The necessary condition for equality is that $\mathbf{X}$ can be reconstructed exactly from $\mathbfcal{Y}$.
\end{theorem}

\begin{proof}
Let $D\geq0$ be given, and let  $p^{*}(\hat{\mathbf{x}}|\mathbf{x}) \in \mathcal{P}_{\mathbf{X}}(D)$ be the conditional distribution that achieves the rate-distortion bound for $\mathbf{X}$, i.e., solves~(\ref{eq:RD_function_X}). This conditional distribution defines a quantized representation $\widehat{\mathbf{X}}$ of the model's input $\mathbf{X}$. 

Now let us draw inputs $\mathbf{X}$ according to the data distribution $p(\mathbf{x})$, quantize $\mathbf{X}$ into $\widehat{\mathbf{X}}$ using $p^{*}(\hat{\mathbf{x}}|\mathbf{x})$ and then pass the quantized input $\widehat{\mathbf{X}}$ through $g(\cdot)$, which is the part of the model that computes the intermediate layers of interest. This will generate $\widetilde{\mathbfcal{Y}}$, which is the intermediate layers' output when the input is $\widehat{\mathbf{X}}$, as shown below.
\begin{equation}
\begin{tikzpicture}[
  inner sep=0pt,
  outer sep=0pt,
  baseline=(x.base),
]
  \tikzstyle{arrow} = [thin,->,>=stealth]
  
  \path[every node/.append style={anchor=base west}]
    (0, 0)
    \foreach \name/\code in {
      x/ \textcolor{white}{\widehat{\mathbf{X}}}\mathbf{X}~,
      tmp/\,\qquad \qquad,
      x_hat/ ~\widehat{\mathbf{X}}~,
      tmp/\,\qquad,
      t_tilde/ ~\widetilde{\mathbfcal{Y}}%
    } {
      node (\name) {$\code$}
      (\name.base east)
    }
  ;
  \path[
    every node/.append style={
      anchor=base,
      font=\scriptsize,
    },
  ]
    (x.base) -- node[below=0.2\baselineskip] (p) {~~~$p^{*}(\hat{\mathbf{x}}|\mathbf{x})$} (x_hat)
    (x_hat.base) -- node[below=0.2\baselineskip] (g) {$g$} (t_tilde)
  ;
  \draw [arrow] (x) -- (x_hat);
  \draw [arrow] (x_hat) -- (t_tilde);
\end{tikzpicture}
\label{eq:producing_T_tilde}
\end{equation}
In parallel, consider the deterministic mapping from $\mathbf{X}$ to $\mathbfcal{Y}$ defined by the model:
\begin{equation}
\begin{tikzpicture}[
  inner sep=0pt,
  outer sep=0pt,
  baseline=(x.base),
]
  \tikzstyle{arrow} = [thin,->,>=stealth]
  
  \path[every node/.append style={anchor=base west}]
    (0, 0)
    \foreach \name/\code in {
      x/ \mathbf{X}~,
      tmp/\,\qquad,
      t/ ~\mathbfcal{Y}%
    } {
      node (\name) {$\code$}
      (\name.base east)
    }
  ;
  \path[
    every node/.append style={
      anchor=base,
      font=\scriptsize,
    },
  ]
    (x.base) -- node[below=0.2\baselineskip] (g) {$g$} (t)
  ;
  \draw [arrow] (x) -- (t);
\end{tikzpicture}
\label{eq:producing_T}
\end{equation}

Together, the last two equations mean that for any \emph{particular} input $\mathbf{X}$, we can obtain a \emph{particular} $\mathbfcal{Y}$ according to~(\ref{eq:producing_T}) and a (possibly degenerate) \emph{distribution} of $\widetilde{\mathbfcal{Y}}$ according to~(\ref{eq:producing_T_tilde}). Pairing up the values of $\mathbfcal{Y}$ with the corresponding distributions of $\widetilde{\mathbfcal{Y}}$ implicitly defines a conditional distribution of $\widetilde{\mathbfcal{Y}}$ given $\mathbfcal{Y}$, which we will denote as $q(\tilde{\mathbf{y}}|\mathbf{y})$.  
In addition, let $q(\mathbf{y})$ be the distribution of $\mathbfcal{Y}$ induced by $p(\mathbf{x})$ through $\mathbfcal{Y}=g(\mathbf{X})$.

First, we show that $q(\tilde{\mathbf{y}}|\mathbf{y}) \in \mathcal{P}_{\mathbfcal{Y}}(D)$, that is to say, $q(\tilde{\mathbf{y}}|\mathbf{y})$ satisfies the distortion constraint for quantizing $\mathbfcal{Y}$.  To see this, note that $p^{*}(\hat{\mathbf{x}}|\mathbf{x}) \in \mathcal{P}_{\mathbf{X}}(D)$, which means that using $p^{*}(\hat{\mathbf{x}}|\mathbf{x})$, we have  $\mathbb{E}[d(f(\mathbf{X}),f(\widehat{\mathbf{X}}))]\leq D$.  But because 
$f(\mathbf{X}) = h(g(\mathbf{X})) = h(\mathbfcal{Y})$ and $f(\widehat{\mathbf{X}}) = h(g(\widehat{\mathbf{X}})) = h(\widetilde{\mathbfcal{Y}})$, we have 
\begin{equation}
\begin{split}
    \mathbb{E}\left[d\left(h(\mathbfcal{Y}),h(\widetilde{\mathbfcal{Y}})\right)\right] &= \sum_{(\mathbf{y},\tilde{\mathbf{y}})} \left[q(\mathbf{y}) \cdot q(\tilde{\mathbf{y}}|\mathbf{y}) \cdot d(h(\mathbf{y}),h(\tilde{\mathbf{y}}))\right] \\
    & = \sum_{(\mathbf{x},\hat{\mathbf{x}})} \left[p(\mathbf{x}) \cdot p^{*}(\hat{\mathbf{x}}|\mathbf{x}) \cdot d(f(\mathbf{x}),f(\hat{\mathbf{x}}))\right] \\
    & = \mathbb{E}\left[d\left(f(\mathbf{X}),f(\widehat{\mathbf{X}})\right)\right] \\
    &\leq D,
    \label{eq:q_distortion_constraint}
\end{split}
\end{equation}
where the second equality follows from the fact that $q(\mathbf{y})$ and $q(\tilde{\mathbf{y}}|\mathbf{y})$ are induced by $p(\mathbf{x})$ and $p^{*}(\hat{\mathbf{x}}|\mathbf{x})$, respectively, through $g(\cdot)$. Therefore, $q(\tilde{\mathbf{y}}|\mathbf{y})$ satisfies the distortion constraint for quantizing $\mathbfcal{Y}$, so $q(\tilde{\mathbf{y}}|\mathbf{y}) \in \mathcal{P}_{\mathbfcal{Y}}(D)$.

Next, we show that $I(\mathbfcal{Y}; \widetilde{\mathbfcal{Y}}) \leq I(\mathbf{X}; \widehat{\mathbf{X}})$. Note that $\mathbf{X} \to \widehat{\mathbf{X}} \to \widetilde{\mathbfcal{Y}}$ is a Markov chain based on~(\ref{eq:producing_T_tilde}), and therefore the reverse chain $\widetilde{\mathbfcal{Y}} \to \widehat{\mathbf{X}} \to \mathbf{X}$ must also be a Markov chain. We now add one more processing step to the end of this  chain, $\mathbfcal{Y}=g(\mathbf{X})$, to obtain the following Markov chain: $\widetilde{\mathbfcal{Y}} \to \widehat{\mathbf{X}} \to \mathbf{X} \to \mathbfcal{Y}$. Applying the DPI to this chain, we conclude $I(\mathbfcal{Y}; \widetilde{\mathbfcal{Y}}) \leq I(\mathbf{X}; \widehat{\mathbf{X}})$. This is true for any $p(\hat{\mathbf{x}}|\mathbf{x})$. But for the optimal $p^{*}(\hat{\mathbf{x}}|\mathbf{x})$, which achieves the rate-distortion bound for $\mathbf{X}$, and its induced distribution $q(\tilde{\mathbf{y}}|\mathbf{y})$, we have 
\begin{equation}
    I(\mathbfcal{Y}; \widetilde{\mathbfcal{Y}}) \leq I(\mathbf{X}; \widehat{\mathbf{X}}) = R_{\mathbf{X}}(D).
    \label{eq:MI_T_X_ineqiuality}
\end{equation}

Together,~(\ref{eq:q_distortion_constraint}) and~(\ref{eq:MI_T_X_ineqiuality}) show that $q(\tilde{\mathbf{y}}|\mathbf{y})$ is one distribution in $\mathcal{P}_{\mathbfcal{Y}}(D)$ that achieves mutual information $I(\mathbfcal{Y}; \widetilde{\mathbfcal{Y}}) \leq R_{\mathbf{X}}(D)$. Therefore, $R_{\mathbfcal{Y}}(D)$, which is the minimum $I(\mathbfcal{Y}; \widehat{\mathbfcal{Y}})$ over all distributions $p(\hat{\mathbf{y}}|\mathbf{y}) \in \mathcal{P}_{\mathbfcal{Y}}(D)$, cannot be larger than $R_{\mathbf{X}}(D)$:
\begin{equation}
    R_{\mathbfcal{Y}}(D) = \min_{p(\hat{\mathbf{y}}|\mathbf{y})~\in~\mathcal{P}_{\mathbfcal{Y}}(D)} I(\mathbfcal{Y};\widehat{\mathbfcal{Y}}) \; \leq \; R_{\mathbf{X}}(D).
    \label{eq:RD_inequality_T}
\end{equation}

We have therefore established the inequality in~(\ref{eq:RD_bound}). We now turn to conditions that are needed for equality  in~(\ref{eq:RD_bound}). The necessary condition for equality to hold in~(\ref{eq:RD_bound}) is that equality holds in the DPI~(\ref{eq:MI_T_X_ineqiuality}). For this to happen, we must be able to recover $\widehat{\mathbf{X}}$ from $\widetilde{\mathbfcal{Y}}$, or in general, recover $\mathbf{X}$ from $\mathbfcal{Y}$. In other words, this is a necessary condition for $R_{\mathbfcal{Y}}(D)=R_{\mathbf{X}}(D)$. But it is not a sufficient condition, because even if we had $I(\mathbfcal{Y}; \widetilde{\mathbfcal{Y}}) = R_{\mathbf{X}}(D)$ in~(\ref{eq:MI_T_X_ineqiuality}), this only holds for $q(\tilde{\mathbf{y}}|\mathbf{y})$, which is induced by $p^{*}(\hat{\mathbf{x}}|\mathbf{x})$; the minimization in~(\ref{eq:RD_inequality_T}) over all conditional distributions in $\mathcal{P}_{\mathbfcal{Y}}(D)$ may still produce a lower $I(\mathbfcal{Y}; \widehat{\mathbfcal{Y}})$. This completes the proof of the theorem.
\end{proof}

The above theorem is a lossy counterpart to Theorem~\ref{thm:lossless}, showing that in the lossy case also, latent representations are at least as compressible as the input for any given distortion level. What is interesting in this case is that even if the mapping from input $\mathbf{X}$ to latent representation $\mathbfcal{Y}$ is perfectly invertible, $\mathbfcal{Y}$ may still be more compressible than $\mathbf{X}$, because it may allow for more efficient quantization. This is different from the lossless case where invertibility of the mapping from $\mathbf{X}$ to $\mathbfcal{Y}$ meant than $\mathbfcal{Y}$ is no more compressible than $\mathbf{X}$.

%








\end{document}